\documentclass[journal]{IEEEtran}
\usepackage{amsmath}
\usepackage{eqparbox}
\usepackage[usenames, dvipsnames]{color}
\usepackage{steinmetz}
\usepackage{mathtools,amssymb,bm,mathabx,nccmath}
\usepackage{amstext}
\usepackage{amssymb}
\usepackage{graphicx}
\usepackage{color}
\usepackage{booktabs}
\usepackage{longtable}
\usepackage{multicol}
\usepackage{algorithmicx}
\usepackage[ruled]{algorithm}
\usepackage{algpseudocode}
\usepackage{algpascal}
\usepackage{algc}
\usepackage{amsfonts}
\usepackage{dsfont}
\usepackage{array}
\usepackage[most]{tcolorbox}
\usepackage{cases}
\usepackage{pgfplots}
\pgfplotsset{compat=newest}
\usetikzlibrary{spy}
\usepackage{accents}
\usepackage{etoolbox}
\usepackage{caption}
\usepackage{subfigmat}
\usepackage{xr-hyper}
\usepackage{hyperref}
\usetikzlibrary{3d}
\usepackage{amsthm}
\DeclareCaptionLabelSeparator{periodspace}{.\quad}
\usetikzlibrary{positioning}
\usepackage{tikz-3dplot}
\usetikzlibrary{matrix}
\usepackage{float}
\usepackage{cite}
\usepackage{epstopdf}

\theoremstyle{remark}

\newcommand\ASTART{\bigskip\noindent\begin{minipage}[b]{0.5\linewidth}}
	
	\newcommand\AENDSKIP{\end{minipage}\bigskip}
\newcommand\AEND{\end{minipage}}
\ifCLASSOPTIONcaptionsoff
\usepackage[nomarkers]{endfloat}
\let\MYoriglatexcaption\caption
\renewcommand{\caption}[2][\relax]{\MYoriglatexcaption[#2]{#2}}
\fi
\theoremstyle{plain}
\newtheorem{thm}{\textbf{Theorem}}
\newtheorem{lem}{\textbf{Lemma}}
\newtheorem{prop}{\textbf{Proposition}}

\theoremstyle{definition}
\newtheorem{defn}{\textbf{Definition}}

\newtheorem{rem}{\textbf{Remark}}

\newcommand*{\rom}[1]{\expandafter\@slowromancap\romannumeral #1@}
\def\change{black}
\def\chang{black}
\newcommand{\RN}[1]{%
\textup{\uppercase\expandafter{\romannumeral#1}}%
}

\usepackage{standalone}
\graphicspath{{./Figures/}}
\usepackage{times}
\usepackage[T1]{fontenc}
\usepackage[english]{babel}
\usepackage{graphicx}
\usepackage{amsmath, amsfonts, dsfont,amssymb, graphicx, array, tabularx, booktabs,multicol}
\usepackage{amsthm}
\usepackage{epsf,epsfig}
\usepackage{setspace}
\usepackage{subfigure}

\usepackage{multirow}
\usepackage{url}
\usepackage{color}
\usepackage[nolist,printonlyused]{acronym}      % Acronym
\usepackage{comment}
\usepackage{cite}
\usepackage{bm}
\usepackage{tikz}
\usetikzlibrary{decorations.pathreplacing}
\usepackage[labelfont=bf,font=footnotesize]{caption}

\usepackage{hyperref}
\usepackage{mathtools}

\usepackage{blindtext}

\usepackage{stfloats}

\newcommand{\gf}[1]{\textcolor{black}{{#1}}}

\newcommand{\mx}[1]{\mathbf{#1}}
\def\change{black}
\def\chang{black}
\newcommand{\bs}[1]{\boldsymbol{#1}}

\definecolor{amber}{rgb}{1.0, 0.49, 0.0}
\definecolor{ao}{rgb}{0.0, 0.5, 0.0}

\def\R2#1{\textcolor{black}{#1}}
\def\R3#1{\textcolor{black}{#1}}

\definecolor{copperrose}{rgb}{0.6, 0.4, 0.4}
\definecolor{azure}{rgb}{0.0, 0.5, 1.0}
\definecolor{ashgrey}{rgb}{0.7, 0.75, 0.71}
\definecolor{chestnut}{rgb}{0.8, 0.36, 0.36}
\definecolor{airforceblue}{rgb}{0.36, 0.54, 0.66}
\definecolor{cadmiumorange}{rgb}{0.93, 0.53, 0.18}
\definecolor{bleudefrance}{rgb}{0.19, 0.55, 0.91}
\definecolor{carolinablue}{rgb}{0.6, 0.73, 0.89}
\definecolor{blue(ncs)}{rgb}{0.0, 0.53, 0.74}
\definecolor{dodgerblue}{rgb}{0.12, 0.56, 1.0}
\definecolor{cssgreen}{rgb}{0.0, 0.5, 0.0}
\definecolor{cadmiumgreen}{rgb}{0.0, 0.42, 0.24}
\definecolor{cadmiumorange}{rgb}{0.93, 0.53, 0.18}
\definecolor{amaranth}{rgb}{0.9, 0.17, 0.31}
\definecolor{bluegray}{rgb}{0.4, 0.6, 0.8}
\definecolor{cerulean}{rgb}{0.0, 0.48, 0.65}
\definecolor{ceil}{rgb}{0.57, 0.63, 0.81}
\definecolor{antiquefuchsia}{rgb}{0.57, 0.36, 0.51}
\definecolor{bronze}{rgb}{0.8, 0.5, 0.2}
\definecolor{carrotorange}{rgb}{0.93, 0.57, 0.13}
\definecolor{coolgrey}{rgb}{0.55, 0.57, 0.67}
\definecolor{corn}{rgb}{0.98, 0.93, 0.36}
\definecolor{frenchbeige}{rgb}{0.65, 0.48, 0.36}
\definecolor{dandelion}{rgb}{0.94, 0.88, 0.19}
\definecolor{cadet}{rgb}{0.33, 0.41, 0.47}

\renewcommand{\triangleq}{\mathbin{\setstackgap{S}{0pt}\stackMath\Shortstack{\smalltriangleup\\ =}}}
\usepackage[usestackEOL]{stackengine}

\begin{document}
%
%\onecolumn
% paper title
% can use linebreaks \\ within to get better formatting as desired
\title{Timely and Painless Breakups: Off-the-\gf{G}rid Blind Message Recovery and Users' Demixing}
%Breaking up without breaking down: Off-the-grid Blind Deconvolution and Demixing
\author{Sajad~Daei, Saeed~Razavikia, Mikael Skoglund, Gabor Fodor, Carlo Fischione
 	\thanks{All the authors are with the School of Electrical Engineering and Computer Science KTH Royal Institute of Technology, Stockholm, Sweden (e-mail: \{sajado, sraz, skoglund, gaborf, carlofi\}@kth.se). M. Skoglund and C. Fischione are also with Digital Futures and SweWin of KTH. 
    Gabor Fodor is also with Ericsson Research, Sweden.
     This work was partly supported by Digital Futures. G. Fodor was also supported by the Swedish Strategic Research (SSF) grant for the FUS21-0004 SAICOM project and the 6G-Multiband Wireless and Optical Signalling for Integrated Communications, Sensing and Localization (6G-MUSICAL) EU Horizon 2023 project, funded by the EU, Project ID: 101139176. S.Daei is supported by STACEY project, Digital Futures.
     S. Razavikia was supported by the Wallenberg AI, Autonomous Systems and Software Program (WASP).
}
 
    \thanks{A preliminary version of this work was presented in part at the IEEE Global Communications Conference, Kuala Lumpur, Malaysia, December 2023, which appears in this manuscript as reference~\cite{Razavikia2023blindDE}.}

}

% make the title area
\maketitle

\begin{abstract}
\gf{The Internet of Things interconnects billions of devices and forms}
a vast network where users sporadically transmit short messages through multi-path wireless channels. 
These channels are characterized by the superposition of a small number of scaled and delayed copies of Dirac spikes. At the receiver, the observed signal is a sum of these convolved signals, and the task is to find the amplitudes, continuous-indexed delays, and transmitted messages from a single signal. 
This task is inherently ill-posed without additional assumptions on the channel or messages. In this work, we assume the channel exhibits sparsity in the delay domain and that \gf{independent and identically distributed} random linear encoding is applied to the messages at the devices. Leveraging these assumptions, we propose a semidefinite programming optimization capable of simultaneously recovering both messages and the delay parameters of the channels from only a single received signal. Our theoretical analysis establishes that the required number of samples at the receiver scales proportionally to the sum-product of sparsity and message length of all users, aligning with the degrees of freedom {\color{\change}in the lifting-type optimization frameworks}. Numerical experiments confirm the efficacy of the proposed method in accurately estimating closely-spaced delay parameters and recovering messages.
% We consider a scenario where there are $r$ transmitters and only one receiver. The waveform corresponding to each transmitter is convolved with a  continuous-time signal formed by the superposition of a small number of scaled and delayed copies of Dirac spikes. The receiver observes the sum of all these convolved signals and does not know the amplitudes, continuous-indexed delays and even the transmitted waveforms, yet intends to find these unknowns from only one single vector. We refer to this problem as \textit{off-the-grid blind deconvolution and demixing}. This problem has numerous applications, specially in multiuser channel estimation and equalization in wireless communications, super resolution microscopy, astronomy, array self-calibration, and the well-known Internet of Things. While it is of great interest to recover the waveforms and the spike signals from only one single receiver, without further assumptions, this problem is highly challenging and ill-posed. In this work, we make two assumptions: the former is that each waveform belongs to a known low-dimensional subspace while the latter puts a separation condition on the Dirac spikes. Under these assumptions, we propose a semidefinite programming that recovers the $r$ transmitted functions as well as the parameters of the spike signals corresponding to different transmitters, simultaneously from only one received signal. We also obtain explicit bounds on the required number of measurements for successful recovery. Our proposed method is verified by numerical experiments.
\end{abstract}

% Note that keywords are not normally used for peerreview papers.

% For peer-reviewed papers, you can put extra information on the cover
% page as needed:
% \ifCLASSOPTIONpeerreview
% \begin{center} \bfseries EDICS Category: 3-BBND \end{center}
% \fi
%
% For peerreview papers, this IEEEtran command inserts a page break and
% creates the second title. It will be ignored for other modes.
\IEEEpeerreviewmaketitle

\section{Introduction}
\IEEEPARstart{T}{h}e proliferation of the Internet of Things (IoT) promises to interconnect billions of wireless devices, \gf{enabled by the capabilities of 5G Advanced and the evolving 6G wireless systems.}
% GF: The current number of IoT devices is already more than 10 billions, with 4G and 5G.
%a scale beyond the capabilities of current 5G wireless systems. 
As the IoT landscape expands, one of the main challenges facing the development of future communication systems lies in efficiently managing the massive number of IoT devices and their sporadic traffic. In fact, these devices are inactive most of the time but regularly access the network for minor updates without human interaction in an uncoordinated way \cite{wunder20145gnow,wunder2015sparse}. A typical solution in these scenarios is to exchange some sort of information between the transmitter and the receiver, known as pilot (training) signals, in order to first estimate the channel and then decode the messages of the devices. However, this makes a severe waste of resources and fails to align with the scalability requirements of the IoT. Aside from this, in dynamic communications channels, the channel rapidly changes and the channel between pilot and data time slots substantially ages \cite{Fodor:23, Daei:24}. 
This, in turn, necessitates %in such scenarios 
to avoid \gf{transmitting pilot signals whose length exceeds those of the short messages sent by IoT devices.} 
%each time a pilot signal that its length might even exceed the actual short messages. 
Thus, the fundamental question \gf{that arises is}: is it possible to design a system that minimizes overhead transmission while efficiently delivering sporadic data from a plethora of IoT devices?

In mathematical terms, we are dealing with the following problem. There are $r$ IoT devices transmitting signals towards a receiver as shown in Figures~\ref{fig:system} and \ref{fig:applications(a)}. The waveform corresponding to the $i$-th transmitter at time $t$ is denoted by $x_i(t)$. The acquired waveform at the receiver denoted by $y(t)$ is considered as a superposition of returns from $r$ transmitters, where the return from each transmitter is the convolution of a channel $h_i(t)$ with $x_i(t)$ given by:
%--------------
 \begin{align}\label{eq.mymodel}
 y(t)=\sum_{i=1}^{r}h_i(t)\ast x_i(t),
 \end{align}
 %--------------
 where 
 %--------------
 \begin{align}\label{eq:channel}
 h_i(t)=\sum_{l=1}^{s_i}c_l^i\delta(t-\overline{\tau}_l^i),
 \end{align}
 %--------------
 is the channel corresponding to the $i$-th transmitter that is formed of $s_i$ paths with continuous-valued delays $\overline{\tau}_l^i\in [0,{\rm T}_{\max})$ and complex-valued amplitudes $c_l^i$. Here, ${\rm T}_{\max}$ is the duration over which the signal is observed at the receiver. Our goal is to estimate the set of delays and amplitudes of the channel $\{h_i(t)\}_{i=1}^r$ as well as the unknown waveforms $\{x_i(t)\}_{i=1}^r$ from the received signal $y(t)$. We refer to this problem as \textit{off-the-grid blind deconvolution and demixing} (OBDD). This model is illustrated in Figure~\ref{fig:system}. In fact, we need to deconvolve the unknown transmitted signals and the channels and simultaneously demix each contribution $h_i(t)\ast x_i(t)$ from the sum  $\sum_{i=1}^{r}h_i(t)\ast x_i(t)$. Moreover, the delay parameters $\overline{\tau}_l^i$s are not confined to lie on a predefined domain of grids and can take any arbitrary continuous values in $[0,{\rm T}_{\max})$. Without having any assumptions on the transmitted signals and the channels, this problem is highly ill-posed. 
 In this paper, we provide some natural assumptions on the transmitted signal of users making the OBDD problem tractable. Moreover, we specify theoretically the required number of samples that the receiver should take to ensure simultaneous message recovery and delay estimation.

\begin{figure*}[!t]
    \centering

\scalebox{1}{

% \tikzset{every picture/.style={line width=0.75pt}} %set default line width to 0.75pt        

\begin{tikzpicture}[x=0.75pt,y=0.75pt,xscale=1]
%uncomment if require: \path (0,175); %set diagram left start at 0, and has height of 175

% %Rounded Rect O dotted 

\draw (-5,115) node    {\footnotesize $\mathbf{F}^{-1}$};

\draw (-5,-5) node    {\footnotesize $\mathbf{F}^{-1}$};

\draw (-35,100) node    {\footnotesize $\mathbf{x}_1$};

\draw (-35,-20) node    {\footnotesize $\mathbf{x}_r$};

\draw [-stealth](-25,100) -- (10,100);

\draw [-stealth](-25,-20) -- (10,-20);

\draw (60,45) node  [font=\Huge] {$\vdots$};

\draw (-88,45) node  [font=\Huge] {$\vdots$};

\draw (180,45) node  [font=\Huge] {$\vdots$};

\draw (-60,115) node    {\footnotesize $\mathbf{B}_1$};

\draw (-60,-5) node    {\footnotesize $\mathbf{B}_r$};

\draw [-stealth](-75,100) -- (-45,100);

\draw [-stealth](-75,-20) -- (-45,-20);

%--------------- h1 --------------
    \begin{axis}[
     title = {\footnotesize $h_1(t)$},
    xlabel = {$t$},
    x label style={at={(axis description cs:1.3,0)},anchor=north},
width=3cm,
height=2.5cm,
label style={font=\tiny},
tick label style={font=\tiny}, 
ytick=\empty,
 yticklabels=none,
 axis line style={draw=none},
 % axis x line*=left,
 xtick=\empty,
  xticklabels=none,
% axis y line*={draw=none},
  % xtick={0,1,2,3,4},
 xshift=5.7cm,yshift=2cm,
 ]
\addplot[ycomb, 
     mark=triangle,
     mark options={scale=0.5, fill=cadmiumgreen},
    % only marks,
    % color=cadmiumgreen,
]
coordinates { (41,49.6) (51,90) (77,80) (95,114)
    };
    % table[x=t,y=Original_Signal_f1]
    % {Data/low_dim_signal.dat}; 
\end{axis}

\draw [-stealth](215,76) -- (290,76);

\draw (220,88) node    {\tiny $c_1^1$};
\draw (227,105) node    {\tiny $c_2^1$};
\draw (250,100) node    {\tiny $c_3^1$};
\draw (263,117) node    {\tiny $c_4^1$};

\draw (220,70) node    {\tiny $\tau_1^1$};
\draw (229,70) node    {\tiny $\tau_2^1$};
\draw (250,70) node    {\tiny $\tau_3^1$};
\draw (263,70) node    {\tiny $\tau_4^1$};

%--------------- h2 --------------
    \begin{axis}[
     title = { \footnotesize $h_r(t)$},
   xlabel = {$t$},
    x label style={at={(axis description cs:1.3,0)},anchor=north},
width=3cm,
height=2.5cm,
label style={font=\tiny},
tick label style={font=\tiny}, 
ytick=\empty,
 yticklabels=none,
 axis line style={draw=none},
 % axis x line*=left,
 xtick=\empty,
  xticklabels=none,
% axis y line*={draw=none},
  % xtick={0,1,2,3,4},
 xshift=5.7cm,yshift=-1.2cm,
 ]
\addplot[ycomb, 
     mark=triangle,
     mark options={scale=0.5, fill=cadmiumgreen},
    % only marks,
    % color=cadmiumgreen,
]
coordinates { (31,86) (25,98) (60,112)
    };
    % table[x=t,y=Original_Signal_f1]
    % {Data/low_dim_signal.dat}; 
\end{axis}

\draw (220,-19) node    {\tiny $c_1^r$};
\draw (227,-33) node    {\tiny $c_2^r$};
\draw (262,-5) node    {\tiny $c_3^r$};

\draw (220,-55) node    {\tiny $\tau_1^r$};
\draw (230,-55) node    {\tiny $\tau_2^r$};
\draw (262,-55) node    {\tiny $\tau_3^r$};

\draw [-stealth](215,-46) -- (290,-46);

\draw (170,100) node    {\footnotesize $\ast$};

\draw [-stealth](175,95) -- (220,50);

\draw (170,-20) node    {\footnotesize $\ast$};

\draw [-stealth](175,-16) -- (220,28);

\draw (225,38) node {\Large $\bigoplus$};

\draw[dash pattern={on 0.84pt off 2.51pt}] (200,40) ellipse (10 and 40);

%--------------- PAM- 5 --------------
%QAM 4

%PAM 
\draw [-stealth](-180,100) -- (-100,100);
\draw (-140,140) node    {{\footnotesize ASK-$4$, $k_1 = 5$}};

\draw  (-170,99) node     {\color{blue(ncs)}\Huge $\cdot$};
\draw (-150,99) node     {\color{carrotorange}\Huge $\cdot$};
\draw (-130,99) node     {\color{coolgrey}\Huge $\cdot$};
\draw (-110,99) node      {\color{chestnut}\Huge $\cdot$};
% \draw (-115,99) node      {\color{frenchbeige}\Huge $\cdot$};

\draw (-170,110) node     {\tiny $00$};
\draw (-150,110) node     {\tiny $01$};
\draw (-130,110) node     {\tiny $10$};
\draw (-110,110) node      {\tiny $11$};
% \draw (-115,110) node      {\tiny $4$};

%--------------- f1 --------------

\draw (-85,145) node      {\footnotesize $\mathbf{f}_1$};

% Rectangle 5 
\draw  [fill= {blue(ncs)} ,fill opacity=0.60 ] (-84,80) -- (-84,90) -- (-90,90) -- (-90,80) -- cycle ;
% Rectangle 4
\draw  [fill={carrotorange}  ,fill opacity=0.7 ] (-84,90) -- (-84,100) -- (-90,100) -- (-90,90) -- cycle ;
% Rectangle 3
\draw  [fill={coolgrey}  ,fill opacity=1 ] (-84,100) -- (-84,110) -- (-90,110) -- (-90,100) -- cycle ;
% Rectangle 2
\draw  [fill={chestnut}  ,fill opacity=0.75 ] (-84,110) -- (-84,120) -- (-90,120) -- (-90,110) -- cycle ;
% Rectangle 1 of size = (6*10) 
\draw  [fill={blue(ncs)}  ,fill opacity=0.6 ] (-84,120) -- (-84,130) -- (-90,130) -- (-90,120) -- cycle ;

%--------------- PAM 4 --------------

\draw [-stealth](-180,-20) -- (-100,-20);

\draw (-140,20) node    {{\footnotesize ASK-$4$, $k_r = 4$}};

\draw (-170,-21) node     {\color{blue(ncs)}\Huge $\cdot$};
\draw (-150,-21) node     {\color{carrotorange}\Huge $\cdot$};
\draw (-130,-21) node     {\color{coolgrey}\Huge $\cdot$};
\draw (-110,-21) node      {\color{chestnut}\Huge $\cdot$};

\draw (-170,-10) node     {\tiny $00$};
\draw (-150,-10) node     {\tiny $01$};
\draw (-130,-10) node     {\tiny $10$};
\draw (-110,-10) node     {\tiny $11$};

%--------------- f2 --------------

\draw (-85,15) node      {\footnotesize $\mathbf{f}_r$};
%Shape: Rectangle 1 
\draw  [fill={carrotorange}  ,fill opacity=0.7 ] (-84,-40) -- (-84,-30) -- (-90,-30) -- (-90,-40) -- cycle ;
%Shape: Rectangle 2 
\draw  [fill={chestnut}  ,fill opacity=0.75 ] (-84,-30) -- (-84,-20) -- (-90,-20) -- (-90,-30) -- cycle ;
%Shape: Rectangle 3 
\draw  [fill={blue(ncs)}  ,fill opacity=0.6 ] (-84,-20) -- (-84,-10) -- (-90,-10) -- (-90,-20) -- cycle ;
%Shape: Rectangle 4 
\draw  [fill={chestnut}  ,fill opacity=0.75 ] (-84,-10) -- (-84,0) -- (-90,0) -- (-90,-10) -- cycle ;

%--------------- x1 --------------
    \begin{axis}[
    title = {\footnotesize$x_1(t)$},
   xlabel = {$t$},
width=5cm,
height=3cm,
label style={font=\tiny},
tick label style={font=\tiny}, 
axis x line*=left,
axis y line*=left,
 xshift=0.8cm,yshift=2cm,
 ]
\addplot[
     % mark=*,
    % only marks,
    color=bluegray,
    line width=0.75pt,
]
    table[x=Nrow,y=Coded_Signal_x1]
    {signal_combined.dat}; 
\end{axis}

%--------------- x2 --------------
    \begin{axis}[
    title = {\footnotesize $x_r(t)$},
   xlabel = {$t$},
width=5cm,
height=3cm,
label style={font=\tiny},
tick label style={font=\tiny}, 
axis x line*=left,
axis y line*=left,
 xshift=0.8cm,yshift=-1cm,
 ]
\addplot[
     % mark=*,
    % only marks,
    line width=0.75pt,
    color=cadmiumorange,
]
    table[x=Nrow,y=Coded_Signal_x2]
    {signal_combined.dat}; 
\end{axis}

\draw [-stealth](238,37) -- (260,37);

\draw (395,45) node    {\footnotesize $\mathbf{F}$};

\draw [-stealth](380,37) -- (405,37);

\draw (425,37) node    {\footnotesize ${Y}(f)$};

%--------------- y --------------
    \begin{axis}[
    title = {\footnotesize ${y}(t)$},
   xlabel = {$t$},
width=4cm,
height=3cm,
label style={font=\tiny},
tick label style={font=\tiny}, 
axis x line*=left,
axis y line*=left,
 xshift=7.6cm,yshift=0.35cm,
 ]
\addplot[
     % mark=*,
    % only marks,
    color=antiquefuchsia,
    line width=0.75pt,
]
    table[x=Nrow,y=Received_Signal]
    {signal_combined.dat}; 
\end{axis}

\draw (480,37) node    {\footnotesize $\mathbf{y}$};

%Straight Lines 
\draw    (440,36) -- (450,36) ;
\draw    (460,36) -- (470,36) ;
\draw    (450,36) -- (460,43) ;
% %Curve Lines  
\draw[-stealth]    (450,45) .. controls (458,40) and (458,40) .. (465,30) ;

% Text Node
\draw (460,55) node     {\footnotesize $1/{\rm B}_{\max}$};

\end{tikzpicture}

}
    
    \caption{An illustration of the mathematical model of OBDD problem. User $i$ transmits a signal $x_i(t)$ to the receiver through the channel $h_i(t)$. The signal transmitted by the $i$-th user comprises a message vector, represented by $\mx{f}_i \in \mathbb{C}^{k_i \times 1}$, where the elements are derived from a constellation (e.g., amplitude-shift keying modulation (ASK)). This is followed by a redundant encoding matrix $\mx{B}_i \in \mathbb{C}^{N \times k_i}$. The channel corresponding to the $i$-th user denoted by $h_i(t)$ is characterized by a sparse number of scaled and delayed Dirac spikes. The receiver observes the contributions of all users in the signal $y(t)$ and takes $N$ samples of its Fourier transform at a rate of $\frac{1}{{\rm B}_{\max}}$ where ${\rm B}_{\max}$ denotes the shared bandwidth utilized by all users. These samples are then collected into a vector denoted by $\mx{y} \in \mathbb{C}^{N \times 1}$.}
    \label{fig:system}
\end{figure*}

 \subsection{Related \gf{W}orks}
Problems of the type \eqref{eq.mymodel} appear in many applications in applied science and engineering, see, e.g., \cite{chi2016guaranteed,wakinsuperres,liu2009time,three-dimensionalsuper,cellfreemassive,chi2013compressive,vetterli2010multichannel,ling2015self,ling2017blind,li2019stable,li2015off,flinth2018sparse}. In the single user case where $r=1$, when the delay parameter~\footnote{In our model in \eqref{eq.mymodel}, the only parameter is the path delays in the channels. Nevertheless, it is straightforward to incorporate additional continuous parameters such as Doppler frequency, angles of arrival and departures as well.} lies in a predefined domain of grids, and after taking the Fourier transform of \eqref{eq.mymodel} and sampling, Equation~\eqref{eq.mymodel} can be expressed as 
%--------------
\begin{align}
    \mx{F}\mx{y}=\mx{F}\mx{x}\odot \mx{F}\mx{h},
\end{align}
%--------------
where $\mx{h}\in\mathbb{C}^{N\times 1}$ is a sparse vector, $\mx{x}\in\mathbb{C}^{N\times 1}$ is the transmitted vector, $\mx{F}\in\mathbb{C}^{N\times N}$ is the discrete Fourier transform and $\odot$ is the point-wise product operator. The blind deconvolution task then refers to the recovery of $\mx{x}$ and $\mx{h}$ from their point-wise product in the frequency domain i.e, $\mx{F}\mx{y}$. The key idea is known as \gf{the} \textit{lifting trick} \gf{that} transforms the inherent bilinear mapping of the blind deconvolution problem into a linear mapping in the outer product of $\mx{h}$ and $\mx{x}$, i.e., $\mx{h}\mx{x}^{\mathsf{H}}$ \cite{ahmed2013blind,ling2015self}. By exploiting different features of $\mx{h}\mx{x}^{\mathsf{H}}$ such as low-rank, sparse, block-sparse, one can then apply the well-established methods in compressed sensing \cite{candes2006robust,daei2019error,daei2019living,daei2019exploiting,daei2019distribution,daei2018sample}, and matrix recovery \cite{candes2012exact,recht2010guaranteed,ardakani2022multi,razavikia2019reconstruction} to the blind convolution problem. For instance, since the matrix $\mx{h}\mx{x}^{\mathsf{H}}$ is of rank one, \cite{ahmed2013blind} proposes nuclear norm minimization to promote low-rankness and shows that $\mathcal{O}(k)$ number of samples is sufficient to recover both $\mx{h}$ and $\mx{x}$ in the case that both $\mx{h}$ and $\mx{x}$ lie in $k<N$-dimensional subspaces.
By assuming that the channel vector $\mx{h}\in\mathbb{C}^{N}$ is sparse i.e, $\|\mx{h}\|_0\le s \ll N$ and that $\mx{x}$ lives in a lower dimensional subspace with dimension $k\ll N$, it has been shown in \cite[Theorem 3.1]{ling2015self} that the blind deconvolution problem can be solved using $\ell_1$ minimization with $\mathcal{O}(s k) $ number of samples. It has also been demonstrated in \cite{li2016identifiability} that either $\mx{x}$ or $\mx{h}$ must adhere to subspace or sparsity assumptions for the blind convolution problem to be identifiable up to a scaling factor.

In the multi-user case where each user has a signal to transmit (say e.g, $\mx{x}_i\in\mathbb{C}^N$) and a channel (say e.g., $\mx{h}_i\in\mathbb{C}^{N}$), the receiver in the frequency domain observes the following signal:
%--------------
\begin{align}
    \mx{F}\mx{y}=\sum_{i=1}^r \mx{F}\mx{x}_i\odot \mx{F}\mx{h}_i.
\end{align}
%--------------
 The task of recovering $\mx{x}_i$ and $\mx{h}_i$ at the recovery is called deconvolution and demixing that has been investigated in \cite{ling2017blind,jung2017blind,flinth2018sparse,strohmer2019painless,mccoy2013achievable}. When $\mx{h}_i\in\mathbb{C}^{N\times 1}$ is an $s$-sparse vector and $\mx{x}_i$ lies in an $k_i$-dimensional subspace, the resulting matrix $\mx{h}_i\mx{x}_i^{\mathsf{H}}$ is a block-sparse matrix and  it has been shown in \cite{flinth2018sparse} that the recovery of both $\mx{x}_i$s and $\mx{h}_i$s is possible by solving $\ell_{1,2}$ minimization problem as long as the number of samples satisfies $\mathcal{O}(k_+\sum_i s_i)$. In our settings, $k_i$ is indeed the message length of user $i$, and the maximum and minimum message lengths are denoted by $k_+\triangleq\max_i k_i$ and $k_{-}\triangleq \min_{i} k_i$, respectively. Leveraging the low-rank structure of the matrices $\mx{h}_i\mx{x}_i^{\mathsf{H}}$, \cite{ling2017blind} has shown that the sum of nuclear norm minimization can solve the blind deconvolution and demixing problem using $\mathcal{O}(r^2\max(k, N))$ samples in the case that all $\mx{x}_i$s lie in the same $k$-dimensional subspace. It is also shown in \cite{jung2017blind} that the latter bound can be improved to have $\mathcal{O}(r(k\log^2(k)+N))$ samples by solving the same problem. When $\mx{x}_i$ is known, and the unknown delay parameters of $\mx{h}_i$s lie on a predefined grid, the problem is called compressive demixing, which is studied in, e.g., \cite{mccoy2013achievable}.

%---------------------
\input{Fig_Apps}
%---------------------

 The aforementioned sample complexity bounds hold only when the delay parameter lies in a predefined domain of grids while in practice, the delay parameter can take any arbitrary values. The difference between the ground-truth values and the assigned parameter on the grid is known as basis mismatch and can substantially degrade the performance of grid-based deconvolution and demixing methods \cite{chi2020harnessing}. The initial idea that considers the continuous sparsity traces back to \cite{candes2014towards} where the authors propose a measure referred to as the total variation measure (TVM), which is a continuous extension of $\ell_1$ norm to the continuous domain. A general definition of TVM is introduced in \cite{chandrasekaran2012convex}, where the authors define a function known as an atomic norm, which is a convex function that promotes the number of atoms required to represent a given signal in an atomic set. Note that the atoms in TVM are Dirac spikes. \cite{candes2014towards} also proposes a convex optimization problem known as TVM optimization or atomic norm minimization (ANM) that minimizes the number of required spikes to represent a given continuous signal in the time domain subject to some measurements in the frequency domain. There are several works that have applied the approach of \cite{candes2014towards} to different signal models such as continuous sparsity in frequency domain \cite{tang2013compressed}, continuous group-sparsity \cite{yang2016exact,li2015off,safari2021off,fernandez2016super}, two-dimensional super-resolution \cite{chi2014compressive,valiulahi2019two}.\\ When $\mx{x}_i$ is known, and the delay parameter in \eqref{eq:channel} is continuous-valued, then OBDD reduces to the continuous demixing problem, which is investigated in several works such as \cite{li2019stable,maskan2023demixing,fernandez2018demixing,seidi2022novel}. \cite{li2019stable} considers the same model as \eqref{eq.mymodel} in the case of $r=2$ and that $\mx{x}_i$s are known and show that $\mathcal{O}(s_+^2\log(s_1+s_2))$ is sufficient for recovering $\mx{h}_i$s and the continuous-valued delay parameters inside $\mx{h}_i$s. Here, $s_+\triangleq \max_i s_i$ is the maximum number of paths for the channels corresponding to all users. In single-user case where $r=1$ and both $\mx{x}$ and $\mx{h}$ are unknown, the problem simplifies to off-the-grid blind deconvolution problem which is investigated in \cite{chi2016guaranteed,wakinsuperres,sayyari2020blind}. In \cite{chi2016guaranteed}, it has been shown that applying a version of ANM can lead to successfully recovering $\mx{h}$ and $\mx{x}$ as long as the number of samples satisfies $\mathcal{O}(s^2 k^2)$. {\color{\change}The derived bound is quadratically proportional to $s k$, which we refer to as the matrix-valued atomic degrees of freedom (DoF) for the ANM problem involving matrix-valued atoms (see Remark \ref{rem.dof} for more details)}. This bound is improved in \cite{wakinsuperres} where the authors show that $\mathcal{O}(sk)$ samples are sufficient for the ANM to recover the unknowns. The authors of \cite{sayyari2020blind} found a general optimization framework to handle multiple-input single-output systems which aligns with the OBDD problem in radar applications. However, they do not provide any performance analysis for their method. 
 {\color{\chang}In \cite{vargas2023dual,jacome2024multi}, the optimization framework introduced in \cite[Eqs. 9,17]{sayyari2020blind} is adapted to tackle the OBDD problem within the context of integrated sensing and communications (ISAC) for a two-transmitter scenario. Under the assumption that the channel amplitudes are known and equal to one,
 \cite{vargas2023dual} derives a sample complexity bound of order $\mathcal{O}(s_+ k)$ where $k=k_1=k_2$.}
 
 % The papers \cite{vargas2023dual,jacome2024multi} adapted the optimization framework from \cite{sayyari2020blind} to address the OBDD problem within the context of integrated sensing and communications (ISAC). In their performance analysis, Vargas et al. \cite{vargas2023dual} extended the work of \cite{wakinsuperres} to scenarios involving $r=2$ transmitters. Their approach includes deriving a concentration bound for a full matrix by leveraging the concentration properties of its sub-matrices. This technique provides valuable insight but can also be challenged since it  results in a sample complexity of $\mathcal{O}(s_+ k_+)$ for an arbitrary number of transmitters. Although informative, this sample complexity result may not fully align with the problem’s true DoF.}
 % In the case of non-convex strategies for single user, \cite{lee2017near} proposed an alternative minimization approach that under some additional constraints on $\mx{x}$ and $\mx{h}$, succeeds with high probability when the number of samples satisfies $\mathcal{O}(s+k)$. 
% In the single user case, it is proved in \cite{kech2017optimal} that using non-convex optimization, one can
% \cite{kech2017optimal}\cite{lee2017near}
 
While the mixture model provided in Equation \eqref{eq.mymodel} finds applications across diverse fields—including audio processing \cite{liu2009time}, neuroscience \cite{neuro}, astronomy \cite{astro}, the IoT \cite{iot1}, super-resolution single-molecule imaging \cite{three-dimensionalsuper,stochasticoptimal}, cell-free massive \gf{multiple-input multiple-output} systems \cite{cellfreemassive}, 
multi-user multi-path channel estimation \cite{chi2013compressive,multipathdcma2001,wang1998blind,liang2001downlink}, and blind calibration in multi-channel sampling systems \cite{vetterli2010multichannel}—the performance guarantee of OBDD has not been explored in the literature. Specifically, the amount of resources required for the OBDD problem to simultaneously decode messages and recover continuous-valued parameters has not been derived.

% in \cite{ahmed2013blind,ling2015self}.
%  Under the known subspace assumption for both the channel and the transmitted signal, it is then shown in \cite{ahmed2013blind} that the problem can be solved using nuclear norm minimization with $\mathcal{O}(k)$ number of samples. Here $k$ is the dimension of the subspaces.
%  In \cite{ling2015self}, the authors have shown that one can solve blind deconvolution problem by exploiting sparsity of the resulting matrix rather than low-rankness by solving $\ell_1$ minimization problem. They show that the recovery of both channel and the transmitted signal is possible by using $\mathcal{O}(s k)$ samples under channel sparsity condition and subspace assumption on the transmitted signal \cite[Theorem 3.1]{ling2015self}. Here, $s$ is the  sparsity level (number of multi-path components in the channel) and $k$ is the subspace dimension. 

 \subsection{Contributions and Key Differences \gf{to} Prior Works}

 To make the OBDD problem tractable, we introduce assumptions regarding the sparsity of the channel in the delay domain \cite{cheng2013channel}. Specifically, we assume that $s_i$ is sufficiently small and that each user transmits a sequence generated by linearly encoded unknown messages. Notably, we assume known codebooks for users and that the unknown messages belong to predefined modulation alphabets. Furthermore, we assume that all transmitted signals $x_i(t)$ from users $i=1,...,r$ are band-limited with a maximum bandwidth ${\rm B}_{\max}$. With these assumptions, we establish a tractable convex optimization framework for recovering the messages of each user and the channel delay parameters. Furthermore, we provide a theoretical framework to obtain the number of samples that one needs to take in the frequency domain to ensure exact message recovery and delay estimation. Equivalently, this sample complexity provides the required amount of resources, i.e., ${\rm B}_{\max}{T}_{\rm max}$ one needs for off-the-grid blind message recovery and users' demixing. 
 The obtained sample complexity is $\mathcal{O}(\sum_{i=1}^r s_ik_i\frac{k_+}{k_-})$.
 With the assumption that the message length of users are not varying too much (i.e., $\frac{k_+}{k_-}$ is small), the proposed bound is precisely proportional to the degrees of freedom in the lifting-type matrix recovery problems. The proposed sample complexity bound simplifies to the sample complexity bound $\mathcal{O}(s k)$ in the single user case ($r=1$) provided in \cite{wakinsuperres}.   \\ However, it is important to emphasize that there are significant differences in theoretical derivations compared to existing works \cite{wakinsuperres,chi2016guaranteed}. Essentially, it will not come as a big surprise to the readers familiar with \cite{wakinsuperres,chi2016guaranteed} that there is no straightforward way to extend the results in \cite{wakinsuperres} to the multi-user setting, particularly without the inclusion of multiple antennas at the receiver. More precisely, the proof steps of our theoretical analysis involve proving the existence of a novel random dual polynomial {\color{\change} which in turn involves developing new concentration inequalities for multi-user settings--a task that was known to be highly challenging by prior approaches}. This ensures users' demixing and simultaneous message recovery and continuous delay estimation of all users rather than message recovery and channel estimation of a single user. 
 Moreover, some incoherence properties between users' channels and correlations between user's codebooks need to be properly taken into account. This makes the multi-user OBDD problem (alternatively demixing and deconvolution problem) highly challenging and even much more complicated compared to the blind deconvolution problem in the single-user case~\cite{wakinsuperres,chi2016guaranteed}. In simple words, our problem formulation is more challenging due to the presence of multiple users, without adding additional equipment (e.g., multiple antennas) at the receiver's end.

\subsection{Applications}\label{sec:applications}
The model that we described in \eqref{eq.mymodel} appears in several applications, including \gf{ISAC},
%integrated sensing and communications (ISAC), 
over-the-air computation \gf{(OAC)} and communications, super-resolution microscopy with unknown point spread functions \cite{storm,storm2017}, and spike sorting in neural recordings \cite{spikesorting}. A detailed discussion of all applications is beyond the scope of this
paper. Below, we only delve into the details of the first two applications:
\begin{enumerate}
    \item \textit{Integrated sensing and communications}: The ISAC common receiver, a joint radar and communication receiver, is designed not only to detect the targets of interest (i.e., sense/monitor the surroundings) but also to decode messages from uplink communication users \cite{liu2020joint,chafii2023twelve} (see Figure~\ref{fig:applications(a)}. In model \eqref{eq.mymodel} that is also depicted in Figure~\ref{fig:applications(a)}, some transmitters may serve as radar transmitters, with $x_i(t)$ representing the waveform transmitted by the $i$-th radar transmitter, or they may function as communication users. Note that the bandwidth is shared between all transmitters. The channel between user $i$ and the common ISAC receiver comprises $s_i$ objects characterized by continuous parameters (such as velocity, range, and angles) that need to be detected. The dual-functional common receiver's objective is to distinguish between radar and communication signals and channels that occupy the same bandwidth. In our simplified model provided in \eqref{eq.mymodel}, we have solely focused on delay parameters, allowing for the determination of the range of targets or objects relative to the receiver. Nevertheless, there exists the potential to extend our model's scope to include velocity and angles as additional parameters. In summary, our proposed method in Section~\ref{sec:proposed_method} empowers single-antenna mobile devices to simultaneously sense their surroundings and decode the uplink messages transmitted by other communication nodes.
    \item \textit{Over-the-air computation}: Unlike the standard transmit-then-compute scheme, in OAC, all the devices simultaneously transmit their data to allow them to access all radio resources~\cite{Nazer2007Comp,Goldenbaum2015Nomographic,Saeed2023ChannelComp}. The main idea of OAC lies in the fact that interference can be harnessed to help computing. Indeed, a network of $r$ transmitters with a receiver as a server aims at computing a function through a shared communication channel. More precisely,  user $i$ use the message vector $\mathbf{f}_i\in \mathbb{R}^{k\times 1}$ to group  $k$ message values $f_{i,l}$ for $l=\{1,..., k\}$, i.e., $\mathbf{f}_i = [f_{i,1},\ldots,f_{i,k}]^{\mathsf{T}}$, where $f_{i,l}$ is the  message $l$ at user~$i$. Then, the receiver aims at computing $k$ functions of the input messages, i.e.,   $g_l(f_{1,l},\ldots,f_{r,l}):\mathbb{R}^r \mapsto \mathbb{R}$ for $l\in\{1,..., k\}$ (see Figure~\ref{fig:applications(b)}). In this scenario, all the transmitters use the same message length $ k_1 =\cdots =k_r$ denoted by $k$. One of the main challenges of OAC in massive IoT networks is its reliance on channel parameters, which can result in high latency and substantial overhead \cite{sahin2023Survey}. Studies by \cite{razavikia2022blind,Dong2020Blind}, have explored OAC in contexts where channel parameters $h_i(t)$ are only partially known—either delay or phase remains uncertain. Consequently, their methods only work under the assumption that either the transmitter or receiver compensates for all $c_l^i$s or $\tau_l^{i}$s.  The formulation presented in \eqref{eq.mymodel} adopts a more general approach, addressing the non-idealities and imperfections inherent in wireless communication systems for OAC.
\end{enumerate}
 \subsection{Outline}
 In Section~\ref{sec:problem_formulation}, we state the problem formulation for the OBDD model and illustrate how the nonlinear OBDD model is transformed into a linear problem in higher dimensions. In Section~\ref{sec:proposed_method}, we present our proposed optimization framework for users' demixing, message recovery, and channel estimation. In Section~\ref{sec.performance_gauranty}, we provide the sample complexity bound for the proposed optimization problem and theoretically specify the amount of resources that are required for the OBDD problem. Section~\ref{sec:simulations} provides some numerical experiments to verify the effectiveness of the proposed strategy in Section~\ref{sec:proposed_method}. Finally, conclusion and future directions are provided in Section~\ref{sec:conclusion}.
 \subsection{Notation}
 Scalars are denoted by italic small letters while vectors and matrices are denoted by bold upright letters. Vectors are considered column vectors throughout the paper. $\mx{I}_N$ denotes the identity matrix of size $N$. $\mx{e}_k\in\mathbb{R}^N$ refers to a vector that has all components equal to zero except for the $k$-th component that is one. $x(i)$ denotes the $i$-th element of a vector $\mx{x}\in\mathbb{C}^{n}$. $[\mx{A}]_{k,l}$ is used for the $(k,l)$-th element of the matrix $\mx{A}$.
 The notation $\mx{A}\preccurlyeq\mx{B}$ implies that $\mx{B}-\mx{A}$ is a positive semidefinite matrix. $\mathds{E}$ and $\mathds{P}$ denote the expectation and probability operators, respectively. The $\ell_q$ norm of a vector $\mx{x}\in\mathbb{C}^{n}$ is denoted by $\|\mx{x}\|_q\triangleq(\sum_{i}|x_i|^q)^{\frac{1}{q}}$. The maximum absolute element of a vector $\mx{x}\in\mathbb{C}^n$ is defined as $\|\mx{x}\|_{\infty}\triangleq \max_i |x_i| $. $\|\mx{A}\|_{\infty\rightarrow \infty}\triangleq \max_{\mx{z}\neq \mx{0}}\frac{\|\mx{A}\mx{z}\|_{\infty}}{\|\mx{z}\|_\infty}\triangleq \max_{i}\sum_{l} |A(i,l)|$ is the maximum absolute row sum of the matrix $\mx{A}$. The spectral norm of a matrix $\mx{X}\in\mathbb{C}^{n_1\times n_2}$ is denoted by $\|\mx{A}\|_{2\rightarrow 2}\triangleq \max_{\mx{z}\neq \mx{0}} {\|\mx{A} \mx{z}\|_2}/{\|\mx{z}\|_2}$. The pseudo-inverse of a matrix $\mx{A}$ is denoted by $\mx{A}^{\dagger}$. The Frobenius norm of a matrix $\mx{A}\in\mathbb{C}^{n_1\times n_2}$ is defined as $\|\mx{A}\|_{\rm F}\triangleq\sqrt{\sum_{i=1}^{n_1}\sum_{l=1}^{n_2}|X(i,l)|^2}$. The element-wise complex conjugate of $\mx{c}\in\mathbb{C}^n$ is denoted by $\mx{c}^\ast\in\mathbb{C}^{n\times 1}$. The set of matrices $\mx{Z}_i\in\mathbb{C}^{n_1\times n_2}, i=1, ...,r$ is denoted by the matrix tuple $\bm{\mathcal{{Z}}}\in \bigoplus_{i=1}^r\mathbb{C}^{n_1\times n_2}$. The sum of the diagonal entries of a matrix $\mx{A}\in\mathbb{C}^{n\times n}$ is denoted by ${\rm trace}(\mx{A})$. The Kronecker product of two matrices $\mx{A}\in\mathbb{C}^{n_1\times n_2}$ and $\mx{B}\in\mathbb{C}^{m_1\times m_2}$ is $\mx{C}=\mx{A}\otimes \mx{B}\in\mathbb{C}^{n_1 m_1\times n_2 m_2}$ where the elements of $\mx{C}$ are shown by $C(m_1 (r_1-1)+q_1,m_2 (r_2-1)+q_2)=A(r_1,r_2)B(q_1,q_2)$ and $r_1=1,..., n_1, r_2=1,..., n_2, q_1=1,...,m_1, q_2=1,...,m_2 $. Also, we use the following notation: 
% ----------------
\begin{align*}
    {\rm diag}(\mx{A}_1,...,\mx{A}_n)\triangleq{\rm diag}(\{\mx{A}_i\}_{i=1}^r)\triangleq\begin{bmatrix}
     \mx{A}_1&\mx{0}&\mx{0}\\
     \vdots&\ddots&\vdots\\
     \mx{0}&\mx{0}&\mx{A}_n
 \end{bmatrix},
\end{align*}
% ----------------
 to show a block diagonal matrix. $g=\mathcal{O}(f)$ shows sample complexity and means that there exists a constant $c$ such that $g$ is bounded above by $c f$. The point-wise product of $\mx{x}\in\mathbb{C}^{n}$ and $\mx{y}\in\mathbb{C}^n$ is denoted by $\mx{z}=\mx{x}\odot \mx{y}$ where $z(i)=x(i)y(i)$. The Frobenius inner product between two matrices $\mx{A}, \mx{B}\in\mathbb{C}^{n_1\times n_2}$ is defined as $\langle \mx{A},\mx{B}\rangle \triangleq\sum_{i=1}^{n_1}\sum_{l=1}^{n_2}A(i,l)B(i,l)$. $[r]$ denotes the set $\{1, \ldots, r\}$ for a scalar $r$. $1_{\mathcal{E}}\triangleq\left\{\begin{array}{cc}
     1 & i\in\mathcal{E} \\
     0 & i\notin\mathcal{E}
 \end{array}\right\}$ denotes the indicator function of the set $\mathcal{E}$. The sign function of $x$ is defined as ${\rm sgn}(x)\triangleq\frac{x}{|x|}$. The real value of the inner product between two matrices is denoted by $\langle \cdot,\cdot\rangle_{\mathbb{R}}\triangleq{\rm Re}\langle \cdot, \cdot\rangle$.

 \section{Problem Formulation}\label{sec:problem_formulation}

 Let $\{x_i(t)\}_{i=1}^r$ be the transmitted signals by users $i=1,..., r$ whose spectrums lie in the interval $[-B_{\rm max},B_{\rm max}]$. Taking the Fourier transform of \eqref{eq.mymodel}, we have
 % --------------
 \begin{align} \label{eq.Yf}
Y(f)=\sum_{i=1}^{r}H_i(f)X_i(f), ~\forall f\in [-B_{\rm max},B_{\rm max}],
 \end{align}   
 % --------------
 where $Y(f)$, $H_i(f)$, and $X_i(f)$ are the Fourier transform of $y(t)$, $h_i(t)$, and $x_i(t)$, respectively. By uniformly sampling \eqref{eq.Yf} at $N\triangleq4M+1$ points $f_n=\frac{B_{\rm max}n}{2M},~n=-2M, \ldots, 2M$, we reach
 % --------------
 \begin{align}\label{eq.sampledmodel1}
 y_n=\sum_{i=1}^rh_n^ix_n^i=\sum_{i=1}^{r}\sum_{k=1}^{s_i}c_k^i{\rm e}^{-j2\pi n \tau_k^i \frac{B_{\rm max}T_{\rm max}}{2M} }x_n^i,
 \end{align}
 % --------------
 where $\tau_k^i\triangleq\frac{\overline{\tau}_k^i}{{\rm T}_{\max}}\in[0,1)$ is called the normalized delay, $y_n\triangleq Y(f_n)$, $h_n^i\triangleq H_i(f_n)$, and $x_n^i\triangleq X_i(f_n)$. To uniquely identify $\{\tau_k^i\}_{i=1}^r$, we must have $B_{\max}{\rm T}_{\max}\le 2M$. Since we would like to use as small $M$ as possible, without loss of generality, we choose $2M=B_{\max}{\rm T}_{\max}$. Hence, \eqref{eq.sampledmodel1} can be rewritten as
 % --------------
 \begin{align}\label{eq.sampledmodel2}
 y_n=\sum_{i=1}^rh_n^ix_n^i=\sum_{i=1}^{r}\sum_{k=1}^{s_i}c_k^i{\rm e}^{-j2\pi n \tau_k^i}x_n^i.
 \end{align}
 % --------------
 The relation \eqref{eq.sampledmodel2} can also be represented in a matrix form as
 % --------------
 \begin{align}
 \mx{y}=\sum_{i=1}^r \mx{h}_i\odot \mx{x}_i\triangleq\sum_{i=1}^r\sum_{k=1}^{s_i}c_k^i\mx{a}(\tau_k^i) \odot \mx{x}_i,
 \end{align}
 % --------------
 where $\mx{y}\triangleq[y_{-2M},...,y_{2M}]^{\mathsf{T}}$, $\mx{h}_i\triangleq[h_{-2M}^i,...,h_{2M}^i]^{\mathsf{T}}$, $\mx{x}_i\triangleq[x_{-2M}^i,...,x_{2M}^i]^{\mathsf{T}}$, and $$\mx{a}(\tau)\triangleq[{\rm e}^{-j2\pi(-2M)\tau},..., {\rm e}^{-j2\pi(0)\tau},... ,{\rm e}^{-j2\pi(2M)\tau}]^{\mathsf{T}}.$$
 Our goal is to recover $\tau_k^i$s, $c_k^i$s, and $\mx{x}_i$s from the observation vector $\mx{y}\in\mathbb{C}^{N}$. Obviously, it is unavoidable to have scaling ambiguities for recovering $\mx{x}_i$s and $\mx{h}_i$s as for any $\alpha_i\in\mathbb{C}\setminus\{0\}$, we have
 % --------------
 \begin{align}
 \mx{y}=\sum_{i=1}^r\alpha_i\mx{h}_i\odot \frac{\mx{x}_i}{\alpha_i}.
 \end{align}
% --------------

 Without further constraints on $\mx{x}_i$s or $\mx{h}_i$s, the task of separating $\mx{x}_i$ and $\mx{h}_i$ 
 from their point-wise product $\mx{h}_i\odot \mx{x}_i$ and simultaneously demixing the terms $\mx{h}_i\odot \mx{x}_i$ from their sum is highly ill-posed because the number of samples $N$ is much smaller than the number of unknowns {\color{\change}$\mathcal{O}(rN+\sum_{i=1}^r s_i)$}. To alleviate this challenge, we make the assumption that each $\mx{x}_i$ is generated by applying a redundant codebook matrix $\mx{B}_i\in\mathbb{C}^{N\times k_i}$ into a message vector $\mx{f}_i\in{\mathbb{C}^{k_i\times 1}}$. This indeed implies that $\mx{x}_i$ lives in a low-dimensional subspace, i.e.,
 % --------------
 \begin{align}\label{eq.subspace_assumption}
 \mx{x}_i=\mx{B}_i\mx{f}_i,
 \end{align} 
 % --------------
 where, 
 \begin{align}
 \mx{B}_i\triangleq[\mx{b}_{-2M}^i, \ldots, \mx{b}_{2M}^i]^{\mathsf{T}}\in\mathbb{C}^{N\times k_i},
 \end{align}
 is a known basis of the subspace with $N\gg k_i$ and $\mx{f}_i\in\mathbb{C}^{k_i}$ is unknown with $\|\mx{f}_i\|_2=1$. Here, $\mx{b}_n^i\triangleq [b_n^i(1),..., b_n^i(k_i)]^{\mathsf{T}}\in\mathbb{C}^{k_i\times 1}$.

 Under the subspace assumption \eqref{eq.subspace_assumption}, \eqref{eq.sampledmodel2} can be rewritten as
 % --------------
 \begin{align}\label{eq.sampledmodel3}
 y_n&=\sum_{i=1}^r\sum_{k=1}^{s_i}c_k^i{\rm e}^{-j2\pi n \tau_k^i}{\mx{b}^i_n}^{\mathsf{T}}\mx{f}_i=
 \sum_{i=1}^r\sum_{k=1}^{s_i}c_k^i\mx{e}_n^{\mathsf{T}}\mx{a}(\tau_k^i){\mx{b}^i_n}^{\mathsf{T}}\mx{f}_i\nonumber\\	
 & = \sum_{i=1}^r\sum_{k=1}^{s_i}c_k^i\mx{e}_n^{\mathsf{T}}\mx{a}(\tau_k^i)\mx{f}_i^{\mathsf{T}}\mx{b}^i_n,
 \end{align}
 % --------------
 where $\mx{e}_n,~ -2M\le n\le 2M$ stands for the $(n+2M+1)$-th column of $\mx{I}_{N}$. Let $\mx{H}_i=\sum_{k=1}^{s_i} c_k^i\mx{f}_i\mx{a}(\tau_k^i)^{\mathsf{T}}\in\mathbb{C}^{k_i\times N}$. Using the lifting trick \cite{ling2015self}, the measurements $y_n, n=-2M, \ldots, 2M$ in \eqref{eq.sampledmodel3} can be written as
 % --------------
 \begin{align}\label{eq:measurement}
 y_n&=\sum_{i=1}^r\sum_{k=1}^{s_i}c_k^i {\rm trace}\Big(\mx{b}^i_n\mx{e}_n^{\mathsf{T}}\mx{a}(\tau_k^i)\mx{f}_i^{\mathsf{T}}\Big)\nonumber\\
 &=\sum_{i=1}^r\big\langle \mx{H}_i, \mx{b}^i_n\mx{e}_n^{\mathsf{T}}\big\rangle.
 \end{align} 
 % --------------
 By writing in matrix form, we have:
 % --------------
 \begin{align}\label{eq:linear_measure}
 \mx{y}=\mathcal{B}(\bm{\mathcal{H}}),
 \end{align}
 % --------------
 where $\mx{y}\triangleq[y_{-2M}, ..., y_{2M}]^{\mathsf{T}}$, $\bm{\mathcal{H}}\triangleq(\mx{H}_i)_{i=1}^r\in \bigoplus_{i=1}^r\mathbb{C}^{k_i\times N}$ is the matrix tuple of interest and $\mathcal{B}$ is the linear mapping defined as
 % --------------
 \begin{align*}
 \bigoplus_{i=1}^r\mathbb{C}^{k_i\times N}\rightarrow \mathbb{C}^N,~~ \bm{\mathcal{H}}\rightarrow \Bigg(\sum_{i=1}^r\big\langle \mx{H}_i, \mx{b}^i_n\mx{e}_n^{\mathsf{T}}\big\rangle  \Bigg)_{n=-2M}^{2M}.	
 \end{align*}
% --------------
In Equation \eqref{eq:linear_measure}, the relationship between the matrix tuple $\bm{\mathcal{{H}}}$ and $\mx{y}$ is linear. In the following section, we assert that the sparsity of channels in the delay domain presents a specific low-dimensional structure for the matrix tuple $\bm{\mathcal{H}}$. We then introduce an optimization framework designed to capture this structure from linear measurements \eqref{eq:linear_measure}.
 
 \section{Proposed Method}\label{sec:proposed_method}
We first describe the proposed method in the noiseless case and then extend it to the noisy case in Section \ref{sec:noisy_sec}.
%--------------------------
\begin{figure*}
\centering

\subfigure[]{
\tdplotsetmaincoords{45}{120} 
\begin{tikzpicture}[scale=2.5, tdplot_main_coords, axis/.style={->,black,thick}]
% Set the viewing angle
 % Adjusts the 3D view angle

% Axes
% \draw[axis] (0, 0, 0) -- (1.2, 0, 0) node [anchor=north east]{};
% \draw[axis] (0, 0, 0) -- (0, 1.2, 0) node [anchor=north west]{};
% \draw[axis] (0, 0, 0) -- (0, 0, 1.5) node [anchor=south]{$\mathcal{A}_1:= \mathbf{f}_1a(\tau)$};

% Vertices of the L1 ball (octahedron)
\coordinate (A) at (0.75, 0, 0);
\coordinate (B) at (-0.75, 0, 0);
\coordinate (C) at (0, 0.75, 0);
\coordinate (D) at (0, -0.75, 0);
\coordinate (E) at (0, 0, 0.75);
\coordinate (F) at (0, 0, -0.75);

\fill[orange!20!white,opacity=0.7] (A) -- (C) -- (E) -- cycle; % Lighter shade for top facets
\fill[orange!60!white,opacity=0.7] (A) -- (D) -- (E) -- cycle; % Intermediate shade
\fill[orange!90!white,opacity=0.6] (B) -- (C) -- (E) -- cycle; % Lighter shade
\fill[orange!60!white,opacity=0.7] (B) -- (D) -- (E) -- cycle; % Intermediate shade
\fill[orange!40!white,opacity=0.7] (A) -- (C) -- (F) -- cycle; % Darker shade for bottom facets
\fill[orange!80!white,opacity=0.7] (A) -- (D) -- (F) -- cycle; % Darkest shade
\fill[orange!80!white,opacity=0.3] (B) -- (C) -- (F) -- cycle; % Darker shade
\fill[orange!30!white,opacity=0.7] (B) -- (D) -- (F) -- cycle; % Darkest shade
% Drawing a horizontal plane close to the top vertex
\fill[blue!50!white, opacity=0.5] (-1,-1,0.8) -- (1,-1,0.8) -- (1,1,0.8) -- (-1,1,0.8) -- cycle;

\draw (0, 0, 0.75) node     {\color{antiquefuchsia}\Huge $\cdot$};
\draw (0, 0.1, 0.97) node     {\color{antiquefuchsia}\large $\widehat{\bm{\mathcal{H}}}=\bm{\mathcal{H}}$};

\draw (1, 0, 2) node     {$\mathbf{y} = \mathcal{B}(\bm{\mathcal{H}})$};
\end{tikzpicture}
}
 % Inserting the provided matrix
\subfigure[]{
\begin{tikzpicture}[scale=2.5, axis/.style={->,black,thick}]
  % Matrix 1
% Loop to replicate the column 6 times
\foreach \x in {-15pt,-13.5pt,...,22.5pt} {
    % Calculate new x position for the shift
    \pgfmathsetmacro\opacityOne{rand*0.5+0.5}
    \pgfmathsetmacro\opacityTwo{rand*0.5+0.5}
    \pgfmathsetmacro\opacityThree{rand*0.5+0.5}
    \pgfmathsetmacro\opacityFour{rand*0.5+0.5}
    \pgfmathsetmacro\opacityFive{rand*0.5+0.5}
    % Rectangle 1
    \draw [fill={blue(ncs)}, fill opacity=\opacityOne] (-7pt + \x, -9pt) -- (-7pt + \x, -6pt) -- (-7pt + \x +1.5pt, -6pt) -- (-7pt + \x +1.5pt, -9pt) -- cycle;
    % Rectangle 2
    \draw [fill={chestnut}, fill opacity=\opacityTwo] (-7pt + \x, -6pt) -- (-7pt + \x , -3pt) -- (-7pt + \x +1.5pt, -3pt) -- (-7pt + \x +1.5pt, -6pt) -- cycle;
    % Rectangle 3
    \draw [fill={coolgrey}, fill opacity=\opacityThree] (-7pt + \x , -3pt) -- (-7pt + \x , 0pt) -- (-7pt + \x +1.5pt, 0pt) -- (-7pt + \x +1.5pt, -3pt) -- cycle;
    % Rectangle 4
    \draw [fill={carrotorange}, fill opacity=\opacityFour] (-7pt + \x , 0pt) -- (-7pt + \x , 3pt) -- (-7pt + \x +1.5pt, 3pt) -- (-7pt + \x +1.5pt, 0pt) -- cycle;
    % Rectangle 5
    \draw [fill={blue(ncs)}, fill opacity=\opacityFive] (-7pt + \x , 3pt) -- (-7pt + \x , 6pt) -- (-7pt + \x +1.5pt, 6pt) -- (-7pt + \x +1.5pt, 3pt) -- cycle;
}

\node at (0,+10pt) {$\mathbf{H}_i$};
\node at (24pt,+10pt) {$\mathbf{f}_i$};
% \node at (27pt,+5pt) {$\mathbf{a}^{\mathsf{T}}(\tau_1)$};

% \node at (41pt,+10pt) {$\mathbf{f}_2$};
% \node at (46pt,+5pt) {$\mathbf{a}^{\mathsf{T}}(\tau_2)$};
  % Equals sign
\node at (20pt,-2pt) {$=$};

\node at (27pt,-2pt) {$\times$};
\node at (30pt,-2pt) {\Huge (};
\node at (78pt,-2pt) {\Huge)};

% \node at (45pt,5pt) {$c_1^1\mathbf{a}(\tau_1^1) + c_2^1\mathbf{a}(\tau_2^1)$};
\node at (71pt,1pt) {\footnotesize $c_2\mathbf{a}^{\mathsf{T}}(\tau_2^i)$};
\node at (50pt,-2pt) {\footnotesize $+$};
\node at (71pt,-5pt) {\footnotesize $c_1\mathbf{a}^{\mathsf{T}}(\tau_1^i)$};
  % Vector 2 

% Rectangle 5 
\draw  [fill= {blue(ncs)} ,fill opacity=0.60 ] (23pt,3pt) -- (23pt,6pt) -- (24.5pt,6pt) -- (24.5pt,3pt) -- cycle ;
% Rectangle 4
\draw  [fill={carrotorange}  ,fill opacity=0.7 ] (23pt,0) -- (23pt,3pt) -- (24.5pt,3pt) -- (24.5pt,0) -- cycle ;
% Rectangle 3
\draw  [fill={coolgrey}  ,fill opacity=1 ] (23pt,-3pt) -- (23pt,0pt) -- (24.5pt,0pt) -- (24.5pt,-3pt) -- cycle ;
% % Rectangle 2
\draw  [fill={chestnut}  ,fill opacity=0.75 ] (23pt,-6pt) -- (23pt,-3pt) -- (24.5pt,-3pt) -- (24.5pt,-6pt) -- cycle ;
% Rectangle 1 of size = (6*10) 
\draw  [fill={blue(ncs)}  ,fill opacity=0.6] (23pt,-9pt) -- (23pt,-6pt) -- (24.5pt,-6pt) -- (24.5pt,-9pt) -- cycle ;

\foreach \x in {10pt,12pt,...,40pt}{
      \pgfmathsetmacro\opacitysix{rand*0.5+0.5}
\draw  [fill={coolgrey}  ,fill opacity=\opacitysix ] (\x +22pt,0pt) -- (\x + 22pt,2pt) -- (\x +24pt,2pt) -- (\x +24pt,0pt) -- cycle ;

}

\foreach \x in {10pt,12pt,...,40pt}{
      \pgfmathsetmacro\opacitysix{rand*0.5+0.5}
\draw  [fill={coolgrey}  ,fill opacity=\opacitysix ] (\x +22pt,-6pt) -- (\x + 22pt,-4pt) -- (\x +24pt,-4pt) -- (\x +24pt,-6pt) -- cycle ;

}
\end{tikzpicture}
}  
\caption{(a): Geometric intuition of the proposed optimization problem \eqref{eq.primalprob}. 
The blue hyperplane represents the linear constraint \eqref{eq:linear_measure} which is the feasible set, while the orange-colored object depicts the convex hull of the atoms corresponding to all users. The solution $\widehat{\bm{\mathcal{H}}}$ provided in \eqref{eq.primalprob} lies at the minimal contour of the convex hull, where it intersects the hyperplane and contains the least number of atoms. (b): An example demonstrating that each matrix $\mx{H}_i\in\mathbb{C}^{k_i\times N}$ comprises $s_i=2$ matrix-valued atoms, denoted as $\mx{f}_i\mx{a}(\tau^i)^{\mathsf{T}}$, formed by the outer product of the message vector $\mx{f}_i\in\mathbb{C}^{k_i\times 1}$ and the steering vector $\mx{a}(\tau^i)\in\mathbb{C}^{N\times 1}$. }
\end{figure*}
 %--------------------------

 In model \eqref{eq.sampledmodel3}, the number of multi-path components  $\{s_i\}_{i=1}^r$ is assumed to be small. We thereby define the atomic norm \cite{chandrasekaran2012convex}
 % --------------
 \begin{align}\label{eq.atomic_def}
 &\|\mx{H}_i\|_{\mathcal{A}_i}\triangleq\inf\{t>0: \mx{H}_i\in t~{\rm conv}(\mathcal{A}_i)\}\nonumber\\
 &=\inf_{c_k, \tau_k}\Big\{\sum_{k}|c_k|\|\mx{f}_i\|_2:~\mx{H}_i=\sum_{k}c_k \mx{f}_i\mx{a}(\tau_k)^{\mathsf{T}}  \in\mathbb{C}^{k_i\times N} \Big\}
 \end{align}
 % --------------
 associated with the atoms
 % --------------
 \begin{align}\label{eq.atoms}
 \mathcal{A}_i=\big\{\mx{f}_i\mx{a}(\tau)^{\mathsf{T}}: \tau\in[0,1), \|\mx{f}_i\|_2=1, \mx{f}_i    \in\mathbb{C}^{k_i\times 1} \big\},
 \end{align}
 % --------------
 where $i\in [r]$. The atomic norm $\|\mx{H}_i\|_{\mathcal{A}_i}$ can be regarded as the best convex alternative for the smallest number of atoms $\mathcal{A}_i$ needed to represent a signal $\mx{H}_i$. Hence, we are interested in recovering the matrix tuple $\bm{\mathcal{H}}\triangleq(\mx{H}_i)_{i=1}^r$ by promoting their atomic sparsity via solving
 % --------------
 \begin{align}\label{eq.primalprob}
 \min_{\bm{\mathcal{Z}}=(\mx{Z}_i)_{i=1}^r} ~\sum_{i=1}^{r}\|\mx{Z}_i\|_{\mathcal{A}_i},\quad \mx{y}_{N\times 1}=\mathcal{B}(\bm{\mathcal{Z}}).	
 \end{align} 
 % --------------
 The dual problem of \eqref{eq.primalprob} is given by (see Appendix \ref{proof.dual} for the proof):
 % --------------
 \begin{align}\label{eq.dualprob}
 \hspace{-3pt}\max_{\substack{\mx{\bs{\lambda}}\in\mathbb{C}^{N}\\\tau^i\in[0,1)}}~{\rm Re}~\langle \mx{\bs{\lambda}}, \mx{y}\rangle~
 {\rm s.t.}~\|(\mathcal{B}^{\rm Adj}\mx{\bs{\lambda}})_i\mx{a}^*(\tau^i)\|_2\le 1,~i\in[r],	
 \end{align}
 % --------------
 where $\mathcal{B}^{\rm Adj}:\mathbb{C}^{N}\rightarrow \bigoplus_{i=1}^r\mathbb{C}^{k_i\times N}$ denotes the adjoint operator of $\mathcal{B}$ and $\mathcal{B}^{\rm Adj}\mx{\bs{\lambda}}\triangleq((\mathcal{B}^{\rm Adj}\mx{\bs{\lambda}})_i)_{i=1}^r$ is a matrix tuple where the $i$-th matrix is given by
 % --------------
 \begin{align}
 (\mathcal{B}^{\rm Adj}\mx{\bs{\lambda}})_i=\sum_{n=-2M}^{2M}\bs{\lambda}_n\mx{b}^i_n\mx{e}_n^{\mathsf{T}}.
 \end{align}
 % --------------
 In the following proposition whose proof is provided in Appendix \ref{proof.optimality}, we obtain sufficient conditions certifying optimality of the matrix tuple $\bm{\mathcal{H}}=(\mx{H}_i)_{i=1}^r$ in the primal problem \eqref{eq.primalprob}.
 \begin{prop}\label{prop.optimality}
Denote the set of delay parameters corresponding to the channel of the $i$-th user as $\mathcal{S}_i\triangleq\{\tau_k^i\}_{k=1}^{s_i}$. The solution $\widehat{\bm{\mathcal{H}}}\triangleq(\widehat{\mx{H}}_i)_{i=1}^r$ of \eqref{eq.primalprob} is unique if there exists a vector $\mx{\bs{\lambda}}\triangleq[{\lambda}({-2M}), ..., {\lambda}({2M})]^{\mathsf{T}}\in\mathbb{C}^N$ such that the vector-valued dual polynomials 
% --------------
\begin{align}\label{eq.Qi}
&\mx{q}_i(\tau)\triangleq(\mathcal{B}^{\rm Adj}\mx{\bs{\lambda}})_i\mx{a}^*(\tau)=\sum_{n=-2M}^{2M}\bs{\lambda}_n {\rm e}^{j2\pi n \tau} \mx{b}_n^i\in\mathbb{C}^{k_i},\nonumber\\
& i\in[r],
\end{align}
% --------------
satisfy the conditions
% --------------
\begin{align}
&\mx{q}_i(\tau_k)={\rm sgn}({c_k^{i}})\frac{\mx{f}_i}{\|\mx{f}_i\|_2}~\forall \tau_k\in \mathcal{S}_i,~ i=1, \ldots, r\label{eq.supp_cond}\\
&\|\mx{q}_i(\tau)\|_2<1~~\forall \tau \in [0,1)\setminus \mathcal{S}_i,~ i=1, \ldots, r\label{eq.offsupp_cond}
\end{align}
% --------------
 \end{prop}
 \begin{rem}(Intuition behind Proposition \ref{prop.optimality})
  The optimization problem \eqref{eq.primalprob} is a convex optimization problem with linear constrains. A matrix tuple $\bm{\mathcal{H}}$ is optimal if and only if the descent cone of the objective function at point $\bm{\mathcal{H}}$ does not have any intersections with the null space of the linear operator $\mathcal{B}(\cdot)$ (see e.g., \cite{amelunxen2014living,daei2019error,daei2019living}). The absence of intersections between two cones implies that their polar cones must intersect at least once. Remarkably, the polar cone of the null space of $\mathcal{B}(\cdot)$ corresponds to the range space of the adjoint operator $\mathcal{B}^{\rm Adj}(\cdot)$. Furthermore, the polar cone of the descent cone is equivalent to the sub-differential of the objective function at point $\bm{\mathcal{H}}$. By applying the chain-rule lemma of sub-differentials, we infer that the sub-differential of atomic-like functions (such as the objective function in \eqref{eq.primalprob}) at point $\bm{\mathcal{H}}$ is a rotated version of the sub-differential of TVM at the ground-truth delay continuous function. Notably, the sub-differential of TVM at a continuous function represents a continuous set where at support locations is equal to the sign of values in the delay domain and its absolute value remains below one at off-support locations, similar to the behavior of the sub-differential of the $\ell_1$ norm at some vector-valued point (see e.g., \cite[Section II]{daei2019exploiting}). In summary, to establish optimality of $\bm{\mathcal{H}}$, we must demonstrate the existence of a polynomial such as $\mx{q}_i(\tau)$, which satisfies the conditions \eqref{eq.supp_cond} and \eqref{eq.offsupp_cond}.
 \end{rem}
Since the $\tau^i$s are continuous-valued, there are an infinite number of constraints in the dual problem \eqref{eq.dualprob}. Based on the theory of trigonometric polynomials, there is an explicit way to transform the infinite dimensional constraints in \eqref{eq.dualprob} to the linear matrix inequalities as follows (see e.g. \cite[Proposition 2.4]{fernandez2016super} or \cite[Section 9.2.2]{dumitrescu2017positive}):
% --------------
\begin{align}\label{eq:trigonometric_rel}
&\|(\mathcal{B}^{\rm Adj}\mx{\bs{\lambda}})_i\mx{a}^*(\tau^i)\|_2\le 1, i\in[r] ~\text{if}~ \text{and}~ \text{only}~ \text{if}\nonumber\\
&\exists \mx{Q}\in\mathbb{C}^{N\times N}: \begin{bmatrix}
        \bm{Q}&(\mathcal{B}^{\rm Adj}(\bm{\lambda}))_i^{\mathsf{H}}\\
        (\mathcal{B}^{\rm Adj}(\bm{\lambda}))_i&\bm{I}_{k_i}
    \end{bmatrix}\succeq \bm{0}, \quad i \in [r],\\
    & ~~~~~~\langle \mathcal{T}({\bm{e}_q}), \bm{Q}  \rangle=1_{q=0}, \quad q=-N+1,..., N-1
\end{align}
% --------------
where $\mathcal{T}(\mx{z})$ for an arbitrary $\mx{z}$ shows the Toeplitz operator whose first row is $\mx{z}$. Plugging the relations \eqref{eq:trigonometric_rel} into the dual problem \eqref{eq.dualprob}, leads to the following semidefinite programming (SDP):
% --------------
\begin{equation}\label{prob.sdp}
    \begin{aligned}
    &\underset{\substack{\bm{\lambda}\in\mathbb{C}^{N}, \bm{Q}\in \mathbb{C}^{N\times N}}}{\rm max}~~{\rm Re}\big\{\langle \bm{\lambda}, \bm{y}\rangle\big\}\\
    &~~~~{\rm s.t.}~~
    \begin{bmatrix}
        \bm{Q}&(\mathcal{B}^{\rm Adj}(\bm{\lambda}))_i^{\mathsf{H}}\\
        (\mathcal{B}^{\rm Adj}(\bm{\lambda}))_i&\bm{I}_{k_i}
    \end{bmatrix}\succeq \bm{0}, \quad i \in [r],\\
    & ~~~~~~\langle \mathcal{T}({\bm{e}_q}), \bm{Q}  \rangle=1_{q=0}, \quad q=-N+1,..., N-1,
    \end{aligned}
\end{equation}	
 %--------------
% \begin{rem}(Active user detection)
%    In Equation \eqref{eq.mymodel}, $r$ represents the number of active IoT devices and is assumed to be known. In practical scenarios characterized by sporadic traffic, these $r$ devices constitute only a small fraction among a vast number of devices (say e.g., $r_{\rm T}\gg r$), the majority of which remain in a silent or inactive state between $[0,{\rm T}_{\max}]$. By imposing additional assumptions on the codebook matrices, $\mx{B}_i$s and employing the blind goal-oriented detection approach proposed in \cite{daei2023blindTSP, daei2023blindWiopt}, we can formulate an SDP problem which tackles active user detection, message recovery, and channel estimation simultaneously. Remarkably, this approach is independent of the total number of inactive devices, i.e., $r_{\rm T}-r$. However, the detailed exploration of this aspect is deferred to future works.
% \end{rem}
  \begin{figure}[!t]
\centering
 \begin{tikzpicture} 

 \begin{scope}[spy using outlines={rectangle, magnification=5,
   width=1.6cm,height=1.6cm,connect spies}]
    \begin{axis}[
        width=0.5\textwidth,
        height=6cm,
        xmin=0, xmax=1,
        ymin=1e-3, ymax=1,
        legend style={nodes={scale=0.6, transform shape}, at={(0.3,0.95)}}, 
        ticklabel style = {font=\footnotesize},
        ymajorgrids=true,
        xmajorgrids=true,
        grid style=dashed,
        grid=both,
        grid style={line width=.1pt, draw=gray!10},
        major grid style={line width=.2pt,draw=gray!30},
    ]
    \addplot[ smooth,
             thin,
        color=chestnut,
        line width=0.9pt,
        ]
    table[x=t,y=D1]
    {PolyDual3.dat};
\addplot[ smooth,
             thin,
        color=airforceblue,
        line width=0.9pt,
        ]
    table[x=t,y=D2]
    {PolyDual3.dat};
    \addplot[ 
        color=cadmiumorange,
        mark=star,
        dashed,
        line width=1pt,
        mark size=1pt,
        ]
    table[x=t,y=X1]
    {PolyDual3.dat};
    \addplot[ 
        color=azure,
        mark=star,
        dashed,
        line width=0.5pt,
        mark size=0.5pt,
        ]
    table[x=t,y=X2]
    {PolyDual3.dat};
    \legend{$\|\mx{q}_1(\tau)\|_2$, $\|\mx{q}_2(\tau)\|_2$,$\tau^1$,$\tau^2$};
     \path (0.639, 0.99) coordinate (X);
    \end{axis}
    \spy [black] on (X) in node (zoom) [left] at ([xshift=2.5cm,yshift=-0.5cm]X);
    \end{scope}
\end{tikzpicture}

  \caption{This figure shows the $\ell_2$ norm of the dual polynomial functions $\mx{q}_i(\tau), i\in[r]$ for two users i.e., $r=2$ and the parameters $N=64$, $s_1=2,s_2=1$ and $k_1=k_2 =5$. The peaks of $\|\mx{q}_i(\tau)\|_2$ specify the delay parameters corresponding to user $i$.}
   \label{fig:Daul}
\end{figure}
The SDP problem \eqref{prob.sdp} can be efficiently solved by the CVX toolbox \cite{grant2014cvx}.   Let $\hat{\bm{\lambda}}$ be the solution to the dual problem in~\eqref{prob.sdp},
then according to Proposition \ref{prop.optimality}, the delay parameters corresponding to the $i$-th user can be estimated by identifying the locations that satisfy $\|\mx{q}_i(\tau)\|_2=1$. This leads to the following estimates for the support sets (delay parameters) for each $i\in[r]$:
% --------------
\begin{align}\label{eq:support_estimate}
    \widehat{\mathcal{S}}_i = \Big\{\tau \in [0,1), \quad  | \big\|  (\mathcal{B}^{\rm Adj} \hat{\bm{\lambda}})_i \bm{a}(\tau) \big\|_2=1 \Big\}.
\end{align}
% --------------
In order to identify $\widehat{\mathcal{S}}_i$, there are two ways: the first one which we used in our simulations is to compute the roots of the following polynomial on the unit circle:
% --------------
\begin{align}
 p_i(z)\triangleq 1-\sum_{k=-4M}^{4M} u_k^i z^{k},   
\end{align}
% --------------
where $u_k^i\triangleq\sum_{l=-2M}^{2M} \widehat{\lambda}(l) \widehat{\lambda}^\ast({l-k}) {\mx{b}_{l-k}^i}^{\mathsf{H}}\mx{b}_{l}^i$ and $z\triangleq {\rm e}^{j2\pi\tau}$. The second way is to discretizing the domain $\tau\in[0,1)$ up to a desired fine grid and then finding the delays that satisfy $\|\mx{q}_i(\tau)\|_2=1$. 
{\color{\change}It is important to note that this approach differs fundamentally from initially discretizing the delay domain and employing compressed sensing methods. In fact, discretizing the delay domain within the dual polynomial plot preserves resolution, whereas initially discretizing the domain destroys resolution.}
Now, by having the estimates $\widehat{\tau}_k^i, k=1, \ldots, \widehat{s}_i, i\in[{r}]$, one can replace these estimates into \eqref{eq:measurement} and form the following over-determined system of equations:
% --------------
\begin{align}\label{eq:overdetermined_equations}
\scalebox{.7}{$
         \begin{bmatrix}
              \mx{e}_{-2M}^{\mathsf{T}}
              \mx{a}(\hat{\tau}_1^1)
              {\mx{b}_{-2M}^1}^{\mathsf{T}}
              &\hdots&\mx{e}_{-2M}^{\mathsf{T}}\mx{a}(\widehat{\tau}_{s_r}^r){\mx{b}_{-2M}^r}^{\mathsf{T}}\\
              \vdots&\ddots&\vdots\\
           \mx{e}_{2M}^{\mathsf{T}}
              \mx{a}(\hat{\tau}_1^1)
              {\mx{b}_{2M}^1}^{\mathsf{T}}
             &\hdots&\mx{e}_{2M}^{\mathsf{T}}\mx{a}(\widehat{\tau}_{s_r}^r){\mx{b}_{2M}^r}^{\mathsf{T}}\\
         \end{bmatrix} \begin{bmatrix}
             c_1^1 \mx{f}_1\\
             \vdots\\
             c_{s_1}^1 \mx{f}_1\\
             \vdots\\
             c_1^r \mx{f}_r\\
             \vdots\\
             c_{s_r}^r \mx{f}_r\\
         \end{bmatrix}=\begin{bmatrix}
             y(-2M)\\
             \vdots\\
             y(2M)
         \end{bmatrix}$}
     \end{align}
% --------------
The solution of the system  \eqref{eq:overdetermined_equations} can be found using the least square method which leads to:
% --------------
\begin{align}\label{eq:least_square}
\scalebox{.8}{$
\begin{bmatrix}
             \widehat{c}_1^1 \widehat{\mx{f}}_1\\
             \vdots\\
             \widehat{c}_{\widehat{s}_1}^1 \widehat{\mx{f}}_1\\
             \vdots\\
             \widehat{c}_1^r \widehat{\mx{f}}_r\\
             \vdots\\
             \widehat{c}_{\widehat{s}_r}^r \widehat{\mx{f}}_r\\
         \end{bmatrix}  = \begin{bmatrix}
              \mx{e}_{-2M}^{\mathsf{T}}
              \mx{a}(\hat{\tau}_1^1)
              {\mx{b}_{-2M}^1}^{\mathsf{T}}
              &\hdots&\mx{e}_{-2M}^{\mathsf{T}}\mx{a}(\widehat{\tau}_{\widehat{s}_r}^r){\mx{b}_{-2M}^r}^{\mathsf{T}}\\
              \vdots&\ddots&\vdots\\
           \mx{e}_{2M}^{\mathsf{T}}
              \mx{a}(\hat{\tau}_1^1)
              {\mx{b}_{2M}^1}^{\mathsf{T}}
             &\hdots&\mx{e}_{2M}^{\mathsf{T}}\mx{a}(\widehat{\tau}_{\widehat{s}_r}^r){\mx{b}_{2M}^r}^{\mathsf{T}}\\
         \end{bmatrix}^{\dagger}\mx{y}$}.
\end{align}
% --------------
From \eqref{eq:least_square}, one can estimate $\widehat{c}_l^i \widehat{\mx{f}}_i$. By knowing that $\|\mx{f}_i\|_2=1$ and $\widehat{c}_l^i \widehat{\mx{f}}_i$, $|\widehat{c}_l^i|$ for each $l=1,..., \widehat{s}_i, i=1,..., \widehat{r}$ can be found. Then, by having $\widehat{c}_l^i \widehat{\mx{f}}_i$ and $|\widehat{c}_l^i|$, one can estimate $|\widehat{{f}}_i(l)|$ for each $l=1,..., k_i$. Thus, the magnitude of the messages is estimated while their phase is overlooked, leading to a persistent phase ambiguity. Consequently, users' message information is constrained to adhere to an amplitude-based modulation scheme rather than a phase-based one. In the subsequent section, we present a mathematical theory that ensures the performance of the aforementioned proposed method.
{\color{\change}
\subsection{OBDD in the noisy case}\label{sec:noisy_sec}
We consider a noisy version of \eqref{eq.sampledmodel1} where the data samples are contaminated by additive noise as follows:
\begin{align}
  \mx{y}=\sum_{i=1}^r \mx{h}_i \odot \mx{x}_i +\mx{w},  
\end{align}
where $\mx{w}$ is the additive noise vector satisfying $\|\mx{w}\|_2\le \eta$. The optimization problem \eqref{eq.primalprob} can be modified to consider the noisy measurements as follows:
\begin{align}\label{eq.primalprob_noisy}
 \min_{\widetilde{\mx{y}}, \bm{\mathcal{Z}}=(\mx{Z}_i)_{i=1}^r} ~\sum_{i=1}^{r}\|\mx{Z}_i\|_{\mathcal{A}_i},\\
 {\rm s.t.}~\widetilde{\mx{y}}=\mx{y}_{N\times 1}-\mathcal{B}(\bm{\mathcal{Z}})\\
 \|\widetilde{\mx{y}}\|_2\le \eta.	
 \end{align} 

The dual problem of \eqref{eq.primalprob_noisy} is given as follows (see Appendix \ref{proof.dual}):
\begin{align}\label{eq.dualprob_noisy}
 &\hspace{-3pt}\max_{\substack{\mx{\bs{\lambda}}\in\mathbb{C}^{N}\\\tau^i\in[0,1)}}~{\rm Re}~\langle \mx{\bs{\lambda}}, \mx{y}\rangle-\eta \|\bs{\lambda}\|_2~\\
 &{\rm s.t.}~\|(\mathcal{B}^{\rm Adj}\mx{\bs{\lambda}})_i\mx{a}^*(\tau^i)\|_2\le 1,~i\in[r].
 \end{align}
 Similar to the discussion in the noise-free case, the optimization problem \eqref{eq.dualprob_noisy} can be transformed to an SDP problem which is given by:
\begin{equation}\label{prob.sdp_noisy}
    \begin{aligned}
    &\underset{\substack{\bm{\lambda}\in\mathbb{C}^{N}, \bm{Q}\in \mathbb{C}^{N\times N}}}{\rm max}~~{\rm Re}\big\{\langle \bm{\lambda}, \bm{y}\rangle\big\}-\eta\|\bs{\lambda}\|_2\\
    &~~~~{\rm s.t.}~~
    \begin{bmatrix}
        \bm{Q}&(\mathcal{B}^{\rm Adj}(\bm{\lambda}))_i^{\mathsf{H}}\\
        (\mathcal{B}^{\rm Adj}(\bm{\lambda}))_i&\bm{I}_{k_i}
    \end{bmatrix}\succeq \bm{0}, \quad i \in [r],\\
    & ~~~~~~\langle \mathcal{T}({\bm{e}_q}), \bm{Q}  \rangle=1_{q=0}, \quad q=-N+1,..., N-1,
    \end{aligned}
\end{equation}

By solving \eqref{eq.dualprob_noisy}, we get the estimate of the dual vector $\widehat{\bs{\lambda}}$ and then we can find the support estimate and the estimate of the channel coefficients $|\widehat{c}_l^i|$ by \eqref{eq:support_estimate} and \eqref{eq:least_square}.

}
% This leads t
% following term For instance, an example of this channel estimation is depicted in Fig~\ref{fig:Daul} for a case with $r=2$. 
 
%  To recover the message vector and the channel amplitudes corresponding to user $k$, we form  $\widehat{\bm{Z}}_k=\sum\nolimits_{\ell}\widehat{g}_{\ell}^k \widehat{\bm{x}}_k\bm{a}(\widehat{\tau}_{\ell}^k)^{\mathsf{T}}$. Let $\widehat{\bm{g}}^k:=[\widehat{g}_1^k, \ldots,\widehat{g}_{P_k}^k]^{\mathsf{T}}$. Then, the rank one matrix $\widehat{\bm{x}}_k$ can be estimated as $\widehat{\bm{x}}_k\widehat{\bm{g}}^{k\mathsf{T}}=\widehat{\bm{Z}}_k\bm{A}^{(k)\dagger}$ where $\bm{A}^{(k)}:=[\bm{a}(\widehat{\tau}_k^1), \ldots,\bm{a}(\widehat{\tau}_k^{P_k})]^\mathsf{T}$. By using the assumption $\|\widehat{\bm{x}}_k\|_2=1$ and taking singular value decomposition, we can find $|\widehat{\bm{x}}_k|$ and $|\widehat{\bm{g}}^k|$ for $k=1, \ldots, K$.

\section{Performance Guarantee}\label{sec.performance_gauranty}

In this section, we provide a performance guarantee for the proposed SDP problem \eqref{prob.sdp}. This, in turn, provides the required amount of resources one needs to perform simultaneous channel estimation and message recovery. Before stating the main result of the paper, we first state the assumptions underlying our analysis below:
%----------------
 \begin{enumerate}\label{item:assumptions}
   \item   We assume the rows of $\mx{B}_i\in\mathbb{C}^{N\times k_i}$, i.e. ${\mx{b}^i_n}^{\mathsf{T}}, n=-2M, \ldots, 2M$ are i.i.d. sampled from a distribution $\mathcal{F}_i$ with the following properties:
 \begin{itemize}
    \item \textbf{Isotropy property}: We assume that the distribution of $\mx{b}_n^i$s satisfy the relation
 	\begin{align}\label{eq.isotropy}
 	\mathds{E}[\mx{b}^i{\mx{b}^i}^{\mathsf{H}}]=\mx{I}_{k_i},~~\mx{b}^i\sim \mathcal{F}_i, i\in[r].
 	\end{align}
 	\item \textbf{Incoherence property}: We assume that there exist upper and lower coherence parameters $\mu_+$ and $\mu_{-}$ such that the distributions $\mathcal{F}_i$s satisfy
 	\begin{align}\label{eq.incoherence}
 	\mu_{-}\le \|\mx{b}^i_n\|_{\infty}^2 \le \mu_{+}, \forall i\in[r].
 	\end{align}
 	almost surely.
 \end{itemize}
%----------------
  \item \textbf{Minimum separation}:  We define the separation between the delay parameters of the $i$-th channel as
	\begin{align}
	\Delta_i\triangleq\min_{k\neq q}|\tau^i_k-\tau^i_q|,
	\end{align} 
	and the minimum separation between all user by $\Delta\triangleq\min_{i} \Delta_i$. The absolute value in the latter definition is evaluated as the wrap-around distance on the unit circle.
	\item \textbf{Random message vectors}: We assume that the coefficient vectors $\mx{f}_i ,i\in[r]$ are i.i.d. random vectors in $\mathbb{R}^{k_i\times 1}$ with zero-mean and satisfy the following spectral norm condition on their covariance matrix
 \begin{align}
     \|\mathds{E}[\frac{\mx{f}_i\mx{f}_i^H}{\|\mx{f}_i\|_2^2}]\|_{2\rightarrow 2} \le \frac{1}{k_i} .
 \end{align}
 This assumption ensures that the expected covariance matrix is well-conditioned.
\end{enumerate}
%----------------

Building upon the aforementioned assumptions, we are prepared to present the main result of the paper. This result precisely outlines the number of samples required by the proposed method in Section \ref{sec:proposed_method} to guarantee precise delay estimation and message recovery.
\begin{thm}\label{thm.main_sample}
Let $\Delta\ge \frac{1}{M}$ and $M>71$. Assume that the codebook matrices of users are independent of each other and that the rows of the codebook $\mx{B}_i$ are distributed according to an i.i.d. distribution $\mathcal{F}_i$ that satisfies the isotropy and incoherence assumptions \eqref{eq.isotropy} and \eqref{eq.incoherence}. Also, assume that the message vector of each user is drawn i.i.d. from a uniform distribution on the complex unit sphere. Then, if the number of samples satisfies
%------------
\begin{align}\label{eq:main_sample}
    \hspace{-6pt}M\ge c_1 \mu_+(\sum_{i=1}^r s_i k_i)\frac{k_+}{k_-} \zeta_1(s_i,k_i)\zeta_2(s_i, k_i)\log^2\hspace{-2pt}\Big(\frac{M k_+}{\delta}\Big), 
\end{align}
%------------
where, 
%------------
\begin{subequations}
\begin{align}
   &  \zeta_1(s_i,k_i) = \log\Big(1+c_2\frac{\mu_+^2\sum_i s_ik_i}{\mu_-}\Big), \\
    & \zeta_2(s_i,k_i) = \log\Big(\frac{M\sum_i s_i k_i}{\delta}\Big),
\end{align}
\end{subequations}
%------------
for some constants $c_1,c_2>0$, then, the solution $\widehat{\bm{\mathcal{H}}}=(\widehat{\mx{H}}_i)_{i=1}^r$ of the optimization problem in \eqref{eq.primalprob} is equal to the ground-truth matrix tuple ${\bm{\mathcal{H}}}$ with probability at least $1-\delta$.

Proof. See Appendix~\ref{proof.thm.main_sample}.
% there exists numerical constants $c_1 , c_2>0$ such that 
\end{thm}

\begin{figure}
    \centering
    \begin{tikzpicture}
\begin{axis}[title={DoF of OBDD problem},
xlabel={$s_i$},
ylabel={$k_i$},
width=0.45\textwidth,
        height=6cm,
        xmin=1, xmax=5,
        ymin=1, ymax=5,
        legend style={nodes={scale=0.42, transform shape}, at={(0.45,1)}}, 
        ticklabel style = {font=\footnotesize},
        ytick={1, 2, 3, 4, 5},
        ymajorgrids=true,
        xmajorgrids=true,
        grid style=dashed,
        grid=both,
        grid style={line width=.1pt, draw=gray!10},
        major grid style={line width=.2pt,draw=gray!30},
]
    \addplot3 [
        color = black,
        surf,
        shader=flat,
        fill = cadmiumorange,
        colormap name=hot,
        samples=20,
        domain=1:5,
    ] {x + y-1};
  \addplot3 [ opacity=0.2,fill opacity=0.5,
        surf,
        color=blue,
        shader=flat,
        colormap name=hot,
        samples=20,
        domain=1:5,
    ] {x*y};
  % \addplot3 [ opacity=0.2,fill opacity=0.5,
  %       surf,
  %       color=blue,
  %       shader=flat,
  %       colormap name=hot,
  %       samples=20,
  %       domain=1:5,
  %   ] {x*y};
    \legend{\Large Independent factor DoF ($s_i + k_i-1$), \Large Atomic DoF ($s_ik_i$)};
\end{axis}
   \end{tikzpicture}
    \caption{{\color{\change}A schematic comparison of DoF in the OBDD problem. The Atomic DoF represents the DoF required to jointly recover the channel and messages by reconstructing their outer product, whereas the Independent Factor DoF treats the channel and message signals independently, using an alternating recovery approach}}
    \label{fig:sample_complexity}
\end{figure}

In what follows, our aim is to delve into different aspects of the bound derived in \eqref{eq:main_sample}. This includes exploring its connection to the degrees of freedom in the proposed convex problem, as well as its relationship with the inherent degrees of freedom in the OBDD problem, as illustrated in Figure \ref{fig:sample_complexity}.

\begin{rem}(Discussion about DoF)\label{rem.dof}
{\color{\change}By assuming that the steering vector can be expressed as a sparse vector in a redundant dictionary $\mx{F}\in\mathbb{C}^{N\times \infty}$, we have $\mx{a}(\tau)=\mx{F}\mx{c}$, where $\mx{c}\in\mathbb{C}^{\infty \times 1}$ is an infinite dimensional sparse vector with sparsity level $s$. Each atom in \eqref{eq.atoms} can be expressed as $\mx{f}\mx{a}^T(\tau)=\mx{f}\mx{c}^T \mx{F}^T$. Now, the DoF in finding each atom is to find $\mx{f}\in\mathbb{C}^{k\times 1}$ and $\mx{c}$ from rank-1 matrix $\mx{f}\mx{c}^T$.

There are two types of recovery methods leading to different effective DoF of the problem:
\begin{enumerate}
    \item Atomic DoF: In the matrix-valued atomic approach, the matrix $\mx{f}\mx{c}^T$ is treated as a structured matrix with block-sparse structure since $\mx{c}$ is a sparse vector. Actually, only $s$ columns of this matrix have non-zero $\ell_2$ norms and the rest have zero $\ell_2$ norms. Each of this non-zero columns contributes $k$ DoF and thus the effective DoF becomes $s k$. This formulation utilizes ANM to leverage the block-sparse structure.
    \item Independent Factor DoF: Instead of treating $\mx{f}\mx{c}^T$ as a single block-sparse matrix, the independent factor approach treats $\mx{f}$ and $\mx{c}$ as separate factors and optimize them alternatively. This formulation uses the sparsity of $\mx{c}$ explicitly without imposing a block-sparse structure. More precisely, the independent alternative approach alternates between solving $\mx{f}$ while fixing $\mx{c}$ and then solving for $\mx{c}$ while fixing $\mx{f}$. As a result, the effective DoF is significantly reduced to $s+k$, with $s$ DoF from the non-zero elements of $\mx{c}$ and $k$ DoF from the elements in $\mx{f}$. If there is additional constrains such as $\|\mx{f}\|_2=1$, then the effective DoF in this case can be reduced to $s+k-1$.
    
\end{enumerate}
Inline with the above explanations, \cite{kech2017optimal} has shown that in the single-use case where $r=1$, the minimum number of samples needed to recover $\mx{x}$ and $\mx{h}$ from their pointwise product $\mx{x}\odot \mx{h}$ is $s+k$ where $s$ is the sparsity of $\mx{h}$ and $k$ is the dimension of the subspace in which $\mx{x}$ lives. This implies that the independent factor DoF in multi-user OBDD problem is $\sum_i (s_i+k_i-1)$ as illustrated in Figure \ref{fig:sample_complexity}. In the single-user case, by using alternative optimization and some additional constraints on $\mx{h}$ and $\mx{x}$, \cite{lee2017near} proposed an algorithm that achieves this amount of sample complexity. However, as of today, to the best of the authors' knowledge, there is no convex programming approach capable of achieving this level of sample complexity.
}
\end{rem}
\begin{rem}(Discussion about sample complexity)
{\color{\change}
The amount of measurements provided in \eqref{eq:main_sample} is more or less what could be expected by convex semidefinite programming and aligns with the atomic DoF of the multi-user OBDD problem. In fact,  when restricted to lifting-type convex problems, OBDD aims to recover a signal from a $\sum_i {s_i k_i}$-dimensional structure embedded in an continuous atomic set by observing $N$ time samples}. Intuitively, based on \cite[Theorem I.1]{tang2013compressed}, this calls for $\mathcal{O}(\sum_i s_i k_i\log(\sum_i s_i k_i))$ samples which is more or less the same with our bound in \eqref{eq:main_sample} if the message length for users does not too much differ. In particular, we have equality for $r=1$ and equal message length ($k_+=k_-$). {\color{\change} The coherence parameters $\mu_+$ and $\mu_-$ in \eqref{eq:main_sample} play essential roles in determining the sample complexity by controlling the distribution of energy across entries in each user’s codebook vector $\mx{b}^i$ for each user $i$. Here, we outline their specific effects:
\begin{itemize}
    \item Upper coherence $\mu_+$: This parameter limits the maximum squared magnitude of entries in $\mx{b}^i$, preventing any single component from dominating. Higher values of $\mu_+$ indicate increased coherence, which raises the sample complexity $M$ necessary for accurate signal recovery. By controlling component dominance, $\mu_+$ promotes a more balanced energy distribution across the entries of $\mx{b}^i$, thereby improving signal separability and reducing sample requirements.
    \item Lower coherence $\mu_-$: This parameter sets a minimum threshold for the squared magnitudes of the entries in $\mx{b}^i$, ensuring that each component contributes meaningfully to the vector. Higher values of $\mu_-$ improve recovery stability and a larger $\mu_-$  results in a tighter, more efficient sample complexity bound by stabilizing energy distribution across components.
\end{itemize}
Together, $\mu_+$ and $\mu_-$ define a coherence range that balances concentration and spread across the entries in $\mx{b}^i$. An optimal balance minimizes redundancy and ensures efficient recovery in multi-user scenarios, directly impacting the sample complexity necessary to achieve reliable separation and accurate signal reconstruction. It is worth noting that due to the isotropy property \eqref{eq.isotropy}, it holds that $\mathds{E}\|\mx{b}^i\|_2^2=k_i$ leading to the relation $1\le \mu_{-}\le \mu_{+}\le k_{+}$. }
\end{rem}
\begin{rem}(Discussion about the assumptions in Theorem \ref{thm.main_sample})
     It is important to emphasize that the assumptions mentioned in the initial part of Section \ref{sec.performance_gauranty} are considered sufficient for our theoretical analysis to hold. However, in Section \ref{sec:simulations}, we encounter practical scenarios where the proposed method in Section \ref{sec:proposed_method} operates with significantly more relaxed assumptions. 
For instance, the proposed method in Section \ref{sec:proposed_method} succeeds even with a minimum separation condition looser than $\Delta > \frac{1}{M}$. The isotropy and incoherence condition can also be extended to encompass correlations between codebooks of different users. Additionally, the randomness of the message vectors does not appear to be a strict requirement in practice, nor is it necessary.
\end{rem}
{\color{\change}
\begin{rem}(Comparison of \eqref{eq:main_sample} with \cite[Theorem III.1]{wakinsuperres} for $r=1$)
For a single-user case $r=1$, the required number of samples in \cite{wakinsuperres} is given as
\begin{align}\label{eq:sample_wakin}
 M\ge c \mu_+ s k   \log\Big(\frac{M s k}{\delta}\Big)\log^2\Big(\frac{M k}{\delta}\Big)
\end{align}
where $c$ is a large constant. In comparison, our bound in \eqref{eq:main_sample} for $r=1$ yields:
\begin{align}\label{eq:main_sample_single}
    M\ge c_1 \mu_+ s k\log\Big(1+c_2\frac{\mu_+^2 s k}{\mu_-}\Big)\log\Big(\frac{M s k}{\delta}\Big)\log^2\Big(\frac{M k}{\delta}\Big)
\end{align}
for some small constant $c_1,c_2>0$.{\color{\chang} We note that our bound includes an extra logarithmic factor, $\log\Big(1+c_2\frac{\mu_+^2 s k}{\mu_-}\Big)$, which explicitly incorporates both the upper and lower coherence parameters $\mu_{+}$ and $\mu_{-}$. Our assumptions in \eqref{eq.isotropy} and \eqref{eq.incoherence} guarantee that $\mu_-\ge 1$ and $\mu_{+}\le k_{+}$. Thus, the extra term remains well-defined and reflects the variation in the codebook entries. In particular, it makes our sample complexity bound adaptive to the actual codebook design: if the codebook is well spread (i.e. ${\mu_{-}}\approx \mu_{+}\approx 1$), then the logarithmic term remains moderate; if the codebook is highly coherent (i.e., $\mu_{-}\approx 1, \mu_{+}\approx k_+$ is large), the term grows accordingly, reflecting the increased difficulty of recovery. While in the worst-case scenario the extra term might result in a slightly looser bound compared to \cite[Theorem III.1]{wakinsuperres}, this adaptivity is advantageous in many practical scenarios where the codebook is designed to be nearly incoherent. Moreover, our analysis extends naturally to the multi-user case ($r>1$), where such an adaptive bound is essential.}

\end{rem}
\begin{rem}(Comparison of \eqref{eq:main_sample} with \cite[Theorem 2]{vargas2023dual} for $r=2$)
{\color{\chang}

In the case of two users ($r=2$) with equal message lengths $k_1=k_2=k$ and known channel coefficients satisfying $|c_l^i|=1$ for $i=1,2$, the sample complexity bound derived in \cite[Theorem 2]{vargas2023dual} is given by:
\begin{align}\label{eq:vargas_bound}
  N\ge c_2 s_+ k  \log^2(\frac{N k}{\delta})\log(\frac{N s_+ k}{\delta}),
\end{align}
where $c_2$ is a constant.
If the result is generalized to the case of $r>2$, then, one expects to have a sample complexity of order $r s_{+} k_{+}$. Since $r$ is treated as a constant in \cite{vargas2023dual}, this multiplicative factor is absorbed into their constant. In contrast, our analysis—although more involved—yields a sample complexity scaling as $\sum_{i} s_i k_i$, thereby adapting to the heterogeneity across users. This refinement is particularly important in practical settings where user parameters vary and the channel coefficients are unknown. In such cases, our bound more accurately captures the intrinsic degrees of freedom of the convex recovery problem.
These features make our result both more general and, in many practical scenarios, more precise than the bound in \cite[Theorem 2]{vargas2023dual}.
}
\end{rem}
}
In the subsequent section, we aim to provide some numerical results to evaluate the performance of the proposed method in Section \ref{sec:proposed_method}.

\section{Numerical Experiments}\label{sec:simulations}
\begin{figure*}[!t]
\centering
\subfigure[]{\label{Fig(a):Circ}
 \begin{tikzpicture} 
 \begin{scope}[spy using outlines={rectangle, magnification=4,
   width=1.6cm,height=1.6cm,connect spies}]
    \begin{axis}[
        width=6cm,
        height=6cm,
        xmin=-1.015, xmax=1.015,
        ymin=-1.015, ymax=1.015,
        legend style={nodes={scale=0.65, transform shape}, at={(0.7,0.92)}}, 
        ticklabel style = {font=\footnotesize},
        ymajorgrids=true,
        xmajorgrids=true,
        grid style=dashed,
        grid=both,
        grid style={line width=.1pt, draw=gray!10},
        major grid style={line width=.2pt,draw=gray!30},
    ]
    \addplot[only marks,
        color=chestnut,
        mark=star,
        line width=0.75pt,
        mark size=2pt,
        ]
    table[x=X1,y=Y1]
    {Sim1.dat};
   \addplot[only marks,
        color=chestnut,
        mark=o,
        line width=0.75pt,
        mark size=2pt,
        ]
    table[x=TX1,y=TY1]
    {Sim2.dat};

    \draw[azure,thick,dashed] (axis cs:0,0) circle [radius=1];
    
    % =============
    \addplot[only marks,
        color=airforceblue,
        mark=star,
        mark options = {rotate = 180},
        line width=0.75pt,
        mark size=2pt,
        ]
    table[x=X2,y=Y2]
    {Sim1.dat};
    \addplot[only marks,
        color=airforceblue,
        mark=o,
        line width=0.75pt,
        mark size=2pt,
        ]
    table[x=TX2,y=TY2]
    {Sim2.dat};
    % =============
    \addplot[only marks,
        color=black,
        mark=star,
        line width=0.75pt,
        mark size=2pt,
        ]
    table[x=X3,y=Y3]
    {Sim1.dat};
    \addplot[only marks,
        color=black,
        mark=o,
        line width=0.75pt,
        mark size=2pt,
        ]
    table[x=TX3,y=TY3]
    {Sim2.dat};
    % =============
    \addplot[only marks,
        color=cadmiumorange,
        mark=star,
        line width=0.75pt,
        mark size=2pt,
        ]
    ttable[x=X4,y=Y4]
    {Sim1.dat};
    \addplot[only marks,
        color=cadmiumorange,
        mark=o,
        line width=0.75pt,
        mark size=2pt,
        ]
    table[x=TX4,y=TY4]
    {Sim2.dat};
    \legend{Estimated delays, Ground-truth delays};
     \path (0.75, -0.62) coordinate (X);
    \end{axis}
    \spy [black] on (X) in node (zoom) [left] at ([xshift=-0.5cm,yshift=1cm]X);
    \end{scope}
\end{tikzpicture}
}
% =====================
\subfigure[]{\label{Fig(b):Circ}
\begin{tikzpicture} 
    \begin{axis}[
       % xlabel={$\cos(t)$},
        %ylabel={$\sin(t)$},
        %label style={font=\tiny},
        %title = {\small $f(\bm{x}) = \prod_{k=1}^Kx_k$},
        % legend cell align={left},
        width=6cm,
        height=6cm,
        xmin=-1.015, xmax=1.015,
        ymin=-1.015, ymax=1.015,
        legend style={nodes={scale=0.65, transform shape}, at={(0.7,0.9)}}, 
        ticklabel style = {font=\footnotesize},
        ymajorgrids=true,
        xmajorgrids=true,
        grid style=dashed,
        grid=both,
        grid style={line width=.1pt, draw=gray!10},
        major grid style={line width=.2pt,draw=gray!30},
    ]
    \addplot[only marks,
        color=chestnut,
        mark=star,
        line width=1pt,
        mark size=3pt,
        ]
    table[x=TEX1,y=TEY1]
    {resultR3L5N128NEW.dat};
   \addplot[only marks,
        color=chestnut,
        mark=o,
        line width=1pt,
        mark size=3pt,
        ]
    table[x=TX1,y=TY1]
    {resultR3L5N128NEW.dat};

    \draw[azure,thick,dashed] (axis cs:0,0) circle [radius=1];
    
    % =============
    \addplot[only marks,
        color=airforceblue,
        mark=star,
        mark options = {rotate = 180},
        line width=1pt,
        mark size=3pt,
        ]
    table[x=TEX2,y=TEY2]
    {resultR3L5N128NEW.dat};
    \addplot[only marks,
        color=airforceblue,
        mark=o,
        line width=1pt,
        mark size=3pt,
        ]
    table[x=TX2,y=TY2]
    {resultR3L5N128NEW.dat};
    % =============
    \addplot[only marks,
        color=black,
        mark=star,
        line width=1pt,
        mark size=3pt,
        ]
    table[x=TEX3,y=TEY3]
    {resultR3L5N128NEW.dat};
    \addplot[only marks,
        color=black,
        mark=o,
        line width=1pt,
        mark size=3pt,
        ]
    table[x=TX3,y=TY3]
    {resultR3L5N128NEW.dat};
    \legend{Estimated delays, Ground-truth delays};
    \end{axis}
\end{tikzpicture}
}
% =====================
\subfigure[]{\label{Fig(c):Circ}
\begin{tikzpicture} 
    \begin{axis}[
       % xlabel={$\cos(t)$},
        %ylabel={$\sin(t)$},
        %label style={font=\tiny},
        %title = {\small $f(\bm{x}) = \prod_{k=1}^Kx_k$},
        % legend cell align={left},
        width=6cm,
        height=6cm,
        xmin=-1.015, xmax=1.015,
        ymin=-1.015, ymax=1.015,
        legend style={nodes={scale=0.65, transform shape}, at={(0.7,0.95)}}, 
        ticklabel style = {font=\footnotesize},
        ymajorgrids=true,
        xmajorgrids=true,
        grid style=dashed,
        grid=both,
        grid style={line width=.1pt, draw=gray!10},
        major grid style={line width=.2pt,draw=gray!30},
    ]
    \addplot[only marks,
        color=chestnut,
        mark=star,
        line width=1pt,
        mark size=3pt,
        ]
    table[x=X1,y=Y1]
    {resultR2L16N64NEW.dat};
   \addplot[only marks,
        color=chestnut,
        mark=o,
        line width=1pt,
        mark size=3pt,
        ]
    table[x=TX1,y=TY1]
    {resultR2L16N64NEW.dat};

    \draw[azure,thick,dashed] (axis cs:0,0) circle [radius=1];
    
    % =============
    \addplot[only marks,
        color=airforceblue,
        mark=star,
        mark options = {rotate = 180},
        line width=1pt,
        mark size=3pt,
        ]
    table[x=X2,y=Y2]
    {resultR2L16N64NEW.dat};
    \addplot[only marks,
        color=airforceblue,
        mark=o,
        line width=1pt,
        mark size=3pt,
        ]
    table[x=TX2,y=TY2]
    {resultR2L16N64NEW.dat};
    % =============
    \legend{Estimated delays, Ground-truth delays};
    \end{axis}
\end{tikzpicture}
}
  \caption{ Delay estimation performance of the proposed method using polar representation. In these figures, $\cos(2\pi \tau)$ is depicted versus $\sin(2\pi \tau)$ for $\tau\in[0,1)$ . Figure~\ref{Fig(a):Circ} shows the delay estimation in the case of $r=4$ users and $s_1=\cdots=s_4=3$, $N=200$ samples and the message size $k_i=5$ for $i\in [4]$. Figure~\ref{Fig(b):Circ} depict the performance of the proposed method for $r=3$ with $s_1=3,s_2=2,s_3 = 1$ from $N=128$ samples. Figure~\ref{Fig(c):Circ} shows the results of an experiment in the case of $k_i=16$ and $N=64$. The delay parameters corresponding to different users are depicted using distinct colors.
  }
   \label{fig:Plor}
\end{figure*}
% \input{Figures/Fig_Message}
% \begin{figure}
%     \centering
%     \includegraphics[scale=0.5]{Figures/NMSE_noisefree.eps}
%     \caption{Message recovery performance versus the number of samples $N$. Here, both users have the same message length $k_1=k_2=3$. Moreover, we consider the number of multi-path components as $s_1 = 3$ and $s_2 = 3$.}
%     \label{fig:Mas}
% \end{figure}
%-----------------------------
\begin{figure*}[!t]
\centering
 \begin{tikzpicture} 
    \begin{axis}[
        xlabel={SNR (dB)},
        ylabel={NMSE},
        width=0.48\textwidth,
        height=6cm,
        xmin=12, xmax=60,
        ymin=-1e-3, ymax=0.6,
        legend style={nodes={scale=0.6, transform shape}, at={(0.7,0.95)}}, 
        ticklabel style = {font=\footnotesize},
        ymajorgrids=true,
        xmajorgrids=true,
        grid style=dashed,
        grid=both,
        grid style={line width=.1pt, draw=gray!10},
        major grid style={line width=.2pt,draw=gray!30},
    ]
    \addplot[thin,
        color=chestnut,
        line width=0.9pt,
        ]
    table[x=N,y=nmse1]
    {nmse_noisyfree_fid.dat};
\addplot[
             thin,
        color=airforceblue,
        line width=0.9pt,
        ]
    table[x=N,y=nmse2]
    {nmse_noisyfree_fid.dat};
    \legend{${\rm NMSE}(1)$, ${\rm NMSE}(2)$};
    \end{axis}
\end{tikzpicture}

   \caption{Message recovery performance versus the number of samples $N$. Here, both users have the same message length $k_1=k_2=3$. Moreover, we consider the number of multi-path components as $s_1 = 3$ and $s_2 = 3$.}
    \label{fig:Mas}
\end{figure*}
%-----------------------------

In this section, we provide some numerical results to evaluate the performance of the proposed method in message recovery and delay estimation. Then numerical experiments are implemented using MATLAB CVX Toolbox \cite{grant2014cvx}.  The delays' locations are generated uniformly at random satisfying the minimum separation $\Delta\geq \frac{1}{N}$, which is slightly smaller than what Theorem \ref{thm.main_sample} suggests. The codebook matrices $\mx{B}_i\in\mathbb{C}^{N\times k_i}, i=1,..., r$ are generated uniformly at random from standard normal distribution $\mathcal{N}(0,1)$. The messages $\mx{f}_i, i=1, \ldots, r$ are generated \textit{i.i.d} and uniformly at random from the complex unit sphere.

In the first experiment depicted in Figure~\ref{Fig(a):Circ}, we examine a scenario with $r=4$ users, each having a message length of $k_i=5$, where $i=1,...,4$, and employing multi-path channels with $s_i=3$, while acquiring $N=200$ samples. The outcomes illustrated in Figure~\ref{Fig(a):Circ} demonstrate the effectiveness of our proposed method in accurately recovering the continuous-valued delay parameters associated with distinct users and its capability to distinguish closely-spaced delay parameters. In the second figure shown in Figure~\ref{Fig(b):Circ}, we repeat the latter experiment for the parameters $r=3$, $k_1=3, k_2=2, k_3=1$ , $N=128$. Figure~\ref{Fig(b):Circ} again shows the capability of the proposed method in exact delay estimation.
Note that in Figure~\ref{Fig(a):Circ}, the delay parameter is represented using polar coordinates, where $\cos(2\pi \tau)$ is plotted versus $\sin(2\pi\tau)$ for $\tau\in[0,1)$. Delays corresponding to the same user are indicated by the same color, while delays for different users are depicted using different colors.
% For the first case,  we set $K=4$ with $P_1=\cdots=P_4=3$ from $N=200$ samples, and the message size $M_k=5$ for $k\in [4]$. Also, the sensing matrix $\bm{D}$ is set to be identity, i.e., $\bm{D}=\bm{I}_N$. Then, the results are depicted in Fig~\ref{Fig(a):Circ}. Fig~\ref{Fig(a):Circ} shows that the \ac{GB2D} can distinguish two closely spaced delays. In Fig~\ref{Fig(b):Circ}, we repeat this experiment for $K=3$ with $M_1=3,M_2=2,M_3=1$ from $N=128$ samples.  

In the second experiment shown in Figure~\ref{Fig(c):Circ}, we examine the scalability of the method by having a larger message length for the users. Specifically, we consider the parameters $r=2$, $k_1=k_2=16$, $s_1=s_2=1$ and the number of samples is set to $N=64$. We observe that the proposed method can successfully estimate the delay parameters.

{\color{\change}
In the third experiment shown in Figure~\ref{fig:Mas}, we examine the message recovery performance of the proposed method. We consider two users with message length $k_1=k_2=3$ and the multi-path channels with $s_1=s_2=3$ and after estimating the delays by solving \eqref{prob.sdp}, we estimate the messages according to \eqref{eq:least_square}. We calculate the 
the normalized mean square error (NMSE) defined as
 \begin{align}
   {\rm NMSE}(i)\triangleq  \frac{\|\widehat{\mx{f}}_i-{\mx{f}}_i\|_2}{\|{\mx{f}}_i\|_2},
 \end{align}
where $\widehat{\mx{f}}_i$ is the estimated message vector.}

The results are shown in Figure~\ref{fig:Mas} and we observe that the proposed method can efficiently decode the magnitude of users' messages.

%----------------------------
\begin{figure*}[!t]
\centering
\subfigure[]{\label{fig:polar_noisy}
 \begin{tikzpicture} 
    \begin{axis}[
        width=6cm,
        height=6cm,
        xmin=-1.015, xmax=1.015,
        ymin=-1.015, ymax=1.015,
        legend style={nodes={scale=0.65, transform shape}, at={(0.7,0.92)}}, 
        ticklabel style = {font=\footnotesize},
        ymajorgrids=true,
        xmajorgrids=true,
        grid style=dashed,
        grid=both,
        grid style={line width=.1pt, draw=gray!10},
        major grid style={line width=.2pt,draw=gray!30},
    ]
    % =============
    \addplot[only marks,
        color=chestnut,
        mark=star,
        mark options = {rotate = 180},
        line width=1pt,
        mark size=3pt,
        ]
    table[x=XD2,y=YD2]
    {Root_estimate_c1.dat};
    \addplot[only marks,
        color=chestnut,
        mark=o,
        line width=1pt,
        mark size=3pt,
        ]
    table[x=XD1,y=YD1]
    {Root_estimate_c1.dat};

    \draw[azure,thick,dashed] (axis cs:0,0) circle [radius=1];
    
    % =============
    \addplot[only marks,
        color=airforceblue,
        mark=star,
        mark options = {rotate = 180},
        line width=1pt,
        mark size=3pt,
        ]
    table[x=XD4,y=YD4]
    {Root_estimate_c1.dat};
    \addplot[only marks,
        color=airforceblue,
        mark=o,
        line width=1pt,
        mark size=3pt,
        ]
    table[x=XD3,y=YD3]
    {Root_estimate_c1.dat};
    \legend{Estimated delays, Ground-truth delays};
    \end{axis}
\end{tikzpicture}
}\subfigure[]{\label{fig:dual_pol_noisy}
 \begin{tikzpicture} 
 \begin{scope}[spy using outlines={rectangle, magnification=5,
   width=1.6cm,height=1.6cm,connect spies}]
    \begin{axis}[
        width=0.5\textwidth,
        height=6cm,
        xmin=0, xmax=1,
        ymin=1e-3, ymax=1,
        legend style={nodes={scale=0.6, transform shape}, at={(0.3,0.3)}}, 
        ticklabel style = {font=\footnotesize},
        ymajorgrids=true,
        xmajorgrids=true,
        grid style=dashed,
        grid=both,
        grid style={line width=.1pt, draw=gray!10},
        major grid style={line width=.2pt,draw=gray!30},
    ]
    \addplot[ smooth,
             thin,
        color=chestnut,
        line width=0.9pt,
        ]
    table[x=xdata1,y=ydata2]
    {PolyDual4.dat};
\addplot[ smooth,
             thin,
        color=airforceblue,
        line width=0.9pt,
        ]
    table[x=xdata1,y=ydata1]
    {PolyDual4.dat};
    \addplot[ 
        color=cadmiumorange,
        mark=star,
        dashed,
        line width=1pt,
        mark size=1.5pt,
        ]
    table[x=t,y=X2]
    {PolyDual5.dat};
    \addplot[ 
        color=azure,
        mark=star,
        dashed,
        line width=1pt,
        mark size=1.5pt,
        ]
    table[x=t,y=X1]
    {PolyDual5.dat};
    \legend{$\|\mx{q}_1(\tau)\|_2$, $\|\mx{q}_2(\tau)\|_2$,$\tau^1$,$\tau^2$};
     \path (0.68, 0.99) coordinate (X);
    \end{axis}
    \spy [black] on (X) in node (zoom) [left] at ([xshift=5cm,yshift=-1.5cm]X);
    \end{scope}
\end{tikzpicture}
}
    \caption{{\color{\change}Delay estimation performance of the proposed method in the noisy case. In Figure \ref{fig:polar_noisy}, $\cos(2\pi \tau)$ is depicted versus $\sin(2\pi \tau)$ for $\tau\in[0,1)$. The ground-truth delay parameters are shown by circles while the estimated delays are shown by the star-like symbol. The delay parameters corresponding to different users are depicted using distinct colors.  Figure~\ref{fig:dual_pol_noisy} shows the $\ell_2$ norm of the dual polynomial function for two users. The used parameters in these two figures are as follows: $s_1=s_2=2$, $N=50$, ${\rm SNR}=5~dB$, $k_1=k_2=3.$}}
    \label{fig:polar+dual_pol_noisy}
\end{figure*}
%----------------------------

%----------------------------
\begin{figure*}[!t]
\centering
\subfigure[]{\label{fig:root_polar_noisy}
 \begin{tikzpicture} 
 \begin{scope}[spy using outlines={rectangle, magnification=4,
   width=1.6cm,height=1.6cm,connect spies}]
    \begin{axis}[
        width=6cm,
        height=6cm,
        xmin=-1.015, xmax=1.015,
        ymin=-1.015, ymax=1.015,
        legend style={nodes={scale=0.65, transform shape}, at={(0.7,0.92)}}, 
        ticklabel style = {font=\footnotesize},
        ymajorgrids=true,
        xmajorgrids=true,
        grid style=dashed,
        grid=both,
        grid style={line width=.1pt, draw=gray!10},
        major grid style={line width=.2pt,draw=gray!30},
    ]
    % =============
    \addplot[only marks,
        color=chestnut,
        mark=star,
        mark options = {rotate = 180},
        line width=1pt,
        mark size=3pt,
        ]
    table[x=XD1,y=YD1]
    {Root_estimate_a2.dat};
    \addplot[only marks,
        color=chestnut,
        mark=o,
        line width=1pt,
        mark size=3pt,
        ]
    table[x=XD1,y=YD1]
    {Root_estimate_a1.dat};

    \draw[azure,thick,dashed] (axis cs:0,0) circle [radius=1];
    
    % =============
    \addplot[only marks,
        color=airforceblue,
        mark=star,
        mark options = {rotate = 180},
        line width=1pt,
        mark size=3pt,
        ]
    table[x=XD2,y=YD2]
    {Root_estimate_a2.dat};
    \addplot[only marks,
        color=airforceblue,
        mark=o,
        line width=1pt,
        mark size=3pt,
        ]
    table[x=XD2,y=YD2]
    {Root_estimate_a1.dat};
    \legend{Estimated delays, Ground-truth delays};
     \path (-0.75, -0.62) coordinate (X);
    \end{axis}
    \spy [black] on (X) in node (zoom) [left] at ([xshift=2.5cm,yshift=1cm]X);
    \end{scope}
\end{tikzpicture}
}\subfigure[]{\label{fig:discretized_polar_noisy}
\begin{tikzpicture} 
    \begin{axis}[
        width=6cm,
        height=6cm,
        xmin=-1.015, xmax=1.015,
        ymin=-1.015, ymax=1.015,
        legend style={nodes={scale=0.65, transform shape}, at={(0.7,0.9)}}, 
        ticklabel style = {font=\footnotesize},
        ymajorgrids=true,
        xmajorgrids=true,
        grid style=dashed,
        grid=both,
        grid style={line width=.1pt, draw=gray!10},
        major grid style={line width=.2pt,draw=gray!30},
    ]
    % =============
    \addplot[only marks,
        color=chestnut,
        mark=star,
        mark options = {rotate = 180},
        line width=1pt,
        mark size=3pt,
        ]
    table[x=XD1,y=YD1]
    {Root_estimate_b2.dat};
    \addplot[only marks,
        color=chestnut,
        mark=o,
        line width=1pt,
        mark size=3pt,
        ]
    table[x=XD1,y=YD1]
    {Root_estimate_b1.dat};

    \draw[azure,thick,dashed] (axis cs:0,0) circle [radius=1];
    
    % =============
    \addplot[only marks,
        color=airforceblue,
        mark=star,
        mark options = {rotate = 180},
        line width=1pt,
        mark size=3pt,
        ]
    table[x=XD2,y=YD2]
    {Root_estimate_b2.dat};
    \addplot[only marks,
        color=airforceblue,
        mark=o,
        line width=1pt,
        mark size=3pt,
        ]
    table[x=XD2,y=YD2]
    {Root_estimate_b1.dat};
    % =============
    \legend{Estimated delays, Ground-truth delays};
    \end{axis}
\end{tikzpicture}
}
    \caption{{\color{\change}Delay estimation performance of the proposed method in the noisy case. In Figure \ref{fig:root_polar_noisy}, $\cos(2\pi \tau)$ is depicted versus $\sin(2\pi \tau)$ for $\tau\in[0,1)$ using the root-finding approach. The ground-truth delay parameters are shown by circles while the estimated delays are shown by the star-like symbol. The delay parameters corresponding to different users are depicted using distinct colors.  Figure~\ref{fig:discretized_polar_noisy} shows the polar representation via discretizing the delay domain and then finding the delays based on $\|\mx{q}_i(\tau)\|_2=1-{\rm th}$, with the threshold th chosen according to the SNR level. The used parameters in these two figures are as follows: $s_1=s_2=2$, $N=20$, ${\rm SNR}=5~dB$, $k_1=k_2=3.$}}
    \label{fig:root+discretized}
\end{figure*}
%----------------------------

{\color{\change}
In the fourth experiment, we consider a noisy case where the measurements are contaminated with additive zero-mean Gaussian noise with variance $\sigma^2_w$. Leveraging the concentration result provided in \cite[Section C]{cai2011orthogonal}, we have
\begin{align}
    \mathds{P}[\|\mx{w}\|_2\ge \sigma_{w}\sqrt{N+\sqrt{2N\log(N)}}]\le \frac{1}{2N}.
\end{align}
Consequently, we set the upper-bound of $\|\mx{w}\|_2$ as $\eta\triangleq\sigma_{w}\sqrt{N+\sqrt{2N\log(N)}}$ which is required in \eqref{prob.sdp_noisy}. The noise variance is generated based on the signal-to-noise ratio (SNR) defined as
\begin{align}
 {\rm SNR}\triangleq  10\log_{10}\frac{\|\sum_{i=1}^r \mx{h}_i \odot \mx{x}_i\|_2^2}{N\sigma^2_w}. 
\end{align}
In Figure \ref{fig:polar+dual_pol_noisy}, we consider a highly noisy setting with $N=50$ noisy measurements at an SNR of $5~ dB$. Then, based on the given SNR, we compute $\sigma_w$ and $\eta$. We consider two users with message sizes $k_1=k_2=3$ and the multi-path channels with the sparsity levels $s_1=s_2=2$. Solving \eqref{prob.sdp_noisy}, we estimate the users' delay parameters. The polar representation of the estimates are shown in Figure \ref{fig:polar_noisy}, while the $\ell_2$ norm of the dual polynomial function is depicted in Figure \ref{fig:dual_pol_noisy}. In this high-noise setting, the proposed method struggles to fully recover the true delay parameters, producing one false estimate. Figure \ref{fig:dual_pol_noisy} suggests using an SNR-dependent threshold ${\rm th}$, whereby estimates with $\|\mx{q}(\tau)\|_2$ close to $1-{\rm th}$ are considered to be valid.  This threshold can be chosen based on the application-specific desired probability of detection and false alarm.

Figure \ref{fig:root+discretized}, presents an experiment that evaluates the sensitivity of delay parameter estimation in a high-noise setting, with SNR set to $5 dB$. Two methods are compared: the root-finding approach, based on \cite[Eq. 4.4]{candes2014towards}, and a discretized delay domain approach.
In the discretized method, delays are identified by the condition $\|\mx{q}_i(\tau)\|_2=1-{\rm th}$ where the threshold 
$t_h$ is chosen according to the SNR level and set to $t_h=0.1$ in this experiment. The results of the root-finding approach are shown in Figure \ref{fig:root_polar_noisy}, and the discretized approach results are displayed in Figure \ref{fig:discretized_polar_noisy}. The root-finding method demonstrates more stability under high noise conditions compared to the discretized approach.

In an additional experiment illustrated in Figure \ref{fig:Message recovery}, we consider $r=2$ users and first generate the source symbol vectors $\mx{f}^S_i\in\mathbb{R}^{k_i\times 1}, i=1,...,r$, with elements chosen from the integer set $\{0,1,2,3\}$ (ASK constellation with order $M_{\rm ASK}=4$). These vectors are normalized to create the message vectors $\mx{f}_i, i=1,..., r$. 
We examine the performance of our method in recovering the source symbol vectors and the message vectors for two users across various SNR levels with $N=50$ samples. At the receiver, we first estimate the transmitted message vectors $\widehat{\mx{f}}_i$ based on \eqref{eq:least_square} and round each element to the nearest integer to recover the source symbol vectors $\widehat{\mx{f}}^S_i\in\mathbb{R}^{k_i\times 1}$. We calculate the symbol error rate (SER) as
% We consider a range of SNR levels and examine the performance of our proposed method in recovering the message $\mx{f}_i, i=1,2$ for two users. The number of samples is set to $N=50$. At the receiver, we first estimate the transmitted message vectors $\widehat{\mx{f}}_i$ based on \eqref{eq:least_square} and then round each element of the estimated vector to the nearest integer to have an estimate of the source symbol vectors $\widehat{\mx{f}}^S_i\in\mathbb{R}^{k_i\times 1}$. Then, we calculate the symbol error rate (SER) defined as
\begin{align}
   {\rm SER}(i)\triangleq \frac{\|{\widehat{\mx{f}}}^S_i-{\mx{f}}^S_i\|_0}{k_i}.
\end{align}
For each SNR level, the codebook matrices $\mx{B}_i, i=1,2$ and the additive noise are generated randomly and recovery is examined over $50$ Monte Carlo iterations to compute average SER and NMSE. The results are displayed in Figures \ref{fig:SER_noisy} and \ref{fig:NMSE_noisy}, showing successful recovery of all source symbols and message vectors in SNR levels higher than 20 dB. 

In summary, our experiments demonstrate that the proposed method consistently provides effective message recovery and accurate estimation of closely-spaced delay parameters across varying multi-path components and message lengths. Figures \ref{fig:Plor} and \ref{fig:polar_noisy} highlight delay estimation performance in both noise-free and noisy conditions, while the accuracy of message and symbol recovery in Figures \ref{fig:Mas}, \ref{fig:Message recovery} show that the performance of the proposed method can reach approximately $10\log_{10}(\frac{1}{\rm NMSE})\approx 50$ dB in noise-free case and performs reliably in different levels of SNR in noisy conditions.
% \gf{\it GF: How can the MSE be 50 dB, when the MSE should be less than 1 in the linear domain ?}

%----------------------------
\begin{figure*}[!t]
\centering
\subfigure[]{\label{fig:SER_noisy}
 \begin{tikzpicture} 
    \begin{axis}[
        xlabel={SNR (dB)},
        ylabel={SER},
        width=0.48\textwidth,
        height=6cm,
        xmin=5, xmax=50,
        ymin=-1e-3, ymax=0.3,
        legend style={nodes={scale=0.6, transform shape}, at={(0.7,0.95)}}, 
        ticklabel style = {font=\footnotesize},
        ymajorgrids=true,
        xmajorgrids=true,
        grid style=dashed,
        grid=both,
        grid style={line width=.1pt, draw=gray!10},
        major grid style={line width=.2pt,draw=gray!30},
    ]
    \addplot[
             thin,
        color=chestnut,
        %airforceblue
        %mark=star,
        %mark options = {rotate = 180},
        line width=0.9pt,
        ]
    table[x=N,y=nmse2]
    {nmse_noisy_fid.dat};
    \addplot[thin,
        color=airforceblue,
        line width=0.9pt,
        ]
    table[x=N,y=nmse1]
    {nmse_noisy_fid.dat};

    \legend{${\rm SER}(1)$, ${\rm SER}(2)$};
    \end{axis}
\end{tikzpicture}
}\subfigure[]{\label{fig:NMSE_noisy}
 \begin{tikzpicture} 
    \begin{axis}[
        xlabel={SNR (dB)},
        ylabel={NMSE},
        width=0.48\textwidth,
        height=6cm,
        xmin=5, xmax=50,
         ymin=0, ymax=0.2,
         ytick={0, 0.05, 0.1, 0.15, 0.2},
        legend style={nodes={scale=0.6, transform shape}, at={(0.7,0.95)}}, 
        ticklabel style = {font=\footnotesize},
        ymajorgrids=true,
        xmajorgrids=true,
        grid style=dashed,
        grid=both,
        grid style={line width=.1pt, draw=gray!10},
        major grid style={line width=.2pt,draw=gray!30},
    ]
    \addplot[
             thin,
        color=chestnut,
        line width=0.9pt,
        ]
    table[x=N,y=nmse3]
    {nmse_noisy_fid.dat};
\addplot[thin,
        color=airforceblue,
        line width=0.9pt,
        ]
    table[x=N,y=nmse4]
    {nmse_noisy_fid.dat};
    \legend{${\rm NMSE}(1)$, ${\rm NMSE}(2)$};
    \end{axis}
\end{tikzpicture}
}

  \caption{{\color{\change}Message and symbol recovery performance versus SNR. Here, both users have the same message length $k_1=k_2=3$. We used the ASK constellation with order $M_{\rm ASK}=4$ to generate integer-valued symbols. Moreover, we consider the number of multi-path components as $s_1 = 1$ and $s_2 = 1$. Figure~\ref{fig:SER_noisy} shows the SER performance of the symbol recovery, while Figure~\ref{fig:NMSE_noisy} shows the NMSE performance of message recovery.}}
   \label{fig:Message recovery}
\end{figure*}
%----------------------------

}

\section{Conclusion and Future Directions}\label{sec:conclusion}

In this paper, we developed a novel approach for simultaneous message recovery and channel estimation tailored for IoT applications, where active devices sporadically transmit short messages to the single-antenna receiver. Addressing the inherently nonlinear nature of the problem, we leverage the lifting trick to transform it into a convex linear problem in higher dimensions. Our method employs semidefinite programming to extract continuous-valued delay parameters corresponding to all users' channels and estimate message magnitudes. Additionally, we propose a sample complexity bound that theoretically specifies the minimum required amount of resources for our solution, directly proportional to the degrees of freedom in the lifting-type matrix recovery formulation. Numerical results are presented to validate our approach's effectiveness in delay estimation and message recovery. Here, we raise several follow-up questions. Exploring scenarios where the receiver is equipped with multiple antennas and examining how the sample complexity changes in such multi-antenna setups would be intriguing. Additionally, a potential future direction could involve exploring non-convex approaches with provable performance guarantees to address the OBDD problem. Answers to these questions could hold significant relevance, especially concerning the future of IoT and ISAC.

% \section{Acknowledgements}

% Saeed Razavikia acknowledges and is grateful to  Mohammad Mahdi Kamjoo for helpful discussions and valuable feedback. 
% \texorpdfstring{\eqref{eq.dualprob}
\appendices
\section{Proof of the dual problem (\ref{eq.dualprob})}\label{proof.dual}
{\color{\change}We prove the dual of the problem provided in \eqref{eq.primalprob_noisy}. For the noise-free case, just put $\eta=0$, leading to \eqref{eq.dualprob}.
By assigning the Lagrangian vector $\mx{\bs{\lambda}}\in\mathbb{C}^N$ to the equality constraint of \eqref{eq.primalprob_noisy} and $\rho$ to the inequality constraint ($\|\widetilde{\mx{y}}\|_2^2\le \eta^2$), we can write the Lagrangian function as
%------------
\begin{align}\label{eq.lagran1}
L(\mx{\bs{\lambda}},\rho)&=\nonumber\\
&\inf_{\widetilde{\mx{y}},\bm{\mathcal{Z}}\in\bigoplus_{i=1}^r\mathbb{C}^{k_i\times N}}\Big[\sum_{i=1}^r\|\mx{Z}_i\|_{\mathcal{A}_i}+\langle \mx{\bs{\lambda}},\mx{y}-\mathcal{B}(\bm{\mathcal{Z}})-\widetilde{\mx{y}} \rangle\nonumber\\
&+\rho(\|\widetilde{\mx{y}}\|_2^2-\eta^2)\Big]=\langle \mx{\bs{\lambda}},\mx{y}\rangle-\eta^2 \rho+\sum_{i=1}^r\nonumber\\
&\inf_{\mx{Z}_i\in\mathbb{C}^{k_i\times N}}\Big[\|\mx{Z}_i\|_{\mathcal{A}_i}-\langle (\mathcal{B}^{\rm Adj}\mx{\bs{\lambda}})_i, \mx{Z}_i \rangle\Big]+\nonumber\\
&\inf_{\widetilde{\mx{y}}} \Big[\rho \|\widetilde{\mx{y}}\|_2^2-\langle \bs{\lambda}, \widetilde{\mx{y}} \rangle\Big]\\
% &\langle \mx{\bs{\lambda}},\mx{y}\rangle+\sum_{i=1}^r\inf_{\mx{Z}_i\in\mathbb{C}^{k_i\times N}}\Big[\|\mx{Z}_i\|_{\mathcal{A}_i}-\langle (\mathcal{B}^{\rm Adj}\mx{\bs{\lambda}})_i, \mx{Z}_i \rangle\Big],
\end{align}
%------------
where we used the fact that 
%------------
\begin{align}
\langle \mathcal{B}^{\rm Adj}\mx{\bs{\lambda}}, \bm{\mathcal{Z}}\rangle=\sum_{i=1}^r\langle (\mathcal{B}^{\rm Adj}\mx{\bs{\lambda}})_i, \mx{Z}_i\rangle.
\end{align}

The expression \eqref{eq.lagran1} in terms of $\widetilde{\mx{y}}$ is convex and the optimal value can be found by setting the derivative to zero which leads to $\widetilde{\mx{y}}=\frac{\bs{\lambda}}{2\rho}$. To find the infimum with respect to $\bm{\mathcal{Z}}$, we use Holder's inequality.
By replacing the optimal $\widetilde{\mx{y}}$ into \eqref{eq.lagran1} and using Holder's inequality , \eqref{eq.lagran1} reads as follows:
% \begin{align}
% \end{align}
%------------
% By using Holder's inequality, \eqref{eq.lagran1} becomes equivalent to
%------------
\begin{align*}
L(\mx{\bs{\lambda}},\rho)&=\nonumber\\
&\langle \mx{\bs{\lambda}},\mx{y}\rangle-\eta^2 \rho-\frac{\|\bs{\lambda}\|_2^2}{4\rho}+\nonumber\\
&\sum_{i=1}^r\inf_{\mx{Z}_i\in\mathbb{C}^{k_i\times N}}\Big[\|\mx{Z}_i\|_{\mathcal{A}_i}(1-\|(\mathcal{B}^{\rm Adj}\mx{\bs{\lambda}})_i\|_{\mathcal{A}_i}^{\mathsf{d}})\Big].
\end{align*}

%------------
It is straightforward to solve the latter optimization problem to reach

%------------
\begin{align*}
&L(\mx{\bs{\lambda}},\rho)=\nonumber\\
&\left\{\begin{array}{lr}
\langle \mx{\bs{\lambda}},\mx{y}\rangle-\eta^2 \rho-\frac{\|\bs{\lambda}\|_2^2}{4\rho}&~~ \|(\mathcal{B}^{\rm Adj}\mx{\bs{\lambda}})_i\|_{\mathcal{A}_i}^{\mathsf{d}}\le 1,~ i\in[r]\\
-\infty&{\rm o.w.}
\end{array}\right\}.	
\end{align*}
%------------
By maximizing over $\rho$ and transforming implicit constraints into explicit ones, the dual problem reads
%------------
\begin{align}\label{eq.lagran2}
\max_{\mx{\bs{\lambda}}\in\mathbb{C}^{N}}\langle  \mx{\bs{\lambda}},\mx{y}\rangle-\eta \|\bs{\lambda}\|_2~,~~\|(\mathcal{B}^{\rm Adj}\mx{\bs{\lambda}})_i\|_{\mathcal{A}_i}^{\mathsf{d}}\le 1,~~ i=1, \ldots, r.
\end{align}}
%------------
The only point that remains is that the
%where the infimum is unbounded unless $\|(\mathcal{B}^{\rm Adj}\mx{\bs{\lambda}})_i\|_{\mathcal{A}_i}^{\mathsf{d}}\le 1$
dual atomic norm $\|\cdot\|_{\mathcal{A}_i}^{\mathsf{d}}$ at an arbitrary point $\mx{Z}\in \mathbb{C}^{k_i\times N}$ is defined as
%------------
\begin{align}\label{eq.dualnorm1}
&\|\mx{Z}\|_{\mathcal{A}_i}^{\mathsf{d}}\triangleq\sup_{\|\mx{H}\|_{\mathcal{A}_i}\le 1}{\rm Re}~\langle \mx{Z}, \mx{H} \rangle=\sup_{\substack{\tau\in [0,1)\\\|\mx{f}\|_2=1}}{\rm Re}~\langle \mx{Z}, \mx{f}\mx{a}(\tau)^{\mathsf{T}} \rangle\nonumber\\
&=\sup_{\substack{\tau\in [0,1)\\\|\mx{f}\|_2=1}}{\rm Re}~\langle \mx{f}, \mx{Z}\mx{a}^*(\tau) \rangle=\sup_{\tau\in [0,1)}\|\mx{Z}\mx{a}^*(\tau)\|_2.
\end{align}
%------------
Lastly, incorporate \eqref{eq.dualnorm1} into \eqref{eq.lagran2} to reach \eqref{eq.dualprob}.

\section{Proof of Proposition \ref{prop.optimality}}\label{proof.optimality}
Any $\mx{\bs{\lambda}}\in\mathbb{C}^N$ satisfying \eqref{eq.supp_cond} and \eqref{eq.offsupp_cond} is a feasible point in the dual problem \eqref{eq.dualprob}. Recall that $\bm{\mathcal{H}}=(\mx{H}_i)_{i=1}^r$ is the matrix tuple of interest where $\mx{H}_i=\sum_{k=1}^{s_i}c_k^i\mx{f}_i\mx{a}(\tau_k^i)^{\mathsf{T}}$. It holds that
%------------
\begin{align}\label{eq.rel1}
&\sum_{i=1}^r\|\mx{H}_i\|_{\mathcal{A}_i}\ge \sum_{i=1}^r\|\mx{H}_i\|_{\mathcal{A}_i}\|(\mathcal{B}^{\rm Adj}\mx{\bs{\lambda}})_i\|_{\mathcal{A}_i}^{\mathsf{d}}\ge\nonumber\\
&\sum_{i=1}^r\langle (\mathcal{B}^{\rm Adj}\mx{\bs{\lambda}})_i, \mx{H}_i \rangle=\sum_{i=1}^r\langle (\mathcal{B}^{\rm Adj}\mx{\bs{\lambda}})_i, \sum_{k=1}^{s_i}c_k^i\mx{f}_i\mx{a}(\tau_k^i)^{\mathsf{T}}  \rangle=\nonumber\\
&\sum_{i=1}^r\sum_{k=1}^{s_i}{\rm Re}~{c_k^i}^*\langle (\mathcal{B}^{\rm Adj}\mx{\bs{\lambda}})_i, \mx{f}_i\mx{a}(\tau_k^i)^{\mathsf{T}}  \rangle\nonumber\\
&=\sum_{i=1}^r\sum_{k=1}^{s_i}{\rm Re}~{c_k^i}^*\langle \mx{q}_i(\tau_k^i), \mx{f}_i \rangle,
\end{align}
%------------
where the second inequality is due to Holder's relation, and the last equality is due to the definition of $\mx{q}_i(\tau_k^i)$ in \eqref{eq.Qi}. We proceed \eqref{eq.rel1} by using the conditions \eqref{eq.supp_cond} and \eqref{eq.offsupp_cond}: 
%------------
\begin{align}\label{eq.rel2}
&\sum_{i=1}^r\|\mx{H}_i\|_{\mathcal{A}_i}\ge \sum_{i=1}^r\sum_{k=1}^{s_i}{\rm Re}~{c_k^i}^*\langle \frac{1}{\|\mx{f}_i\|_2}{\rm sgn}(c_k^i)\mx{f}_i, \mx{f}_i \rangle=\nonumber\\
&\sum_{i=1}^r\sum_{k=1}^{s_i}|c_k^i|\|\mx{f}_i\|_2\ge\sum_{i=1}^r\|\mx{H}_i\|_{\mathcal{A}_i},
\end{align}
%------------
where we used the definition of atomic norm \eqref{eq.atomic_def} in the last step. From \eqref{eq.rel1} and \eqref{eq.rel2}, we find that 
$\langle \mx{\bs{\lambda}}, \mathcal{B}\bm{\mathcal{H}}\rangle=\sum_{i=1}^r\|\mx{H}_i\|_{\mathcal{A}_i}$. Since the pair $(\bm{\mathcal{H}}, \mx{\bs{\lambda}})$ is primal-dual feasible, we reach to the conclusion that $\bm{\mathcal{H}}$ is an optimal solution of \eqref{eq.primalprob} and $\mx{\bs{\lambda}}$ is an optimal solution of \eqref{eq.dualprob} by strong duality. For proving uniqueness, suppose $\widehat{\bm{\mathcal{H}}}\triangleq(\widehat{\mx{H}}_i)_{i=1}^r$ is another optimal solution of \eqref{eq.primalprob} where $\widehat{\mx{H}}_i=\sum_{\widehat{\tau}_k^i\in \widehat{\mathcal{S}}_i}\widehat{c}_k^i\widehat{\mx{f}}_i\mx{a}(\widehat{\tau}_k^i)^{\mathsf{T}}$. If $\widehat{\bm{\mathcal{H}}}$ and $\bm{\mathcal{H}}$ have the same set of spike locations (support), i.e. $\widehat{\mathcal{S}}_i=\mathcal{S}_i,~\forall i\in[r]$, we then have $\widehat{\bm{\mathcal{H}}}=\bm{\mathcal{H}}$ since the set of atoms building $\bm{\mathcal{H}}$ are linearly independent. If there exists some $\widehat{\tau}_k^i\notin \mathcal{S}_i$, then we have 
%------------
\begin{align}\label{eq.rel3}
&\langle \mx{\bs{\lambda}}, \mathcal{B}\widehat{\bm{\mathcal{H}}}\rangle=\sum_{i=1}^r\langle (\mathcal{B}^{\rm Adj}\mx{\bs{\lambda}})_i, \widehat{\mx{H}}_i\rangle=\nonumber\\
&\sum_{i=1}^r \sum_{k}{\rm Re}~{\overset{*}{\widehat{c}_k^i}}\langle (\mathcal{B}^{\rm Adj}\mx{\bs{\lambda}})_i, \widehat{\mx{f}}_i\mx{a}(\widehat{\tau}_k^i)^{\mathsf{T}} \rangle=\nonumber\\
&	\sum_{i=1}^r\Big[\sum_{\widehat{\tau}_k^i\in \mathcal{S}_i}{\rm Re}~{\overset{*}{\widehat{c}_k^i}}
\langle\mx{q}_i(\widehat{\tau}_k^i),\widehat{\mx{f}}_i \rangle+\sum_{\widehat{\tau}_k^i\notin \mathcal{S}_i}{\rm Re}~{\overset{*}{\widehat{c}_k^i}}
\langle\mx{q}_i(\widehat{\tau}_k^i),\widehat{\mx{f}}_i \rangle\Big]\le\nonumber\\
&\sum_{i=1}^r\Big[\sum_{\widehat{\tau}_k^i\in \mathcal{S}_i}|\widehat{c}_k^i|
\|\mx{q}_i(\widehat{\tau}_k^i)\|_2\|\widehat{\mx{f}}_i\|_2 +\sum_{\widehat{\tau}_k^i\notin \mathcal{S}_i}|\widehat{c}_k^i|
\|\mx{q}_i(\widehat{\tau}_k^i)\|_2\|\widehat{\mx{f}}_i\|_2\Big]\nonumber\\
&<\sum_{i=1}^r\Big[\sum_{\widehat{\tau}_k^i\in \mathcal{S}_i}|\widehat{c}_k^i|\|\widehat{\mx{f}}_i\|_2
 +\sum_{\widehat{\tau}_k^i\notin \mathcal{S}_i}|\widehat{c}_k^i|\|\widehat{\mx{f}}_i\|_2\Big]=\sum_{i=1}^r|\widehat{c}_k^i|\|\widehat{\mx{f}}_i\|_2=\nonumber\\
& \sum_{i=1}^r\|\widehat{\mx{H}}_i\|_{\mathcal{A}_i},
\end{align}
%------------
where we used the conditions \eqref{eq.supp_cond} and \eqref{eq.offsupp_cond}. The relation \eqref{eq.rel3} contradicts strong duality, hence $\bm{\mathcal{H}}$ is the unique optimal solution of \eqref{eq.primalprob}.

\section{Proof of Theorem \ref{thm.main_sample}}\label{proof.thm.main_sample}
We first provide a road-map of the proof of Theorem \ref{thm.main_sample}. The proof details is divided into three subsections listed below:
\begin{enumerate}
    \item \textbf{Constructing explicit vector-valued dual polynomials}: In this step, we construct random vector-valued dual polynomials denoted by $\mx{q}_i(\tau), i\in[r]$ that satisfy the conditions \eqref{eq.supp_cond} and \eqref{eq.offsupp_cond} with high probability when the number of samples is sufficiently large. The proof of this step is provided in Appendix \ref{proof.construct_dual}.
	\item \textbf{Certifying the support constraint \eqref{eq.supp_cond}}: In this step, we prove that as long as the number of samples satisfies \ref{eq:main_sample}, the constructed dual polynomials $\mx{q}_i(\tau), i\in[r]$ satisfy the support condition \eqref{eq.supp_cond} with high probability. The proof of this step is provided in Appendix \ref{proof.supp_cond}.
	\item \textbf{Certifying the off-support constraint \eqref{eq.offsupp_cond}}: In this step, we prove that the constructed dual polynomials $\mx{q}_i(\tau), i\in[r]$ satisfy the off-support condition $\|\mx{q}_i(\tau)\|_2<1$ provided in \eqref{eq.offsupp_cond} for $\tau\in[0,1)\setminus \mathcal{S}_i$. The proof of this step is given in Appendix \ref{proof.offsupp_cond}.
\end{enumerate}
By combining the required samples that each of the aforementioned steps requires, we reach to the sample complexity given in \eqref{eq:main_sample} and the result of the theorem is concluded. It is worth noting that each step outlined above comes with its own individual proof road-map, facilitating a clearer understanding for the reader.
%
%The remainder of the proof consists
%of three steps.
%1)
%2) Show , the random perturbations introduced
%by the random observation process, are small on
%a set of discrete points with high probability, implying the
%random dual polynomial satisfies the constraints in Proposition
%II.4 on the grid. This step is proved using a modification
%of the idea in [48].
%3) Extend the result to using Bernstein’s polynomial inequality
%[50] and eventually show for .

% use section* for acknowledgement
\subsection{Constructing the dual certificate}\label{proof.construct_dual}
{Proof road-map:} We construct the dual polynomials that satisfy the alternative sufficient conditions  \eqref{eq.supp_cond_alt} and \eqref{eq.offsupp_cond_alt}. The dual polynomials $\mx{q}_i(\tau)$ in \eqref{eq.Qi} depends on the dual vector $\bs{\lambda}\in\mathbb{C}^N$. We aim to find an explicit construction of $\mx{q}_i(\tau)$ which equivalently leads to an explicit construction of $\bs{\lambda}$ in \eqref{eq.lambdastar}. The explicit form of $\bs{\lambda}$ depends on some free coefficients $\bs{\alpha}^i$ and $\bs{\beta}^i$ which are specified later. Finally, by replacing this $\lambda$ into \eqref{eq.Qi}, the specific form of the random dual polynomials are provided in \eqref{eq.qii_main} which depend on some known matrix-valued Fe\'{j}er kernels. The randomness of the kernels comes from the randomness of codebook matrices $\mx{B}_i, i\in[r]$.

{Proof details:} Our goal here is to construct explicit dual polynomials $\{q_i(\tau)\}_{i=1}^r$ satisfying the conditions \eqref{eq.supp_cond} and \eqref{eq.offsupp_cond} with high probability when the number of samples is sufficiently large. To proceed, we require that the dual polynomials $\{\mx{q}_i(\tau)\}_{i=1}^r$ satisfy the conditions
\begin{align}
&\mx{q}_i(\tau_k)={\rm sgn}({c_k^{i}})\frac{\mx{f}_i}{\|\mx{f}_i\|_2}~\forall \tau_k\in \mathcal{S}_i,~ i=1, ..., r\label{eq.supp_cond_alt}\\
&\frac{\partial \mx{q}_i(\tau)}{\partial \tau}\Big|_{\tau=\tau_k}=\mx{0}_{k_i\times 1}~~\forall \tau_k\in \mathcal{S}_i,~ i=1, ..., r.	\label{eq.offsupp_cond_alt}
\end{align}
The constraint \eqref{eq.supp_cond_alt} is the same as the condition \eqref{eq.supp_cond}, whereas \eqref{eq.offsupp_cond_alt} is used to ensure that $\|\mx{q}_i(\tau)\|_2$ achieves its local maximum at $\tau_k\in\mathcal{S}_i$ for any $i=1, ..., r$. Alternatively, the constraints \eqref{eq.supp_cond_alt} and \eqref{eq.offsupp_cond_alt} can be written in a matrix form as
\begin{align}
\mx{A}\mx{\bs{\lambda}}=\mx{u},
\end{align}
where
\begin{align*}
&\mx{A}=\begin{bmatrix}
\mx{A}^{1}\\
\vdots\\
\mx{A}^{r}
\end{bmatrix}\in\mathbb{C}^{2\sum_{i=1}^{r}s_ik_i\times N},
\mx{u}=\begin{bmatrix}
\mx{u}_1\\
\mx{0}_{s_i k_i \times 1}\\
\vdots\\
\mx{u}_{r}\\
\mx{0}_{s_r k_r\times 1}
\end{bmatrix}\in \mathbb{C}^{2\sum_{i=1}^{r}s_ik_i\times 1}
\end{align*}
with
\begin{align*}
&\mx{A}^i=\scalebox{.8}{$\begin{bmatrix}
\mx{b}_{-2M}^i\mx{e}_{-2M}^{\mathsf{T}}\mx{a}^*(\tau_1^i)& \cdots & \mx{b}_{2M}^i\mx{e}_{2M}^{\mathsf{T}}\mx{a}^*(\tau_1^i)\\
\vdots&\ddots&\vdots\\
\mx{b}_{-2M}^i\mx{e}_{-2M}^{\mathsf{T}}\mx{a}^*(\tau_{s_i}^i)& \cdots & \mx{b}_{2M}^i\mx{e}_{2M}^{\mathsf{T}}\mx{a}^*(\tau_{s_i}^i)\\
j2\pi(-2M)\mx{b}_{-2M}^i\mx{e}_{-2M}^{\mathsf{T}}\mx{a}^*(\tau_1^i)&\cdots& j2\pi(2M)\mx{b}_{2M}^i\mx{e}_{2M}^{\mathsf{T}}\mx{a}^*(\tau_1^i)\\
\vdots&\ddots&\vdots\\
j2\pi(-2M)\mx{b}_{-2M}^i\mx{e}_{-2M}^{\mathsf{T}}\mx{a}^*(\tau_{s_i}^i)&\cdots& j2\pi(2M)\mx{b}_{2M}^i\mx{e}_{2M}^{\mathsf{T}}\mx{a}^*(\tau_{s_i}^i)
\end{bmatrix}$}\nonumber\\
&\in \mathbb{C}^{2s_ik_i\times N},
\end{align*}
and
\begin{align}\label{ui}
\mx{u}_i=\begin{bmatrix}
{\rm sgn}(c_1^i)\tfrac{\mx{f}_i}{\|\mx{f}_i\|_2}\\
\vdots\\
{\rm sgn}(c_{s_i}^i)\tfrac{\mx{f}_i}{\|\mx{f}_i\|_2}
\end{bmatrix}\in\mathbb{C}^{s_ik_i\times 1}.
\end{align}
To construct the dual polynomial $\mx{q}_i(\tau)$ in \eqref{eq.Qi}, we first find a suitable $\mx{\bs{\lambda}}\in\mathbb{C}^N$ by solving 
\begin{align}\label{eq.lambda_prob}
&\min_{\mx{\bs{\lambda}}}\frac{1}{2}\|\bs{\Omega}\mx{\bs{\lambda}}\|_2^2\nonumber\\
&{\rm s.t.}~ \mx{A}\mx{\bs{\lambda}}=\mx{u},
\end{align}
where $\bs{\Omega}\triangleq{\rm diag}(\omega_{-2M}, ..., \omega_{2M})$ is a diagonal matrix to be specified later. By Karush-Kuhn-Tucker (KKT) condition, it is straightforward to see that the optimal solution of \eqref{eq.lambda_prob} reads as 
\begin{align}\label{eq.lambdastar}
&\mx{\bs{\lambda}}=(\bs{\Omega}^{\mathsf{H}}\bs{\Omega})^{-1}\mx{A}^{\mathsf{H}}[{\bs{\alpha}^1}^{\mathsf{H}},{\bs{\beta}^1}^{\mathsf{H}}, ..., {\bs{\alpha}^r}^{\mathsf{H}},{\bs{\beta}^r}^{\mathsf{H}}]^{\mathsf{H}}=\nonumber\\
&{\rm diag}(\frac{1}{\omega_{-2M}^2}, ..., \frac{1}{\omega_{2M}^2} )\sum_{i=1}^r{\mx{A}^i}^{\mathsf{H}}\begin{bmatrix}
\bs{\alpha}^i\\
\bs{\beta}_i
\end{bmatrix}=\nonumber\\
&{\rm diag}(\frac{1}{\omega_{-2M}^2}, ..., \frac{1}{\omega_{2M}^2} )\sum_{i=1}^r\Bigg(\sum_{p=1}^{s_i}\begin{bmatrix}
\mx{a}(\tau_p^i)^{\mathsf{T}}\mx{e}_{-2M}{\mx{b}_{-2M}^{i}}^{\mathsf{H}}\\
\vdots\\
\mx{a}(\tau_p^i)^{\mathsf{T}}\mx{e}_{2M}{\mx{b}_{2M}^{i}}^{\mathsf{H}}
\end{bmatrix}\bs{\alpha}_p^i\nonumber\\
&+\sum_{p=1}^{s_i}\begin{bmatrix}
-j2\pi(-2M)\mx{a}(\tau_p^i)^{\mathsf{T}}\mx{e}_{-2M}{\mx{b}_{-2M}^{i}}^{\mathsf{H}}\\
\vdots\\
-j2\pi(2M)\mx{a}(\tau_p^i)^{\mathsf{T}}\mx{e}_{2M}{\mx{b}_{2M}^{i}}^{\mathsf{H}}
\end{bmatrix}\bs{\beta}_p^i
\Bigg),
\end{align}
for some 
\begin{align*}
&\bs{\alpha}^i\triangleq\begin{bmatrix}
\bs{\alpha}_1^i\\
\vdots\\
\bs{\alpha}_{s_i}^i
\end{bmatrix}~\bs{\beta}^i\triangleq\begin{bmatrix}
\bs{\beta}_1^i\\
\vdots\\
\bs{\beta}_{s_i}^i
\end{bmatrix},
&\bs{\alpha}_p^i, \bs{\beta}_p^i \in\mathbb{C}^{k_i\times 1},
\end{align*}
that satisfies 
\begin{align}
    \mx{A}(\bs{\Omega}^{\mathsf{H}}\bs{\Omega})^{-1}\mx{A}^{\mathsf{H}}[{\bs{\alpha}^1}^{\mathsf{H}},{\bs{\beta}^1}^{\mathsf{H}}, ..., {\bs{\alpha}^r}^{\mathsf{H}},{\bs{\beta}^r}^{\mathsf{H}}]^{\mathsf{H}}=\mx{u}.
\end{align}
By replacing this $\mx{\bs{\lambda}}$ in \eqref{eq.Qi}, we obtain the form of $\mx{q}_m(\tau)$ for each $m=1, ..., r$ as
\begin{align}\label{eq.qii1}
&\mx{q}_m(\tau)=\sum_{n=-2M}^{2M}\bs{\lambda}(n){\rm e}^{j2\pi n\tau}\mx{b}_n^m=
\sum_{i=1}^r\sum_{p=1}^{s_i}\Bigg[\nonumber\\
&\sum_{n=-2M}^{2M}\frac{1}{\omega_n^2}{\rm e}^{-j2\pi n \tau_p^i} {\mx{b}_n^i}^{\mathsf{H}}\bs{\alpha}_p^i{\rm e}^{j2\pi n\tau}\mx{b}_n^m\nonumber\\
&+\sum_{n=-2M}^{2M}\frac{1}{\omega_n^2}(-j2\pi n){\rm e}^{-j2\pi n \tau_p^i} {\mx{b}_n^i}^{\mathsf{H}}\bs{\beta}_p^i{\rm e}^{j2\pi n\tau}\mx{b}_n^m\Bigg]=\nonumber\\
&\sum_{i=1}^r\sum_{p=1}^{s_i}\Bigg[\sum_{n=-2M}^{2M}\frac{1}{\omega_n^2} {\rm e}^{-j2\pi n (\tau-\tau_p^i)}\mx{b}_n^m{\mx{b}_n^i}^{\mathsf{H}}
\bs{\alpha}_p^i+\nonumber\\
&\sum_{n=-2M}^{2M}\frac{1}{\omega_n^2}(-j2\pi n){\rm e}^{-j2\pi n (\tau-\tau_p^i)}\mx{b}_n^m{\mx{b}_n^i}^{\mathsf{H}}
\bs{\beta}_p^i\Bigg].
\end{align}
Consider the function
\begin{align}\label{eq.gn}
g(n)\triangleq\frac{1}{M}\sum_{l=\max\{n-M, -M\}}^{\min\{n+M, M\}}\big(1-\frac{|l|}{M}\big)\big(1-\frac{|n-l|}{M}\big)
\end{align}
which is the discrete convolution of two triangular functions. Using this function, we define the random matrix kernels
\begin{align}\label{eq.kernel_mat}
&\mx{K}_{m,i}(\tau)=\frac{1}{M}\sum_{n=-2M}^{2M}g(n){\rm e}^{-j2\pi n \tau}\mx{b}_n^m{\mx{b}_n^i}^{\mathsf{H}}\in\mathbb{C}^{k_m\times k_i},\nonumber\\
& m, i=1, ..., r
\end{align}
which are matrix versions of the scalar deterministic kernel
\begin{align}\label{eq.fejer_kernel}
K(\tau)=\frac{1}{M}\sum_{n=-2M}^{2M}g(n){\rm e}^{-j2\pi n \tau}=\Big[\frac{\sin(\pi M\tau)}{M\sin(\pi \tau)}\Big]^4,
\end{align}
known as the squared Fe\'{j}er kernel. From the independence and isotropy assumption in \eqref{eq.isotropy}, it holds that
\begin{align}\label{eq.EKl}
&\mathds{E}\mx{K}^{\ell}_{m,i}(\tau)=\frac{1}{M}\sum_{n=-2M}^{2M}g(n)(-j2\pi n)^{\ell}{\rm e}^{-j2\pi n\tau}\mathds{E}\big[\mx{b}_n^m{\mx{b}_n^i}^{\mathsf{H}}\big]\nonumber\\
&=K^{\ell}(\tau)\mx{I}_{k_i}1_{m=i},
\end{align}
where ${K}^{(\ell)}(\tau)\triangleq\frac{\partial {K}^{\ell}(\tau)}{\partial\tau^{\ell}}$ is the $\ell$-th derivative of $K(\tau)$ with respect to $\tau$. To proceed, we set $\omega_n\triangleq\sqrt{\frac{M}{g(n)}}$ in \eqref{eq.qii1} and using the definitions in \eqref{eq.kernel_mat}, we rewrite the vector-valued random dual polynomial in \eqref{eq.qii1} as
\begin{align}\label{eq.qii_main}
&\mx{q}_m(\tau)=\sum_{i=1}^r\sum_{p=1}^{s_i}\Big[\mx{K}_{m,i}(\tau-\tau_p^i)\bs{\alpha}_p^i+\mx{K}'_{m,i}(\tau-\tau_p^i)\bs{\beta}_p^i\Big],\nonumber\\
&m=1, ..., r,
\end{align} 
where $\mx{K}'(\tau)=\frac{\partial \mx{K}(\tau)}{\partial\tau}$ is the first entry-wise derivative of $\mx{K}(\tau)$ with respect to $\tau$. We also define the deterministic dual polynomials
\begin{align}\label{eq.qi_main_deterministic}
&\overline{\mx{q}}_m(\tau)=\sum_{i=1}^r\sum_{p=1}^{s_i}\Big[\mathds{E}\mx{K}_{m,i}(\tau-\tau_p^i)\overline{\bs{\alpha}}_p^i+\mathds{E}\mx{K}_{m,i}'(\tau-\tau_p^i)\overline{\bs{\beta}}_p^i\Big]\nonumber\\
&=\sum_{p=1}^{s_m}\Big[K(\tau-\tau_p^m)\overline{\bs{\alpha}}_p^m+K'(\tau-\tau_p^m)\overline{\bs{\beta}}_p^m\Big],\nonumber\\
& \quad m=1, \ldots, r,
\end{align}
where we used the relation \eqref{eq.EKl}, and $\overline{\bs{\alpha}}_p^i$ and $\overline{\bs{\beta}}_p^i$ are chosen such that the equations
\begin{align}\label{eq.q_bar_conditions}
&\overline{\mx{q}}_i(\tau_k)={\rm sgn}({c_k^{i}})\frac{\mx{f}_i}{\|\mx{f}_i\|_2}~\forall \tau_k\in \mathcal{S}_i,~ i=1, ..., r\\
&\overline{\mx{q}}'_i(\tau_k)=\mx{0}_{k_i\times 1}~~\forall \tau_k\in {\mathcal{S}}_i,~ i=1, ..., r,
\end{align}	
are satisfied. Our random dual polynomials $\mx{q}_m(\tau)$ in \eqref{eq.qii_main} can be regarded as random versions of the deterministic dual polynomials $\overline{\mx{q}}_m(\tau)$ in \eqref{eq.qi_main_deterministic} where the random parts are introduced by the vectors $\{\mx{b}_n^i\}_{i=1}^r$.
%\begin{align}\label{eq.EKl}
%\mathds{E}\mx{K}'_{m,i}(\tau)=\frac{1}{M}\sum_{n=-2M}^{2M}(-j2\pi n){\rm e}^{-j2\pi n\tau}\mathds{E}\big[\mx{b}_n^m{\mx{b}_n^i}^{\mathsf{H}}\big],
%\end{align}
%is used to denote the expected value of the first derivative of $\mx{K}_{m,i}(\tau)$. By the isotropy and independence assumptions(refer), we have 
%\begin{align}
%\mathds{E}\mx{K}'_{m,i}(\tau)=K'(\tau)\mx{I}_{k_i}1_{m=i}, 
%\end{align}
%where $K'(\tau)=\frac{1}{M}\sum_{n=-2M}^{2M}g(n)(-j2\pi n){\rm e}^{-j2\pi n \tau}$. 
\subsection{Certifying (\ref{eq.supp_cond})}\label{proof.supp_cond}
Proof road-map: In this part, we aim to prove that there exists specific coefficients $\bs{\alpha}^i$ and $\bs{\beta}^i$ such that the dual polynomials $\mx{q}_i(\tau), i\in[r]$ satisfy the support condition \eqref{eq.supp_cond_alt} with high probability. By combining the dual polynomial construction \eqref{eq.qii_main} and the condition \eqref{eq.supp_cond_alt}, we reach to the linear system of equations provided in \eqref{eq.D_equation}. To show the existence of the coefficients $\bs{\alpha}^i$ and $\bs{\beta}^i$, we need to prove that $\mx{D}$ is invertible with high probability when the number of samples is sufficiently large.  To show this, we show that $\mathds{E}[\mx{D}]$ is invertible in Lemma \ref{lem.ED_rel1}. Then, we show that $\mx{D}$ is concentrated around $\mathds{E}[\mx{D}]$ in Lemma \ref{lemma.concentrate1} with high probability given sufficient number of samples. We also show that $\mx{D}^{-1}$ is close to $\mathds{E}[\mx{D}]^{-1}$ in Lemma \ref{lem.EDrel2}.

Proof details:
In the previous subsection, we obtained the form of $\mx{q}_m(\tau)$ for any $m=1, ..., r$ in \eqref{eq.qii_main}. In this subsection, we aim at finding $\bs{\alpha}_p^i$ and $\bs{\beta}_p^i$ in \eqref{eq.qii_main} such that the conditions \eqref{eq.supp_cond_alt} and \eqref{eq.offsupp_cond_alt} hold. By applying the constraints \eqref{eq.supp_cond_alt} and \eqref{eq.offsupp_cond_alt} to \eqref{eq.qii_main}, we reach the equations
\begin{align}
&\mx{q}_m(\tau_k^m)=\sum_{i=1}^r\sum_{p=1}^{s_i}\Big[\mx{K}_{m,i}(\tau_k^m-\tau_p^i)\bs{\alpha}_p^i+\mx{K}'_{m,i}(\tau_k^m-\tau_p^i)\bs{\beta}_p^i\Big],\nonumber\\
&={\rm sgn}({c_k^{m}})\frac{\mx{f}_m}{\|\mx{f}_m\|_2}\nonumber\\
&\mx{q}'_m(\tau_k^m)=\sum_{i=1}^r\sum_{p=1}^{s_i}\Big[\mx{K}'_{m,i}(\tau_k^m-\tau_p^i)\bs{\alpha}_p^i+\mx{K}''_{m,i}(\tau_k^m-\tau_p^i)\bs{\beta}_p^i\Big]\nonumber\\
&=\mx{0}_{k_m\times 1},\nonumber\\
&\forall \tau_k^m\in \mathcal{S}_m, m=1, ..., r.
\end{align}
The latter relation can be expressed in a matrix form as
\begin{align}\label{eq.D_equation}
\underbrace{
\begin{bmatrix}
\mx{D}^{1,1}&\cdots&\mx{D}^{1,r}\\
\vdots&\ddots&\vdots\\
\mx{D}^{r,1}&\cdots&\mx{D}^{r,r}
\end{bmatrix}}_{\triangleq\mx{D}}
\begin{bmatrix}
\bs{\alpha}^{1}\\
\kappa^{-1}\bs{\beta}^{1}\\
\vdots\\
\bs{\alpha}^{r}\\
\kappa^{-1}\bs{\beta}^{r}
\end{bmatrix}
=\mx{u},
\end{align}
where 
\begin{align}
\mx{D}^{m,i}=\begin{bmatrix}
\mx{D}_0^{m,i}&\kappa\mx{D}_1^{m,i}\\
-\kappa\mx{D}_1^{m,i}&-\kappa^2\mx{D}_2^{m,i}
\end{bmatrix}\in\mathbb{C}^{2s_mk_m\times 2s_ik_i},
\end{align}
$\mx{D}^{m,i}_{\ell}(k,p)=\mx{K}_{m,i}^{(\ell)}(\tau_k^m-\tau_p^i)$ is the $(k,p)$th sub-matrix of $\mx{D}^{m,i}_{\ell}$, and $\kappa\triangleq\frac{1}{\sqrt{K''(0)}}$ with $K''(0)=\frac{-4\pi^2(M^2-1)}{3}$.
By defining 
\begin{align}\label{eq.nu-i}
\bs{\nu}_n^i\triangleq\begin{bmatrix}
{\rm e}^{-j2\pi\tau_1^i n}\\
\vdots\\
{\rm e}^{-j2\pi \tau_{s_i}^i n}\\
(j2\pi n \kappa){\rm e}^{-j2\pi \tau_1^i n}\\
\vdots\\
(j2\pi n \kappa){\rm e}^{-j2\pi \tau_{s_i}^i n}
\end{bmatrix}\in\mathbb{C}^{2s_i\times 1},
\end{align}
%and
%\begin{align}
%\Phi\triangleq\frac{1}{M}\sum_{n=-2M}^{2M}g(n)\bs{\nu}_n^m{\bs{\nu}_n^i}^{\mathsf{H}}\otimes\mx{b}_n^m{\mx{b}_n^i}^{\mathsf{H}}
%\end{align}
it is straightforward to see that the matrix $\mx{D}^{m,i}$ can be expressed as
\begin{align}\label{eq.Dmi}
\mx{D}^{m,i}=\frac{1}{M}\sum_{n=-2M}^{2M}g(n)\bs{\nu}_n^m{\bs{\nu}_n^i}^{\mathsf{H}}\otimes\mx{b}_n^m{\mx{b}_n^i}^{\mathsf{H}}.
\end{align}
Denote
\begin{align}
\widetilde{\mx{D}}^i=\begin{bmatrix}
\widetilde{\mx{D}}_0^i&\kappa\widetilde{\mx{D}}_1^i\\
-\kappa\widetilde{\mx{D}}_1^i&-\kappa^2\widetilde{\mx{D}}_2^i
\end{bmatrix}\in\mathbb{C}^{2s_i\times 2s_i},
\end{align}
where $\widetilde{\mx{D}}^{i}_{\ell}(k,p)={K}^{(\ell)}(\tau_k^i-\tau_p^i)$ is the $(k,p)$ element of the matrix $\widetilde{\mx{D}}^{i}_{\ell}$. The expected value of $\mx{D}^{m,i}$ for any $m,i\in [r]$ can be obtained by
\begin{align}
\nonumber
\mathds{E}\mx{D}^{m,i}&= \frac{1}{M}\sum_{n=-2M}^{2M}g(n)\bs{\nu}_n^m{\bs{\nu}_n^i}^{\mathsf{H}}\otimes \mx{I}_{k_i}1_{m=i},\\
&=\widetilde{\mx{D}}^i\otimes \mx{I}_{k_i} 1_{m=i}.\label{eq.EDmi}
\end{align}
Our approach for finding $\bs{\alpha}^{i}$ and $\bs{\beta}^i$ is to first show that $\mx{D}$ in \eqref{eq.D_equation} is invertible with high probability. To accomplish this, we show that 
\begin{align}\label{eq.ED}
&\mathds{E}\mx{D}={\rm diag}(\widetilde{\mx{D}}^1\otimes \mx{I}_{k_1}, \ldots, \widetilde{\mx{D}}^r\otimes \mx{I}_{k_r})\triangleq\nonumber\\
&\begin{bmatrix}
\widetilde{\mx{D}}^1\otimes \mx{I}_{k_1}&\mx{0}&\mx{0}\\
\vdots&\ddots&\vdots\\
\mx{0}&\mx{0}&\widetilde{\mx{D}}^r\otimes \mx{I}_{k_r}
\end{bmatrix}
\end{align}
is invertible in the following lemma:
\begin{lem}\label{lem.ED_rel1}
Suppose that the minimum separation satisfies $\Delta\ge \frac{1}{M}$. Then, $\mathds{E}\mx{D}$ is invertible and 
\begin{align*}
&\|\mx{I}-\mathds{E}\mx{D}\|_{2\rightarrow 2}\le 0.3623, \\
&\|\mathds{E}\mx{D}\|_{2\rightarrow 2}\le 1.3623, \\
&\|(\mathds{E}\mx{D})^{-1}\|_{2\rightarrow 2}\le 1.568.
\end{align*}
\end{lem}
\begin{proof}\label{proof.lemma_EDrel1}
 First, we provide a lemma that describes the properties of $\widetilde{\mx{D}}^{i}\triangleq\frac{1}{M}\sum_{n=-2M}^{2M}g(n){\bs{\nu}_n^i}{\bs{\nu}_n^i}^{\mathsf{H}}$.
\begin{lem}\cite[Proposition IV.1]{tang2013compressed}\label{lem.tang}
Suppose that $\Delta>\frac{1}{M}$. Then, $\widetilde{\mx{D}}^{i}$ is invertible and the following relations hold.
\begin{subequations}
    \begin{align}
&\|\mx{I}_{s_i}-\widetilde{\mx{D}}^{i}\|_{2\rightarrow 2}\le 0.3623,\\
&\|\widetilde{\mx{D}}^{i}\|_{2\rightarrow 2}\le 1.3623,\\
&\|(\widetilde{\mx{D}}^{i})^{-1}\|_{2\rightarrow 2}\le 1.568.
\end{align}
\end{subequations}
\end{lem}
According to \eqref{eq.EDmi} and Lemma~\ref{lem.tang}, it holds that
\begin{align}
&\|\mathds{E}\mx{D}^{i,i}\|_{2\rightarrow 2}=\|\widetilde{\mx{D}}^i\otimes \mx{I}_{k_i}\|_{2\rightarrow 2}=\|\widetilde{\mx{D}}^i\|_{2\rightarrow 2}\le 1.3623, \nonumber\\ \nonumber
&  \|({\mathds{E}\mx{D}^{i,i}})^{-1}\|_{2\rightarrow 2}=\|{(\widetilde{\mx{D}}^i})^{-1}\otimes \mx{I}_{k_i}\|_{2\rightarrow 2}
\\ 
& \qquad =\|(\widetilde{\mx{D}}^i)^{-1}\|_{2\rightarrow 2}\le 1.568,\nonumber\\
&\|\mx{I}_{2s_ik_i}-\mathds{E}\mx{D}^{i,i}\|_{2\rightarrow 2}=\|(\mx{I}_{s_i}-\widetilde{\mx{D}}^i)\otimes \mx{I}_{k_i}\|_{2\rightarrow 2}\nonumber\\
&=\|\mx{I}_{s_i}-\widetilde{\mx{D}}^i\|_{2\rightarrow 2}\le 0.3623.
\end{align}
Consequently by \eqref{eq.ED}, we have
\begin{align}
&\|\mathds{E}\mx{D}\|_{2\rightarrow 2}=\max_{i}\|\widetilde{\mx{D}}^i\otimes \mx{I}_{k_i}\|_{2\rightarrow 2}=\max_{i}\|\widetilde{\mx{D}}^i\|_{2\rightarrow 2}\le 1.3623\nonumber\\
& \nonumber  \|({\mathds{E}\mx{D}})^{-1}\|_{2\rightarrow 2}=\max_i\|(\widetilde{\mx{D}}^i)^{-1}\otimes \mx{I}_{k_i}\|_{2\rightarrow 2}\\ 
& =\max_i\|(\widetilde{\mx{D}}^i)^{-1}\|_{2\rightarrow 2}
\le 1.568.
\end{align}

\end{proof}

The inverse of the block diagonal matrix $\mathds{E}\mx{D}$ is given by
\begin{align}\label{eq.EDinv}
(\mathds{E}\mx{D})^{-1}=\begin{bmatrix}
(\widetilde{\mx{D}}^1)^{-1}\otimes \mx{I}_{k_1}&\mx{0}&\mx{0}\\
\vdots&\ddots&\vdots\\
\mx{0}&\mx{0}&(\widetilde{\mx{D}}^r)^{-1}\otimes \mx{I}_{k_r}
\end{bmatrix}.
\end{align}
Then, we investigate the invertibility of $\mx{D}$ by showing
 that $\mx{D}$ is concentrated around $\mathds{E}\mx{D}$ with high probability given enough measurements and under the minimum separation condition. To proceed, we move toward a concentration result by combining \eqref{eq.Dmi} and \eqref{eq.EDmi} to write
\begin{align}\label{eq.D-ED}
\mx{D}-\mathds{E}\mx{D}=\sum_{n=-2M}^{2M}\mx{S}_n,
\end{align}
where 
\begin{align}\label{eq.Sn}
\mx{S}_n=\begin{bmatrix}
\mx{S}_n^{1,1}&\cdots&\mx{S}_n^{1,r}\\
\vdots&\ddots&\vdots\\
\mx{S}_n^{r,1}&\cdots&\mx{S}_n^{r,r}
\end{bmatrix}\in\mathbb{C}^{2\sum_{i=1}^rs_ik_i\times 2\sum_{i=1}^rs_ik_i}
\end{align}
and
\begin{align}\label{eq.Smi}
\mx{S}_n^{m,i}=\frac{1}{M}g(n)\bs{\nu}_n^{m}{\bs{\nu}_n^i}^{\mathsf{H}}\otimes \big(\mx{b}_n^m{\mx{b}_n^i}^{\mathsf{H}}-\mx{I}_{k_i}1_{m=i}\big)\in\mathbb{C}^{2s_mk_m\times 2s_ik_i}
\end{align}
Now, we are ready to establish the concentration of $\mx{D}$ around $\mathds{E}\mx{D}$. First, define the event $\mathcal{E}_{1,\epsilon_1}\triangleq\{\|\mx{D}-\mathds{E}\mx{D}\|_{2\rightarrow 2}\le \epsilon\}$  for $\epsilon_1>0$. The probability of occurring $\mathcal{E}_{1,\epsilon_1}$ is high if there is enough number of samples. This is provided in the following lemma whose proof is given in Appendix \ref{proof.lemma.concentrate1}
\begin{lem}\label{lemma.concentrate1}
Let $0<\delta<1$ and $\Delta\ge \frac{1}{M}$. 
 For any $0<\epsilon_1 <0.6376$,
 as long as
 \begin{align}\label{eq.measure_bound_concent}
 M\ge \frac{c_1 \mu_{+} \sum_i s_i k_i}{\epsilon_1^2}\log\Big(1+\frac{c_2\mu_{+}^2\sum_i s_i k_i}{\mu_{-}}\Big)\log\Big(\frac{4\sum_i s_i k_i}{\delta}\Big)
 \end{align}	
\gf{for some constants} $c_1, c_2>0$, the event $\mathcal{E}_{1,\epsilon_1}$ holds and $\mx{D}$ is invertible with probability at least $1-\delta$.
 
 Proof. See Appendix \ref{proof.lemma.concentrate1}.
\end{lem}
The following properties of matrix $\mx{D}$ are also helpful in the reminding of the paper.
\begin{lem}\label{lem.EDrel2}
Suppose that the event $\mathcal{E}_{1,\epsilon}$ holds with $0<\epsilon\le \frac{1}{4	}$. Then, we have
\begin{align}
&\|\mx{D}^{-1}-{(\mathds{E}\mx{D})^{-1}}\|_{2\rightarrow 2}\le 2\|(\mathds{E}\mx{D})^{-1}\|_{2\rightarrow 2}^2\epsilon\nonumber\\
&\|\mx{D}^{-1}\|_{2\rightarrow 2}\le 2\|(\mathds{E}\mx{D})^{-1}\|_{2\rightarrow 2}
\end{align}	
\end{lem} 
% Proof. See Appendix \ref{proof.lemma.EDrel2}
\begin{proof}\label{proof.lemma.EDrel2}
 Since $\|\mx{D}-\mathds{E}\mx{D}\|_{2\rightarrow 2}\le 0.31$ and $\|(\mathds{E}\mx{D})^{-1}\|_{2\rightarrow 2}\le 1.
568$, we have $\|\mx{D}-\mathds{E}\mx{D}\|_{2\rightarrow 2}\|(\mathds{E}\mx{D})^{-1}\|_{2\rightarrow 2}\le \frac{1}{2}$ and by leveraging \cite[Appendix E]{tang2013compressed}, we have
\begin{align}
&\|\mx{D}^{-1}\|_{2\rightarrow 2}\le 2\|(\mathds{E}\mx{D})^{-1}\|_{2\rightarrow 2}\nonumber\\
&\|\mx{D}^{-1}-{(\mathds{E}\mx{D})^{-1}}\|_{2\rightarrow 2}\le 2\|(\mathds{E}\mx{D})^{-1}\|_{2\rightarrow 2}^2\|\mx{D}-\mathds{E}\mx{D}\|_{2\rightarrow 2}\nonumber\\
&\le 
2\|(\mathds{E}\mx{D})^{-1}\|_{2\rightarrow 2}^2\epsilon
\end{align}   
\end{proof}

\subsection{Certifying (\ref{eq.offsupp_cond})}\label{proof.offsupp_cond}
Proof road-map: In this section, we aim to prove that the constructed dual polynomial in \eqref{eq.qii_main} satisfies the condition \eqref{eq.offsupp_cond}. We first express $\mx{q}_i(\tau)$ based on its expected value, i.e., $\overline{\mx{q}}_i(\tau)$ in \eqref{q-qbar} plus some residual terms. Then, we show in Lemmas \ref{lem.I1Concentrate} and \ref{lem.I2concentrate} that these residuals get sufficiently small on a discrete domain of grids when the number of samples becomes sufficiently large. This shows that $\mx{q}_i(\tau)$ concentrates around $\overline{\mx{q}}_i(\tau)$ with high probability on a discrete domain of grids. Next, we extend the result to the continuous domain $[0,1]$ using Bernstein's polynomial inequality \cite{bernstein_polynomial_inequalities} and prove in Lemmas \ref{lem.Tfar} and \ref{lem.Tnear} that $\mx{q}_i(\tau)$ concentrates around $\overline{\mx{q}}_i(\tau)$ and that $\|\mx{q}_i(\tau)\|_2<1$ everywhere in $[0,1]\setminus \mathcal{S}_i$ with high probability.

Proof details: In this section, we prove that $\|\mx{q}_i(\tau)\|_2<1$ for $\tau \in [0,1)\setminus \mathcal{S}_i \forall i=1,\ldots, r$. 
\begin{itemize}
	\item Showing that $\mx{q}_i(\tau)$ concentrates around $\overline{\mx{q}}_i(\tau)$ on a discrete domain of grids $\mathcal{T}_{\rm grid}$.
\item Extending the result to the continuous domain $[0,1]$ using Bernstein's polynomial inequality \cite{bernstein_polynomial_inequalities}.
\end{itemize}

%\subsubsection{ $\mx{q}_i(\tau)$ concentrates around $\overline{\mx{q}}_i(\tau)$ in}
First, we define the vector
\begin{align}\label{eq.V_l(tau)}
&\mx{V}_{\ell}^{i,m}(\tau)\triangleq\kappa^{\ell}\begin{bmatrix}
{{\mx{K}_{m,i}^{(\ell)}}^{\mathsf{H}}(\tau-\tau_1^i)}\\
\vdots\\
{\mx{K}_{m,i}^{(\ell)}}^{\mathsf{H}}(\tau-\tau_{s_i}^i)\\
\kappa{\mx{K}_{m,i}^{(l+1)}}^{\mathsf{H}}(\tau-\tau_{1}^i)\\
\vdots\\
\kappa{\mx{K}_{m,i}^{(\ell+1)}}^{\mathsf{H}}(\tau-\tau_{s_i}^i)
\end{bmatrix}=\nonumber\\
&\frac{1}{M}\sum_{n=-2M}^{2M}g(n)(j2\pi n\kappa)^\ell { \rm e}^{-j2\pi n \tau} \bs{\nu}_n^i\otimes {\mx{b}_n}^i {{\mx{b}_n}^m}^{\mathsf{H}}  \in\mathbb{C}^{2 s_i k_i\times k_m}
\end{align}
and 
\begin{align}
\mx{V}_{\ell}^m(\tau)\triangleq\begin{bmatrix}
\mx{V}_{\ell}^{1,m}(\tau)\\
\vdots\\
\mx{V}_{\ell}^{r,m}(\tau)
\end{bmatrix}
\end{align}
By taking the expectation from the latter expression, we reach
\begin{align}
\mathds{E}\mx{V}_{\ell}^{i,m}(\tau)=\underbrace{\frac{1}{M}\sum_{n=-2M}^{2M}g(n)(j2\pi n\kappa)^\ell { \rm e}^{-j2\pi n \tau} \bs{\nu}_n^i}_{=:\mx{v}_{\ell}^{i}(\tau)}\otimes \mx{I}_{k_i} 1_{m=i},
\end{align}
and 
\begin{align}\label{eq.EV}
\mathds{E}\mx{V}_{\ell}^m(\tau)=\begin{bmatrix}
   \mx{0}\\
   \vdots\\
   \mx{v}^{1}_{\ell}(\tau)\otimes \mx{I}_{k_1}\\
   \vdots\\
   \mx{0}
\end{bmatrix}
\end{align}
where 
\begin{align}
\mx{v}_l^i(\tau)=\kappa^{\ell}\begin{bmatrix}
{{{K}^{(\ell)}}^*(\tau-\tau_1^i)}\\
\vdots\\
{{K}^{(\ell)}}^*(\tau-\tau_{s_i}^i)\\
\kappa{{K}^{(l+1)}}^*(\tau-\tau_{1}^i)\\
\vdots\\
\kappa{{K}^{(\ell+1)}}^*(\tau-\tau_{s_i}^i)
\end{bmatrix}.
\end{align}
Set 
\begin{align}\label{eq.Dinv}
\mx{D}^{-1}=\begin{bmatrix}
\mx{L}^{1,1} & \mx{R}^{1,1}&\cdots&\mx{L}^{1,r} &\mx{R}^{1,r}\\
\vdots&\vdots&\ddots&\vdots&\vdots\\
\mx{L}^{r,1}&\mx{R}^{r,1}&\cdots&\mx{L}^{r,r}&\mx{R}^{r,r}
\end{bmatrix}\in\mathbb{C}^{2\sum_i s_i k_i \times 2\sum_i s_i k_i},
\end{align}
\begin{align}
\mx{L}\triangleq\begin{bmatrix}
\mx{L}^{1,1} &\cdots&\mx{L}^{1,r}\\
\vdots&\vdots&\vdots\\
\mx{L}^{r,1}&\cdots&\mx{L}^{r,r}
\end{bmatrix}\in\mathbb{C}^{2\sum_{i}s_ik_i\times \sum_{i}s_ik_i},
\end{align}
and
\begin{align}
\mx{R}\triangleq\begin{bmatrix}
\mx{R}^{1,1} &\cdots&\mx{R}^{1,r}\\
\vdots&\vdots&\vdots\\
\mx{R}^{r,1}&\cdots&\mx{R}^{r,r}
\end{bmatrix}\in\mathbb{C}^{2\sum_{i}s_ik_i\times \sum_{i}s_ik_i},
\end{align}
where $\mx{L}^{i,p}, \mx{R}^{i,p} \in\mathbb{C}^{2s_i k_i \times s_p k_p}$.
Also, consider $(\widetilde{\mx{D}}^i)^{-1}=\begin{bmatrix}
\overline{\mx{L}}^i &\overline{\mx{R}}^i
\end{bmatrix}\in\mathbb{C}^{2s_i\times 2 s_i}$ where $\overline{\mx{L}}^i, \overline{\mx{R}}^i\in\mathbb{C}^{2 s_i \times s_i}$. Leveraging \eqref{eq.D_equation} and \eqref{eq.Dinv}, leads to
\begin{align}\label{eq.alpha_ibeta_i}
\begin{bmatrix}
\bs{\alpha}\\
\kappa^{-1}\bs{\beta}
\end{bmatrix}=\mx{D}^{-1}\mx{u}=\mx{L}\widetilde{\mx{u}},
\end{align} 
where $\bs{\alpha}\triangleq[{\bs{\alpha}^1}^{\mathsf{T}},...,{\bs{\alpha}^{r}}^{\mathsf{T}}]^{\mathsf{T}}$, $\bs{\beta}\triangleq[{\bs{\beta}^1}^{\mathsf{T}},...,{\bs{\beta}^{r}}^{\mathsf{T}}]^{\mathsf{T}}$ and $\widetilde{\mx{u}}\triangleq[\mx{u}_1^{\mathsf{T}},..., \mx{u}_r^{\mathsf{T}}]^{\mathsf{T}}$.
Then, by exploiting \eqref{eq.V_l(tau)} and \eqref{eq.alpha_ibeta_i}, we may rewrite the $\ell$-th derivatives of $\mx{q}_m(\tau)$ in \eqref{eq.qii_main} as
\begin{align}\label{eq.kappal q_l(tau)}
\kappa^{\ell}\mx{q}_m^{\ell}(\tau)={{{\mx{V}_{\ell}}^m}}^{\mathsf{H}}(\tau)\mx{D}^{-1}\mx{u}={{\mx{V}_{\ell}}^m}^{\mathsf{H}}(\tau)\mx{L}\widetilde{\mx{u}},
\end{align}
where $\mx{q}_m^{\ell}(\tau)\in\mathbb{C}^{k_m\times 1}.$
% where $\mx{q}^{(\ell)}(\tau)=[{\mx{q}_1^{(\ell)}}^T(\tau),..., {\mx{q}_r^{(\ell)}}^T(\tau)]^T\in\mathbb{C}^{(\sum_{i=1}^r k_i) \times 1}$.
The above expression could be stated in the form of
\begin{align}
&\scalebox{.9}{${{{{\mx{V}}}}_{\ell}^m}^{\mathsf{H}}(\tau)\mx{L}\widetilde{\mx{u}}=\Big({{\mx{V}}}_{\ell}^m(\tau)-\mathds{E}{{\mx{V}}}_{\ell}^m(\tau)+\mathds{E}{{\mx{V}}}_{\ell}^m(\tau)\Big)^{\mathsf{H}}\Big(\mx{L}
-\overline{\mx{L}}+\overline{\mx{L}}\Big)\widetilde{\mx{u}}$}\nonumber\\
&=\mathds{E}{{{\mx{V}}}_{\ell}^m}^{\mathsf{H}}(\tau)\overline{\mx{L}}\widetilde{\mx{u}}+\Big( {{\mx{V}}}_{\ell}^m(\tau)-\mathds{E}{{\mx{V}}}_{\ell}^m(\tau)\Big)^{\mathsf{H}}\mx{L}\widetilde{\mx{u}}+\nonumber\\
&\mathds{E}{{{\mx{V}}}_{\ell}^m}^{\mathsf{H}}(\tau)
\Big(\mx{L}-\overline{\mx{L}}\Big)\widetilde{\mx{u}},
\end{align}
where $\overline{\mx{L}}\triangleq{\rm diag}\Big(\overline{\mx{L}}^{1,1}\otimes \mx{I}_{k_1},\cdots, \overline{\mx{L}}^{r,r}\otimes \mx{I}_{k_r} \Big)$.
%\begin{align}
%\overline{\mx{L}}\triangleq\begin{bmatrix}
%\overline{\mx{L}}^1\otimes \mx{I}_{k_1}&&\\
%&\ddots&\\
%&&\overline{\mx{L}}^r\otimes \mx{I}_{k_r}
%\end{bmatrix}
%\end{align}
Also, by \eqref{eq.EV}, we have: 
\begin{align}\label{re}
&\mathds{E}{{\mx{V}_{\ell}}^m}^{\mathsf{H}}(\tau)\overline{\mx{L}}\widetilde{\mx{u}}=\begin{bmatrix}
\mx{0}&\hdots& {\mx{v}_{\ell}^m}^{\mathsf{H}}(\tau)\otimes \mx{I}_{k_m}&\hdots&\mx{0}   
\end{bmatrix}\nonumber\\
% {\rm diag}\Big({\mx{v}_{\ell}^1}^{\mathsf{H}}(\tau)\otimes \mx{I}_{k_1},..., {\mx{v}_{\ell}^r}^{\mathsf{H}}(\tau)\otimes \mx{I}_{k_r} 
% \Big)\nonumber\\
&\underbrace{\begin{bmatrix}
\overline{\mx{L}}^{1,1}\otimes \mx{I}_{k_1}&\overline{\mx{R}}^{1,1}\otimes \mx{I}_{k_1}&\mx{0}\\
\mx{0}&\ddots&\mx{0}\\
\mx{0}&\overline{\mx{L}}^{r,r}\otimes \mx{I}_{k_r}&\overline{\mx{R}}^{r,r}\otimes \mx{I}_{k_r} 
\end{bmatrix}}_{=(\mathds{E}\mx{D})^{-1}}\begin{bmatrix}
\mx{u}_1\\
\mx{0}\\
\vdots\\
\mx{u}_r\\
\mx{0}
\end{bmatrix}
=\nonumber\\
&{\mx{v}_{\ell}^m}^{\mathsf{H}}(\tau)\otimes \mx{I}_{k_m} \begin{bmatrix}
\overline{\bs{\alpha}}^m\\
\kappa^{-1}\overline{\bs{\beta}}^m
\end{bmatrix}
% {\rm diag}\Big({\mx{v}_{\ell}^1}^{\mathsf{H}}(\tau)\otimes \mx{I}_{k_1},...,{ \mx{v}_{\ell}^r}^{\mathsf{H}}(\tau)\otimes \mx{I}_{k_r} 
% \Big)\begin{bmatrix}
% \overline{\bs{\alpha}}\\
% \kappa^{-1}\overline{\bs{\beta}}
% \end{bmatrix}
=\kappa^{\ell}\overline{\mx{q}}_m^{(\ell)}(\tau)
\end{align}
where in the last line, we use 
$\overline{\bs{\alpha}}=[{\overline{\bs{\alpha}}^1}^{\mathsf{T}},...,{\overline{\bs{\alpha}}^r}^{\mathsf{T}}]^{\mathsf{T}}$, $\overline{\bs{\beta}}=[{\overline{\bs{\beta}}^1}^{\mathsf{T}},...,{\overline{\bs{\beta}}^r}^{\mathsf{T}}]^{\mathsf{T}}$ and the fact that
\begin{align}\label{eq.alpha_bar}
\begin{bmatrix}
\overline{\bs{\alpha}}\\
\kappa^{-1}\overline{\bs{\beta}}
\end{bmatrix}=(\mathds{E}\mx{D})^{-1}\mx{u}
\end{align}
obtained from \eqref{eq.q_bar_conditions}, \eqref{eq.qi_main_deterministic} and \eqref{eq.EDinv}.
Consequently, by using \eqref{eq.alpha_bar}, we conclude that
\begin{align}\label{q-qbar}
&\kappa^{\ell}\mx{q}_m^{(\ell)}(\tau)=\kappa^{\ell} \overline{\mx{q}}_m^{(\ell)}(\tau)+\mx{I}_{1,m}^{\ell}(\tau)+\mx{I}_{2,m}^{\ell}(\tau)
\end{align}
where
\begin{align}
&\mx{I}_{1,m}^{\ell}(\tau)=\Big( {{\mx{V}}}_{\ell}^m(\tau)-\mathds{E}{{\mx{V}}}_{\ell}^m(\tau)\Big)^{\mathsf{H}}\mx{L}\widetilde{\mx{u}}\nonumber\\
&\mx{I}_{2,m}^{\ell}(\tau)=\mathds{E}{{{\mx{V}}}_{\ell}^m}^{\mathsf{H}}(\tau)
\Big(\mx{L}-\overline{\mx{L}}\Big)\widetilde{\mx{u}}
\end{align}
and $\overline{\mx{q}}_m^{(\ell)}(\tau)\in\mathbb{C}^{k_m\times 1}$.
% and $\overline{\mx{q}}^{(\ell)}(\tau)\triangleq[{\overline{\mx{q}}_1^{(\ell)}}^T(\tau),..., {\overline{\mx{q}}_r^{(\ell)}}^T	(\tau)]^T\in\mathbb{C}^{(\sum_{i=1}^r k_i) \times 1}$.
\begin{lem}\label{lem.V_concentrate}
Let $m\in\{1,..r\}$ be an integer. Fix $\epsilon_2$ and $\tau$ to be in $(0,1)$. Define the event $\mathcal{E}_{2,\epsilon_2}\triangleq\Big\{\|\mx{V}_{\ell}^m(\tau)-\mathds{E}\mx{V}_{\ell}^m(\tau)\|_{2\rightarrow2}\le \epsilon_2, \ell=0,1,2,3\Big\}$. If the number of samples satisfies
\begin{align}\label{eq.measure_V_concentrate}
&M\ge c_1 \tfrac{4^{\ell+1}}{\epsilon_2^2}\mu_+\sqrt{k_m}\sqrt{\sum_{i}s_i k_i}\log\Big(1+c_2\tfrac{4^{2\ell+2}\mu_{+}^2\sum_{i}s_ik_i}{\mu_{-}}\Big)\nonumber\\
&\log\Big(\tfrac{2\sum_is_ik_i+k_m}{\delta_2}\Big),
\end{align}
then, the event $\mathcal{E}_{2,\epsilon_2}$ holds with probability at least $1-4 \delta_2$. 
\end{lem}
Proof. See Appendix \ref{proof.lem.V_concentrate}.
\begin{lem}\label{lem.I1Concentrate}
Assume that $\mx{f}_i\in\mathbb{C}^{k_i\times 1}, i=1,..., r$ are i.i.d. and distributed symmetrically on the complex unit circle. Then, provided that the number of samples satisfies
\begin{align}
&M\ge c_5 \mu_{+} (\sum_i s_i k_i )\frac{k_+}{k_-}\log\Big(1+c_6\tfrac{\mu_{+}^2\sum_{i}s_ik_i}{\mu_{-}}\Big) \nonumber\\&\max\Big\{ \frac{1}{\epsilon_4^2} \log\Big(\tfrac{|\mathcal{T}_{\rm grid}|(\sum_is_ik_i)}{\delta}\Big)\times 
\log^2(\frac{|\mathcal{T}_{\rm grid}|(k_+)}{\delta})\nonumber\\&, \log(\frac{\sum_i s_i k_i}{\delta})\Big\}
\end{align}	
for some certain constants $c_5, c_6>0$, 
we have
\begin{align}
\mathds{P}\{\sup_{\tau_d\in\mathcal{T}_{\rm grid}}\|\mx{I}_{1,m}^{(\ell)}(\tau_d)\|_2\ge \epsilon_4, \ell=0,1,2,3\}\ge 1-12 \delta.
\end{align}
\end{lem}
Proof. See Appendix \ref{proof.lem.I1Concentate}.
\begin{lem}\label{lem.I2concentrate}
Let $m\in[r]$. Assume that $\mx{b}_i\in\mathbb{C}^{k_i\times 1}, i=1,..., r$ are i.i.d. and distributed symmetrically on the complex unit circle. Then, if the number of samples satisfies
\begin{align}
&M\ge \frac{c_1 \mu_{+} }{\epsilon_5^2}(\sum_i s_ik_i)\tfrac{k_+}{k_-}\log\Big(1+\frac{c_2\mu_{+}^2\sum_i s_i k_i}{\mu_{-}}\Big)\nonumber\\&\log(\frac{\sum_i s_i k_i}{\delta})\log(\frac{|\mathcal{T}_{\rm grid}|( k_+)}{\delta})
\end{align}
for some certain constants $c_1, c_2>0$, we have:
\begin{align}
\mathds{P}\{\sup_{\tau_d\in\mathcal{T}_{\rm grid}}\|\mx{I}_{2,m}^{(\ell)}(\tau_d)\|_2\ge \epsilon_5, \ell=0,1,2,3\}\ge 1-8 \delta.
\end{align}
\end{lem}
Proof. See Appendix \ref{proof.lem.I2}.

Next, we aim to show that $\mx{q}_m^{(\ell)}(\tau)$ is close to $\overline{\mx{q}}_m^{(\ell)}(\tau)$ in \eqref{q-qbar} for all $m=1,..., r$. First, we need to define the event
\begin{align}
\mathcal{E}=\Bigg\{\sup_{\tau_d\in\mathcal{T}_{\rm grid}}\|\kappa^{\ell} \mx{q}_m^{\ell}(\tau_d)-\kappa^{\ell}\overline{\mx{q}}_m^{\ell}(\tau_d)\|_2\le \frac{\epsilon}{3}, \ell=0,1,2,3 \Bigg\}.
\end{align}

Combining Lemmas \ref{lem.I1Concentrate}, \ref{lem.I2concentrate} and \eqref{q-qbar} with appropriate redefinition of $\epsilon$ and $\delta$ leads to the following proposition.
\begin{prop}\label{prop.closeness on grid}
Let $\mathcal{T}_{\rm grid}\subset [0,1]$ be a finite set of grid points and $\delta>0$ be a parameter to control probability. Then, there exist constants $c_1,c_2>0$ such that
 \begin{align}
 &M\ge c_1 \mu_{+} (\sum_i s_i k_i )\tfrac{k_+}{k_-}\log\Big(1+c_2\tfrac{\mu_{+}^2\sum_{i}s_ik_i}{\mu_{-}}\Big) \nonumber\\&\max\Big\{ \frac{1}{\epsilon^2} \log\Big(\tfrac{|\mathcal{T}_{\rm grid}|(\sum_is_ik_i)}{\delta}\Big)\times 
 \log^2(\frac{|\mathcal{T}_{\rm grid}|k_+}{\delta})\nonumber\\&, \log(\frac{\sum_i s_i k_i}{\delta})\Big\},
 \end{align}
is sufficient to guarantee the event $\mathcal{E}$ holds with probability at least $1-\delta$.
\end{prop}
In what follows, we aim to show that $\mx{q}_m^{(\ell)}(\tau)$ is close to $\overline{\mx{q}}_m^{(\ell)}(\tau)$ anywhere in $\tau\in [0,1)$.
\begin{lem}\label{lem.everywhere_off_suppoort}
Suppose the minimum separation between spikes satisfies $\Delta\ge\frac{1}{M}$. Then, if
\begin{align}
&M\ge c_1 \mu_{+} (\sum_i s_i k_i )\tfrac{k_{+}}{k_{-}}\log\Big(1+c_2\tfrac{\mu_{+}^2\sum_{i}s_ik_i}{\mu_{-}}\Big) \nonumber\\&\max\Big\{ \frac{1}{\epsilon^2} \log\Big(\tfrac{M(\sum_is_ik_i)}{\epsilon\delta}\Big)\times 
\log^2\big(\frac{M k_+}{\epsilon \delta}\big)\nonumber\\&, \log(\frac{\sum_i s_i k_i}{\delta})\Big\},
\end{align}  
for some constants $c_1, c_2>0$,
it holds that for any $\tau\in [0,1)$,
\begin{align}
\Big\|\kappa^{\ell}\mx{q}_m^{\ell}(\tau)-\kappa^{\ell}\overline{\mx{q}}_m^{\ell}(\tau)\Big\|_2\le \epsilon,
\end{align}
with probability at least $1-\delta$.
\end{lem}
Proof. See Appendix \ref{proof.lem.everywhere_off_suppoort}.

In what follows, we  aim to show that $\|\mx{q}_m(\tau)\|_2< 1$ anywhere in $\tau\in \mathcal{S}_m$ for each $m=1,..., r$. To this end, we define
\begin{align}
&\mathcal{T}^m_{\rm near}=\bigcup_{j=1}^{s_m}[\tau_j^m-\tau_{b,1},\tau_j^m+\tau_{b,1}],\nonumber\\
&\mathcal{T}^m_{\rm far}=[0,1)\setminus \mathcal{T}^m_{\rm near}, m=1,..., r
\end{align}
where $\tau_{b,1}\triangleq\frac{8.245\times 10^{-2}}{M}$. 
\begin{lem}\label{lem.Tfar}
Assume that the minimum separation satisfies $\Delta\ge \frac{1}{M}$. If the number of samples satisfies
\begin{align}
&M\ge c_1 \mu_{+} (\sum_i s_i k_i )\tfrac{k_+}{k_-}\log\Big(1+c_2\tfrac{\mu_{+}^2\sum_{i}s_ik_i}{\mu_{-}}\Big) \nonumber\\&\max\Big\{ \frac{1}{\epsilon^2} \log\Big(\tfrac{M(\sum_is_ik_i)}{\epsilon\delta}\Big)\times 
\log^2\Big(\frac{M k_+}{\epsilon \delta}\Big)\nonumber\\&, \log(\frac{\sum_i s_i k_i}{\delta})\Big\},
\end{align}
for some $c_1, c_2>0$, then we have:
\begin{align}
\|\mx{q}_m(\tau)\|_2<1, \forall \tau\in \mathcal{T}_{\rm far}^m,
\end{align}
with probability at least $1-\delta$.
\end{lem}
Proof. See Appendix \ref{proof.lem.Tfar}.

The following lemma proves $\|\mx{q}_m(\tau)\|_2<1$ for $\tau \in \mathcal{T}_{\rm near}^m$ for all $m=1,..., r$.
\begin{lem}\label{lem.Tnear}
Assume that $\tau\in \mathcal{T}_{\rm near}$ and the number of samples satisfies
\begin{align}
    &M\ge c_1 \mu_{+} (\sum_i s_i k_i )\tfrac{k_+}{k_-}\log\Big(1+c_2\tfrac{\mu_{+}^2\sum_{i}s_ik_i}{\mu_{-}}\Big) \nonumber\\
    &  \log\Big(\tfrac{M(\sum_is_ik_i)}{\delta}\Big)\times 
\log^2(\frac{M k_+}{ \delta}).
\end{align}
Then, it holds that
\begin{align}
    \|\mx{q}_m(\tau)\|_2<1, ~~\forall \tau\in \mathcal{T}_{\rm near}^m, m=1,..., r
\end{align}
with probability at least $1-\delta$.
\end{lem}
Proof. See Appendix \ref{proof.lem.T_near}.

Combining Lemmas \ref{lem.Tfar} and \ref{lem.Tnear} shows that $\|\mx{q}_m(\tau)\|_2<1 \forall \tau \in (0,1)\setminus \mathcal{S}_m$ for all $m=1,..., r$.

\section{Proof of Lemma \ref{lemma.concentrate1}}\label{proof.lemma.concentrate1}
 Proof road-map:
In this section, we aim to prove that $\mx{D}$ is concentrated around $\mathds{E}[\mx{D}]$. To this end, we use a strong version of the matrix Bernstein inequality known as Kolchinskii inequality \cite[Theorem 4]{koltchinskii2013remark}. Similar to the Bernstein inequality, this theorem also involves two parameters: $R$ and $\sigma^2$ (see Theorem \ref{thm.koltchinskii} and \eqref{eq.R-parameter} and \eqref{eq.sigma2-parameter}). However, in contrast to the traditional matrix Bernstein inequality, which depends solely on upper-bounds for $R$ and $\sigma^2$, our approach takes into account both upper and lower bounds. This distinctive characteristic enhances its precision and tightness compared to relying solely on upper-bounds. The upper-bound for $R$ parameter is provided in \eqref{eq.R_upper_bound}. The upper and lower bounds on $\sigma^2$ parameter are respectively provided in \eqref{eq.sigma_upper} and \eqref{eq.sigma_lower}.

 Proof details:
In this section, we use a concentration result to show that $\mx{D}$ and $\mathds{E}\mx{D}$ are close to each other with high probability given enough measurements. First, we need to define the Orlicz norm.  
\begin{defn}
Let $\Psi:\mathbb{R}_{+}\rightarrow\mathbb{R}_{+}$ be a non-decreasing function with $\Psi(0)=0$. For a random variable $H\in\mathbb{C}$, the Orlicz norm is defined as
\begin{align}
\|H\|_{\Psi}\triangleq\inf\Big\{C>0:\mathds{E}\Psi\Big(\frac{|H|}{C}\Big)\le 1\Big\}.
\end{align}
\end{defn}
The common choices of Orlicz norm are 
\begin{align*}
&\Psi(t)=t^p, p\ge 1 \rightarrow L_p~ \text{norm},\\
&\Psi(t)=\Psi_1(t)\triangleq{\rm e}^{t}-1\rightarrow \text{sub-exponential norm},\\
&\Psi(t)=\Psi_2(t)\triangleq{\rm e}^{t^2}-1\rightarrow \text{sub-Gaussian norm}.\\
\end{align*}
With this definition at hand, we provide an adaptation of Koltchinskii inequality \cite[Theorem 4]{koltchinskii2013remark} in the following theorem.
\begin{thm}\label{thm.koltchinskii}
Consider a set of zero-mean, and independent random matrices $\{\mx{H}_i\}_{i=1}^n$ of dimension $d_1\times d_2$. Let $\Psi(t)=\Psi_{\alpha}(t)\triangleq{\rm e}^{t^{\alpha}}-1$ with $\alpha \ge 1$.  Set 
\begin{align}
&R\triangleq\max_{i}\|\|\mx{H}_i\|_{2\rightarrow 2}\|_{\Psi_{\alpha}},\label{eq.R-parameter}\\
&\sigma^2\triangleq\max\Bigg\{\|\sum_{i=1}^{n}\mathds{E}\mx{H}_i^{\mathsf{H}}\mx{H}_i\|_{2\rightarrow 2}, \|\sum_{i=1}^{n}\mathds{E}\mx{H}_i\mx{H}_i^{\mathsf{H}}\|_{2\rightarrow 2}\Bigg\}.\label{eq.sigma2-parameter}
\end{align}
Then, there exists a universal constant $c$ such that
\begin{align}\label{eq.KolProbability}
&\mathds{P}\Big(\|\sum_{i=1}^{n}\mx{H}_i\|_{2\rightarrow 2}\ge 
c\max \Bigg\{\sigma\sqrt{t+\log(d_1+d_2)}, \nonumber\\
&R\Big(\log\big(1+\frac{n R^2}{\sigma^2}\big)\Big)^{\frac{1}{\alpha}}(t+\log(d_1+d_2))\Bigg\}\Big)\le {\rm e}^{-t}.
\end{align}
\end{thm}
Our aim is to apply Theorem \ref{thm.koltchinskii} to prove that $\|\mx{D}-\mathds{E}\mx{D}\|_{2\rightarrow 2}$ becomes small by taking enough measurements. It is straightforward to observe that the random matrix variables $\mx{S}_n, n=-2M, \ldots, 2M$ in \eqref{eq.Sn} are independent, and zero mean (by the isotropy assumption \eqref{eq.isotropy} ). In order to apply Theorem \ref{thm.koltchinskii}, we first need to calculate the $\Psi_{\alpha}$ norm of them. For any fixed
\begin{align}\label{eq.fixed_z}
\mx{z}=
\begin{bmatrix}
\mx{z}^1\\
\vdots\\
\mx{z}^{r}
\end{bmatrix},~\mx{z}^i\in\mathbb{C}^{2s_ik_i\times 1},
\end{align}
we have
\begin{align}\label{eq.rell1}
&\|\mx{S}_n\mx{z}\|_2^2\stackrel{(\RN{1})}{=} \sum_{m=1}^{r}\|\sum_{i=1}^{r}\mx{S}_n^{m,i}\mx{z}^i\|_2^2\stackrel{(\RN{2})}{\le} \sum_{m=1}^{r}(\sum_{i=1}^{r}\|\mx{S}_n^{m,i}\mx{z}^i\|_2)^2\stackrel{(\RN{3})}{\le}\nonumber\\
&\sum_{m=1}^{r}(\sum_{i=1}^{r}\|\mx{S}_n^{m,i}\|_{2\rightarrow 2}\|\mx{z}^i\|_2)^2\stackrel{(\RN{4})}{\le} \sum_{m=1}^{r}\sum_{i=1}^{r}\|\mx{S}_n^{m,i}\|_{2\rightarrow 2}^2\sum_{i=1}^{r}\|\mx{z}^i\|_2^2\nonumber\\
&=\sum_{m=1}^{r}\sum_{i=1}^{r}\|\mx{S}_n^{m,i}\|_{2\rightarrow 2}^2\|\mx{z}\|_2^2,
\end{align}
where $(\RN{1})$ comes from the definition of $\mx{S}_n$ in \eqref{eq.Sn}, $(\RN{2})$ is because of triangle inequality of norms, $(\RN{3})$ comes from the definition of operator norm $\|\cdot\|_{2\rightarrow 2}$ of a matrix, and $(\RN{4})$ is the result of Cauchy Schwartz inequality. Hence, according to \eqref{eq.rell1}, we have
\begin{align}\label{eq.rell2}
\|\mx{S}_n\|_{2\rightarrow 2}\le \sqrt{\sum_{m=1}^{r}\sum_{i=1}^{r}\|\mx{S}_n^{m,i}\|_{2\rightarrow 2}^2},
\end{align}
where $\|\mx{S}_n^{m,i}\|_{2\rightarrow 2}$, regarding \eqref{eq.Smi}, is given by
\begin{align}\label{eq.rell3}
&\|\mx{S}_n^{m,i}\|_{2\rightarrow 2}=\frac{1}{M}|g(n)|\|\bs{\nu}_n^m{\bs{\nu}_n^i}^{\mathsf{H}}\|_{2\rightarrow 2}\|\mx{b}_n^m{\mx{b}_n^i}^{\mathsf{H}}-\mx{I}_{k_i}1_{m=i}\|_{2\rightarrow 2}\nonumber\\
&\le \frac{1}{M}|g(n)|\|\bs{\nu}_n^m\|_2\|\bs{\nu}_n^i\|_2\max\{\|\mx{b}_n^m\|_2\|\mx{b}_n^i\|_2, 1\},
\end{align}
where we used the fact that
\begin{align}\label{eq.bnbn^H_spectral}
\|\mx{b}_n^m{\mx{b}_n^i}^{\mathsf{H}}-\mx{I}_{k_i}1_{m=i}\|_{2\rightarrow 2}=\left\{\begin{array}{lr}
\max\{\|\mx{b}_n^i\|_2^2-1, 1\},&~m=i\\
\|\mx{b}_n^i\|_2\|\mx{b}_n^m\|_2. &m\neq i
\end{array}\right.
\end{align}
Regarding \eqref{eq.nu-i},
\begin{align}\label{eq.rell4}
\|\bs{\nu}_n^i\|_2=\sqrt{s_i}\sqrt{1+|2\pi n \kappa|^2}.
\end{align}
We know that \cite[Proof of Proposition 4.12]{tang2013compressed}:
\begin{align}\label{eq.rell5}
\max_{|n|\le 2M}|2\pi n \kappa|^2\le 13~\text{when}~M\ge 4,
\end{align}
and 
\begin{align}\label{eq.rell6}
\max_{|n|\le 2M} |g(n)| \le 1.
\end{align}
Therefore, combining \eqref{eq.rell3}, \eqref{eq.rell4}, \eqref{eq.rell5}, and \eqref{eq.rell6} and using the assumption \eqref{eq.incoherence}, $\|\mx{S}_n^{m,i}\|_{2\rightarrow 2}$ can be bounded as
\begin{align}
\|\mx{S}_n^{m,i}\|_{2\rightarrow 2}\le \frac{14}{M}\mu_{+}\sqrt{s_i s_m k_i k_m}.
\end{align}
Consequently, using \eqref{eq.rell2}, 
\begin{align}
\|\mx{S}_n\|_{2\rightarrow 2}\le \frac{14}{M}\mu_{+}\sum_{i=1}^{r}s_ik_i.
\end{align}
By using the sub-exponential norm, i.e. $\Psi=\Psi_1$, the $R$ parameter is finally given by
\begin{align}\label{eq.R_upper_bound}
\|\|\mx{S}_n\|_{2\rightarrow 2}\|_{\Psi_1}\le \frac{14}{M}\mu_{+}\sum_{i=1}^{r}s_ik_i.
\end{align}
We now turn to calculate the $\sigma^2$ parameter. For a fixed $\mx{z}$ (as in \eqref{eq.fixed_z}), we have
\begin{align}\label{eq.rell7}
(\mx{S}_n\mx{S}_n^{\mathsf{H}}\mx{z})_i=\sum_{l=1}^{r}\sum_{p=1}^{r}\mx{S}_n^{i,p}{\mx{S}_n^{l,p}}^{\mathsf{H}}\mx{z}^{l},
\end{align}
where $\mx{S}_n^{i,p}{\mx{S}_n^{l,p}}^{\mathsf{H}}$, using \eqref{eq.Smi}, reads as
\begin{align}\label{eq.rell8}
&\mx{S}_n^{i,p}{\mx{S}_n^{l,p}}^{\mathsf{H}}=\frac{1}{M^2}g^2(n)\Big[\bs{\nu}_n^{i}{\bs{\nu}_n^p}^{\mathsf{H}}\otimes \big(\mx{b}_n^i{\mx{b}_n^p}^{\mathsf{H}}-\mx{I}_{k_p}1_{i=p}\big)\Big].\nonumber\\
&\Big[\bs{\nu}_n^{p}{\bs{\nu}_n^l}^{\mathsf{H}}\otimes \big(\mx{b}_n^p{\mx{b}_n^l}^{\mathsf{H}}-\mx{I}_{k_p}1_{l=p}\big)\Big]=\frac{1}{M^2}g^2(n)
\|\bs{\nu}_n^p\|_2^2\bs{\nu}_n^i{\bs{\nu}_n^l}^{\mathsf{H}}\otimes\nonumber\\
&\Big[\|\mx{b}_n^p\|_2^2\mx{b}_n^i{\mx{b}_n^l}^{\mathsf{H}} -\mx{b}_n^{i}{\mx{b}_n^p}^{\mathsf{H}}1_{l=p}-\mx{b}_n^{p}{\mx{b}_n^l}^{\mathsf{H}}1_{i=p}+\mx{I}_{k_p}1_{i=p}1_{l=p}\Big].
\end{align}
Plug the latter relation into \eqref{eq.rell7} to form
\begin{align}\label{eq.rell9}
&(\mx{S}_n\mx{S}_n^{\mathsf{H}}\mx{z})_i=\sum_{l=1}^{r}\sum_{p=1}^{r}\frac{1}{M^2}g^2(n)
\|\bs{\nu}_n^p\|_2^2\bs{\nu}_n^i{\bs{\nu}_n^l}^{\mathsf{H}}\otimes\nonumber\\
&\scalebox{.9}{$\Big[\|\mx{b}_n^p\|_2^2\mx{b}_n^i{\mx{b}_n^l}^{\mathsf{H}} -\mx{b}_n^{i}{\mx{b}_n^p}^{\mathsf{H}}1_{l=p}-\mx{b}_n^{p}{\mx{b}_n^l}^{\mathsf{H}}1_{i=p}+\mx{I}_{k_p}1_{i=p}1_{l=p}\Big]\mx{z}^{l}.$}
\end{align}
Taking the expectation of \eqref{eq.rell9} leads to
\begin{align}\label{eq.rell11}
&\mathds{E}(\mx{S}_n\mx{S}_n^{\mathsf{H}}\mx{z})_i=\sum_{l=1}^{r}\sum_{p=1}^{r}\frac{1}{M^2}g^2(n)
\|\bs{\nu}_n^p\|_2^2\bs{\nu}_n^i{\bs{\nu}_n^l}^{\mathsf{H}}\otimes\nonumber\\
&\scalebox{.9}{$\Big[\mathds{E}\{\|\mx{b}_n^p\|_2^2\mx{b}_n^i{\mx{b}_n^l}^{\mathsf{H}}\} -\mx{I}_{k_i}1_{i=p}1_{l=p}-\mx{I}_{k_p}1_{l=p}1_{i=p}+\mx{I}_{k_p}1_{i=p}1_{l=p}\Big]$}\mx{z}^{l}\nonumber\\
&=\Bigg[\frac{1}{M^2}g^2(n)\|\bs{\nu}_n^i\|_2^2\bs{\nu}_n^i{\bs{\nu}_n^i}^{\mathsf{H}}\otimes\Big(\mathds{E}\{\|\mx{b}_n^i\|_2^2\mx{b}_n^i{\mx{b}_n^i}^{\mathsf{H}}\} -\mx{I}_{k_i}\Big)+\nonumber\\
&\sum_{\substack{p=1\\p\neq i}}^{r}\frac{1}{M^2}g^2(n)
\|\bs{\nu}_n^p\|_2^2\bs{\nu}_n^i{\bs{\nu}_n^i}^{\mathsf{H}}\otimes\Big(\mathds{E}\{\|\mx{b}_n^p\|_2^2\}\mathds{E}\{\mx{b}_n^i{\mx{b}_n^i}^{\mathsf{H}}\}\Big)\Bigg]\mx{z}^{i},
\end{align}
where we used
%---------------
\begin{align}\label{eq.rell10}
    \mathds{E}\{\|\mx{b}_n^p\|_2^2\mx{b}_n^i{\mx{b}_n^l}^{\mathsf{H}}\} =  \begin{cases}
        \mathds{E}\|\mx{b}_n^p\|_2^2\mathds{E}\mx{b}_n^i{\mx{b}_n^i}^{\mathsf{H}}&~~ i=l\neq p\\
\mx{0},&i\neq l.
    \end{cases}
\end{align}
%---------------
In fact, for the second relation above, when $i\neq l=p$,
\begin{align}
\mathds{E}\{\|\mx{b}_n^p\|_2^2\mx{b}_n^l{\mx{b}_n^i}^{\mathsf{H}}\}=\mathds{E}\mx{b}_n^i\mathds{E}\{ \|\mx{b}_n^p\|_2^2{\mx{b}_n^i}^{\mathsf{H}}\}=\mx{0}.
\end{align}
The case of $i=p\neq l$ follows a similar argument. Using \eqref{eq.nu-i}, it is beneficial to express $\mathds{E}\mx{S}_n\mx{S}_n^{\mathsf{H}}$ in a matrix form as  
\begin{align}\label{eq.Gamma_relation}
\sum_{n=-2M}^{2M}\mathds{E}\mx{S}_n\mx{S}_n^{\mathsf{H}}={\rm diag}([\mx{\Gamma}_1, ..., \mx{\Gamma}_r]),
\end{align}
where 
\begin{align}
&\mx{\Gamma}_i=\sum_{n=-2M}^{2M}\Bigg[\frac{1}{M^2}g^2(n)(1+|2\pi n\kappa|^2)s_i\bs{\nu}_n^i{\bs{\nu}_n^i}^{\mathsf{H}}\otimes\nonumber\\
&\Big(\mathds{E}\{\|\mx{b}_n^i\|_2^2\mx{b}_n^i{\mx{b}_n^i}^{\mathsf{H}}\} -\mx{I}_{k_i}\Big)+\sum_{\substack{p=1\\p\neq i}}^{r}\frac{1}{M^2}g^2(n)
(1+|2\pi n\kappa|^2)s_p\nonumber\\
&\bs{\nu}_n^i{\bs{\nu}_n^i}^{\mathsf{H}}\otimes\Big(\mathds{E}\{\|\mx{b}_n^p\|_2^2\}\mx{I}_{k_i}\Big)\Bigg]
\end{align}
Benefiting the assumptions \eqref{eq.isotropy} and \eqref{eq.incoherence}, we can obtain upper and lower bounds for $\mx{\Gamma}_i$ as follows:
\begin{align}\label{eq.Gamma_i}
&\mx{\Gamma}_i\overset{\preccurlyeq}{\succcurlyeq}\sum_{n=-2M}^{2M}\Bigg[\frac{1}{M^2}g^2(n)(1+|2\pi n\kappa|^2)s_i\bs{\nu}_n^i{\bs{\nu}_n^i}^{\mathsf{H}}\otimes\nonumber\\
&(\mu k_i-1)\mx{I}_{k_i}+\sum_{\substack{p=1\\p\neq i}}^{r}\frac{1}{M^2}g^2(n)
(1+|2\pi n\kappa|^2)s_p\nonumber\\
&\bs{\nu}_n^i{\bs{\nu}_n^i}^{\mathsf{H}}\otimes \mu k_p\mx{I}_{k_i}\Bigg]=\Big(s_i(\mu k_i-1)+\sum_{\substack{p=1\\p\neq i}}^{r}\mu s_p k_p\Big).\nonumber\\
&\Bigg(\sum_{n=-2M}^{2M}\frac{1}{M^2}g^2(n)(1+|2\pi n\kappa|^2)\bs{\nu}_n^i{\bs{\nu}_n^i}^{\mathsf{H}}\Bigg)\otimes \mx{I}_{k_i}\nonumber\\
&\triangleq\Big(\frac{s_i(\mu k_i-1)+\sum_{\substack{p=1\\p\neq i}}^{r}\mu s_p k_p}{M}\Big)\mx{\Phi}_i\otimes \mx{I}_{k_i},
\end{align}
where $\mx{\Phi}_i\triangleq\frac{1}{M}\sum_{n=-2M}^{2M}g^2(n)(1+|2\pi n\kappa|^2)\bs{\nu}_n^i{\bs{\nu}_n^i}^{\mathsf{H}}$, and $\mu$ is meant to be interpreted as $\mu_{+}$ for $\preccurlyeq$ and $\mu_{-}$ for $\succcurlyeq$.

Now, we are ready to calculate the $\sigma^2$ parameter by taking the spectral norm from $\mx{\Gamma}_i$:
\begin{align}\label{eq.gamma_bound}
\|\mx{\Gamma}_i\|_{2\rightarrow 2}\lessgtr \frac{ \Big|\mu\sum_{p=1}^{r}s_p k_p-s_i\Big|}{M}\|\mx{\Phi}_i\|_{2\rightarrow 2},
\end{align}
where we used $\|\mx{\Phi}_i\otimes \mx{I}_{k_i}\|_{2\rightarrow 2}=\|\mx{\Phi}_i\|_{2\rightarrow 2}$. The following lemma whose proof is given in Appendix \ref{proof.lem.phi_bound}, is useful in obtaining upper- and lower-bounds regarding the operator norm of $\mx{\Phi}_i$.
\begin{lem}\label{lem.phi_bound}
Suppose that $\Delta>\frac{1}{M}$ and that $M>70$. Then, the matrices $\mx{\Phi}_i, i=1,..., r$ are invertible and satisfy the following relations:
\begin{subequations}
\begin{align}
&\| \mathbf{I} - \boldsymbol{\Phi}_{i}	\|_{2 \rightarrow 2} \leq 1.39\times 10^{-1},\\
&\| \boldsymbol{\Phi}_{i}	\|_{2 \rightarrow 2} \leq 1.139 ,\\
&\|\boldsymbol{\Phi}_{i}^{-1}	\|_{2 \rightarrow 2} \leq 1.161,
\end{align}
\end{subequations}
for $i=1, \ldots, r$.
\end{lem}
\begin{proof}
    See Appendix~\ref{proof.lem.phi_bound}. 
\end{proof}
As a consequence, \eqref{eq.gamma_bound} becomes
\begin{align}
&\|\mx{\Gamma}_i\|_{2\rightarrow 2}\le\frac{ \Big|\mu\sum_{p=1}^{r}s_p k_p-s_i\Big|}{M} 1.139\label{eq.Gamma_upper}\\
&\|\mx{\Gamma}_i\|_{2\rightarrow 2}\ge\frac{ \Big|\mu\sum_{p=1}^{r}s_p k_p-s_i\Big|}{M} 0.86.\label{eq.Gamma_lower}
\end{align}
Hence, the upper-bound for $\sigma^2=\max_{i}\|\mx{\Gamma}_i\|_{2\rightarrow 2}$ parameter reads as
\begin{align}\label{eq.sigma_upper}
&\sigma^2\le \max_i\Big[\frac{ \Big(\mu_+\sum_{p=1}^{r}s_p k_p-s_i\Big)}{M} 1.139\Big]<\nonumber\\
&\frac{ \Big(\mu_+\sum_{p=1}^{r}s_p k_p\Big)}{M} 1.139
\end{align}
For the lower-bound on $\sigma^2$, the following lemma whose proof is provided in Appendix \ref{proof.lem.sigma_lower} is useful.

\begin{lem}\label{lem.sigma_lower}
There always exists a constant $0<c_1\le 1-\frac{s_{-}}{\mu_{+}\sum_i s_i k_i}$ such that 
\begin{align}\label{eq.sigma_lower}
&\sigma^2\ge \max_{i} \Big[\frac{ \Big|\mu\sum_{p=1}^{r}s_p k_p-s_i\Big|}{M} 0.86\Big]\ge \frac{c_1 \mu_{-}\sum_p s_p k_p}{M}
%=\nonumber\\
%& \frac{\max\{|\mu_{+}\sum_p k_ps_p-\min_i s_i|,|\mu_{-}\sum_p k_ps_p-\max_i s_i|\}}{M}.86\label{eq.sigma_lower}
\end{align}
\end{lem}
By replacing bounds on $R$ and $\sigma^2$ in \eqref{eq.R-parameter}, \eqref{eq.sigma_upper} and \eqref{eq.sigma_lower} into the Bernstein inequality in Theorem \ref{thm.koltchinskii}, we have
\begin{align}\label{eq.rel11}
&R\log(1+\frac{NR^2}{\sigma^2})(t+\log(4 \sum_i s_ik_i))\le \nonumber\\
&\frac{14}{M}\mu_{+}\sum_i s_ik_i \log\Big(1+\frac{N 14^2 \mu_{+}^2 \sum_i s_i k_i}{M \mu_{-}}\Big)\le \nonumber\\
&\frac{14}{M}\mu_{+}\sum_i s_ik_i \log\Big(1+\frac{c_2 \mu_{+}^2 \sum_i s_i k_i}{\mu_{-}}\Big),
\end{align}
where the last inequality comes from the fact that there exists a certain constant $c_2$ which $14^2 N=196(4M+1)$ can be bounded from above by $c_2 M$. We also have
\begin{align}\label{eq.rel22}
\sigma\sqrt{t+\log(4\sum_i s_ik_i)}\le \sqrt{1.139\frac{\mu_{+}\sum_i s_i k_i}{M}}
\end{align} 
By taking enough number of measurements according to \eqref{eq.measure_bound_concent} and setting $t=-\log(\delta)$, it holds that
\begin{align}
&\max \Bigg\{\sigma\sqrt{t+\log(2d)}, R\Big(\log\big(1+\frac{n R^2}{\sigma^2}\big)\Big)^{\frac{1}{\alpha}}(t+\log(2d))\Bigg\}\nonumber\\
\le \epsilon.
\end{align}
By using Theorem \ref{thm.koltchinskii}, the event $\mathcal{E}_{1,\epsilon}\triangleq\{\|\mx{D}-\mathds{E}\mx{D}\|_{2\rightarrow 2}\le \epsilon\}$ holds with probability at least $1-\delta$. This implies that
\begin{align}
&\|\mx{I}-\mx{D}\|_{2\rightarrow 2}\le \|\mx{I}-\mathds{E}\mx{D}\|_{2\rightarrow 2}+\|\mx{D}-\mathds{E}\mx{D}\|_{2\rightarrow 2}\le 0.3623+\epsilon_1\nonumber\\
&<1,
\end{align}
with probability at least $1-\delta$. Here, $\|\mx{I}-\mathds{E}\mx{D}\|_2\le 0.3623$ comes from Lemma \ref{lem.ED_rel1}
‌\section{Sub-proofs}
\subsection{Proof of Lemma \ref{lem.sigma_lower}}\label{proof.lem.sigma_lower}
Since $\mathds{E}\|\mx{b}^i\|_2^2=k_i$ due to \eqref{eq.isotropy} and $k_i\ge 1$, we can conclude from \eqref{eq.incoherence} that $\mu_{+}k_i\ge 1 ~\forall i\in [r]$ almost surely. By \eqref{eq.Gamma_relation} and \eqref{eq.Gamma_lower}, it is simple to show that the lower-bound is equal to
\begin{align}
&\sigma^2=\max_i \|\mx{\Gamma}_i\|_{2\rightarrow 2}>\max_{i} \Big[\frac{ \Big|\mu\sum_{p=1}^{r}s_p k_p-s_i\Big|}{M} 0.86\Big]=\nonumber\\
&\frac{\max\{|\mu_{+}\sum_p k_ps_p-\min_i s_i|,|\mu_{-}\sum_p k_ps_p-\max_i s_i|\}}{M}.86\nonumber\\
&\ge \frac{\mu_{+}\sum_p k_ps_p-\min_i s_i}{M}.86
\end{align}
It is now sufficient to show that
\begin{align}
\mu_{+}\sum_i s_ik_i-\min_i s_i\ge c_1\mu_{+} \sum_i s_i k_i.
\end{align}
This gives the condition:
\begin{align}
    c_1\le 1- \frac{s_{-}}{\mu_{+}\sum_{i} s_i k_i}.
\end{align}
which concludes the result.

\subsection{Proof of Lemma \ref{lem.V_concentrate}}\label{proof.lem.V_concentrate}
For each $m=1,..., r$, we can write 
\begin{align}\label{V_l(tau)}
&\mx{V}_{\ell}^m(\tau)-\mathds{E}\mx{V}_{\ell}^m(\tau)=\nonumber\\
&\frac{1}{M}\sum_{n=-2M}^{2M}g(n)(j2\pi n\kappa)^{\ell}e^{-j2\pi n\kappa\tau}(\mx{B}_n^m-\overline{\mx{B}}_n^m)=:\nonumber\\
&\sum_{n=-2M}^{2M}\mx{Y}_n^m\in\mathbb{C}^{2\sum_{i}s_ik_i \times k_m},
%
%\bs{\nu}_n^i\otimes \Big(\mx{b}_n^i{\mx{b}_n^m}^{\mathsf{H}}-\mx{I}_{k_i}1_{m=i}\Big)\nonumber\\
%&=:\sum_{n=-2M}^{2M}\mx{Y}_n^{m,i}.
\end{align}
where 
\begin{align}
\mx{B}_n^m\triangleq\begin{bmatrix}
\bs{\nu}^1\otimes \mx{b}_n^1{\mx{b}_n^m}^{\mathsf{H}}\\
\vdots\\
 \bs{\nu}^r\otimes \mx{b}_n^r{\mx{b}_n^m}^{\mathsf{H}} 
\end{bmatrix}
\end{align}
and
\begin{align}
\overline{\mx{B}}_n^m\triangleq\begin{bmatrix}
\mx{0}\\
\vdots\\
\bs{\nu}^m\otimes \mx{I}_{k_m}\\
\vdots\\
\mx{0} 
\end{bmatrix}
\end{align}

$\mx{Y}_n^m$s are independent zero mean matrices due to \eqref{eq.isotropy} and can be stated as
\begin{align}
\mx{Y}_n^m=\begin{bmatrix}
\mx{Y}_n^{1,m}\\
\vdots\\
\mx{Y}_n^{r,m}
\end{bmatrix}
\end{align}
where \begin{align}
    {\mx{Y}_n^{i,m}=\frac{1}{M}g(n)(j2\pi n\kappa)^{\ell} e^{-j2\pi n\kappa \tau} \bs{\nu}_n^i\otimes (\mx{b}_n^{i}{\mx{b}_n^{m}}^{\mathsf{H}}-\mx{I}_{k_m}1_{i=m})}.
\end{align}
 To apply Bernstein inequality, we need $R$ and $\sigma^2$ parameters. For $R$ parameter, regarding \eqref{eq.rell4}, \eqref{eq.rell5}, \eqref{eq.rell6} and \eqref{eq.bnbn^H_spectral}, we have:
\begin{align}\label{eq.Rparam}
&R=\|{\mx{Y}_n^m}\|_{2\rightarrow 2}\le\frac{1}{M}|2\pi n\kappa|^{\ell}\|\mx{B}_n^m-\overline{\mx{B}}_n^m\|_{2\rightarrow 2}\nonumber\\
&\frac{1}{M}4^{\ell}\sqrt{\sum_{i=1}^r\|\bs{\nu}_n^i\otimes (\mx{b}_n^i{\mx{b}_n^m}^{\mathsf{H}}-\mx{I}_{k_i}1_{m=i})\|_{2\rightarrow 2}^2}\nonumber\\
&=\frac{1}{M}4^{\ell}\sqrt{\sum_{i=1}^r\|\bs{\nu}_n^i\|_2^2 \|\mx{b}_n^i{\mx{b}_n^m}^{\mathsf{H}}-\mx{I}_{k_i}1_{m=i}\|_{2\rightarrow 2}^2}\nonumber\\
&= \frac{1}{M}4^{\ell}\sqrt{\sum_{i=1}^r\|\bs{\nu}_n^i\|_2^2 \big(\max\{\|\mx{b}_n^i\|_2\|\mx{b}_n^m\|_2,1\}\big)^2}\nonumber\\
&\le \frac{1}{M}4^{\ell} \sqrt{14}\mu_{+}\sqrt{\sum_{i}s_ik_i}\sqrt{k_m}\le\nonumber\\
&\frac{1}{M}4^{\ell+1}\mu_{+}\sqrt{\sum_{i}s_ik_i}\sqrt{ k_m}.
\end{align}
The first inequality comes from \eqref{eq.rell2}. The third line stems from the spectral norm properties of Kronecker products. The fourth line uses \eqref{eq.bnbn^H_spectral}, and the fifth line is the result of incoherence property \eqref{eq.incoherence} and the fact that $\mu_{+}k_i \ge 1 \forall i=1, \ldots, r$. For the variance term, it follows that
\begin{align}
&\mx{Y}_n^{i,p}{\mx{Y}_n^{l,p}}^{\mathsf{H}}=\frac{1}{M^2}|g(n)|^2|2\pi n\kappa|^{2\ell}\bs{\nu}_n^i{\bs{\nu}_n^l}^{\mathsf{H}}\otimes\nonumber\\
&\Big[\|\mx{b}_n^p\|_2^2\mx{b}_n^i{\mx{b}_n^l}^{\mathsf{H}} -\mx{b}_n^{i}{\mx{b}_n^p}^{\mathsf{H}}1_{l=p}-\mx{b}_n^{p}{\mx{b}_n^l}^{\mathsf{H}}1_{i=p}+\mx{I}_{k_p}1_{i=p}1_{l=p}\Big].
\end{align}
Taking expectation and leveraging the same arguments as in the relations \eqref{eq.rell9}, \eqref{eq.rell11} and \eqref{eq.rell10}, we reach
\begin{align}
\sum_{n=-2M}^{2M}\mathds{E}\mx{Y}_n\mx{Y}_n^{\mathsf{H}}=\begin{bmatrix}
\mx{\Xi}_1&&\\
&\ddots&\\
&&\mx{\Xi}_r
\end{bmatrix}
\end{align}
where 
\begin{align}
    &\mx{\Xi}_m=\sum_{n=-2M}^{2M}\Bigg[\frac{1}{M^2}g^2(n)|2\pi n\kappa|^{2\ell}\bs{\nu}_n^m{\bs{\nu}_n^m}^{\mathsf{H}}\otimes\nonumber\\
&\Big(\mathds{E}\{\|\mx{b}_n^m\|_2^2\mx{b}_n^m{\mx{b}_n^m}^{\mathsf{H}}\} -\mx{I}_{k_m}\Big)
\end{align}
and for $i\neq m$
\begin{align}
&\mx{\Xi}_i=\sum_{n=-2M}^{2M}\Bigg[\frac{1}{M^2}g^2(n) |2\pi n\kappa|^{2\ell}\bs{\nu}_n^i{\bs{\nu}_n^i}^{\mathsf{H}}\otimes\Big(\mathds{E}\{\|\mx{b}_n^p\|_2^2\}\mx{I}_{k_i}\Big)\Bigg]
\end{align}
Define $\mx{\Psi}_i^{\ell}\triangleq\sum_{n=-2M}^{2M}\frac{1}{M}|g(n)|^2|2\pi n\kappa|^{2\ell}\bs{\nu}_n^i{\bs{\nu}_n^i}^{\mathsf{H}}, \ell=0,1,2,3$.
Thus, due to the same arguments as in \eqref{eq.Gamma_i}, it holds that
\begin{align}
\scalebox{.95}{$\sum_{n=-2M}^{2M}\mathds{E}\mx{Y}_n{\mx{Y}_n}^{\mathsf{H}}\overset{\preccurlyeq}{\succcurlyeq}{\rm diag}(\{\frac{(\mu \sum_{m=1}^r k_m  -1_{i=m})}{M}\mx{\Psi}_i^{\ell} \otimes 
 \mx{I}_{k_i}\}_{i=1}^r)$}
\end{align}
and
\begin{align}
&\sigma^2=\|\sum_{n=-2M}^{2M}\mathds{E}\mx{Y}_n{\mx{Y}_n}^{\mathsf{H}}\|_{2\rightarrow 2}\lessgtr \nonumber\\
&
\max_i\frac{|\mu \sum_i k_i -1_{i=m}|}{M}{\|\mx{\Psi}_i^{\ell}\|_{2\rightarrow 2}}
\end{align}
Similar to the arguments provided in \eqref{eq.sigma_upper} and \eqref{eq.sigma_lower}, one can show that there exist certain constants $c_1, c_2>0$ such that
{
\begin{align}\label{eq.sigma2param}
&\sigma^2< c_1 \frac{\mu_{+}}{M}k_m\nonumber\\
&\sigma^2>c_2 \frac{\mu_{-}}{M}k_m
\end{align}
}To use Bernstein inequality, we integrate $R$ and $\sigma^2$ parameters obtained in \eqref{eq.Rparam} and \eqref{eq.sigma2param} into  Theorem \ref{thm.koltchinskii} to reach
\begin{align}\label{eq.relll1}
&R\log(1+\frac{N R^2}{\sigma^2})(t+\log(2\sum_i s_i k_i+k_m))\le\nonumber\\
&\frac{1}{M}4^{\ell+1}\mu_{+}\sqrt{\sum_{i}s_ik_i}\sqrt{ k_m}\log\Big(1+\frac{c 4^{2\ell+2}\mu_{+}^2 \sum_i s_i k_i}{\mu_{-}}\Big)\nonumber\\
&(t+\log(2\sum_i s_i k_i+k_m))
\end{align}
for some certain constant $c$.
Making the failure probability no more than $\delta_2$ by setting $t=-\log(\delta_2)$ yields to
\begin{align}
&\max\Big\{R\log(1+\frac{N R^2}{\sigma^2})(t+\log(2\sum_i s_i k_i+k_m)), \nonumber\\
&\sigma\sqrt{t+\log(2\sum_i s_i k_i+k_m)}\Big\}\le \epsilon_2
\end{align}
which by \eqref{eq.relll1} leads to \eqref{eq.measure_V_concentrate} for certain constants $c_1$ and $c_2$.

\subsection{Proof of Lemma \ref{lem.I1Concentrate} }\label{proof.lem.I1Concentate}
For any $\tau_d\in\mathcal{T}_{\rm grid}$, $\mx{I}^{\ell}_{1,m}(\tau_d)$ can be stated in the form of
\begin{align}\label{eq.I1rel}
&\mx{I}^{\ell}_{1,m}(\tau_d)=(\mx{V}_{\ell}^m(\tau_d)-\mathds{E}\mx{V}_{\ell}^m(\tau))^{\mathsf{H}}\mx{L}\widetilde{\mx{u}}=:\mx{Q}\widetilde{\mx{u}}=\nonumber\\
&\sum_{i=1}^r\sum_{j=1}^{s_i}\mx{Q}_j^i{\rm sgn}(c_j^i)\frac{\mx{f}_i}{\|\mx{f}_i\|_2}=:\sum_{i=1}^r\sum_{j=1}^{s_i}\mx{z}_j^i \in \mathbb{C}^{k_m \times 1},
\end{align}
where we defined 
\begin{align}
&(\mx{V}_{\ell}^m(\tau_d)-\mathds{E}\mx{V}_{\ell}^m(\tau_d))^{\mathsf{H}}\mx{L}=:\mx{Q}=\nonumber\\
&[\underbrace{\mx{Q}^1_{1},...,\mx{Q}_{s_1}^1}_{\mx{Q}^1},\underbrace{\mx{Q}_{1}^2,...,\mx{Q}_{s_2}^2}_{\mx{Q}^2},...,
\underbrace{\mx{Q}_{1}^r,...,\mx{Q}_{s_r}^r}_{\mx{Q}^r}]
\end{align}
and $\mx{z}_j^i\triangleq\mx{Q}_j^i{\rm sgn}(c_j^i)\frac{\mx{f}_i}{\|\mx{f}_i\|_2}$.
Before proving that $\mx{I}_{1,m}^{(\ell)}(\tau)$ is small on the set of grid points $\mathcal{T}_{\rm grid}$, we need to prove that $\|\mx{Q}\|_{2\rightarrow 2}$ is small conditioned on $\mathcal{E}_{2,\epsilon_2}$ and provided that the number of samples satisfy \eqref{eq.measure_V_concentrate}. To do this, we first define the event
\begin{align}\label{eq.E3cond}
&\scalebox{.75}{$\mathcal{E}_{3}\triangleq\Bigg\{ \sup_{\tau_d\in\mathcal{T}_{\rm grid}}\|(\mx{V}_{\ell}^m(\tau_d)-\mathds{E}\mx{V}_{\ell}^m(\tau_d))^{\mathsf{H}}\mx{L}\|_{2\rightarrow 2}\le 4\epsilon_2 , \ell=0,1,2,3\Bigg\}$}
\end{align}
which includes the events $\mathcal{E}_{1,\epsilon_1}$  and $\mathcal{E}_{2,\epsilon_2}$.
As is shown in Lemma \ref{lem.V_concentrate}, when the number of samples satisfies \eqref{eq.measure_V_concentrate}, it follows that $\mathcal{E}_{2,\epsilon_2}$ holds with high probability. Conditioned on this and $\mathcal{E}_{1,\epsilon_1}$ with $0<\epsilon_1\le\frac{1}{4}$, we have
\begin{align}
&\scalebox{.8}{$\|(\mx{V}_{\ell}^m(\tau_d)-\mathds{E}\mx{V}_{\ell}^m(\tau_d))^{\mathsf{H}}\mx{L}\|_{2\rightarrow 2}\le 
\|\mx{V}_{\ell}^m(\tau_d)-\mathds{E}\mx{V}_{\ell}^m(\tau_d)\|_{2\rightarrow 2}\|\mx{L}\|_{2\rightarrow 2}$}\nonumber\\
&\le \epsilon_2 2 \|(\mathds{E}\mx{D})^{-1}\|_{2\rightarrow 2}\le 4\epsilon_2
\end{align} 
where the last line comes from Lemmas \ref{lem.ED_rel1}, \ref{lem.EDrel2} and that $\mx{L}$ is a sub-matrix of a permuted version of $\mx{D}^{-1}$ by noting that permutation does not change the spectral norm. Using the union bound, we have
\begin{align}\label{eq.prob.E3c}
\mathds{P}\{\mathcal{E}_3^c\}\le 4 |\mathcal{T}_{\rm grid}|\delta_2+\mathds{P}(\mathcal{E}_{1,\epsilon_1}^c),
\end{align}

In what follows, we aim to show that $\mx{I}^{(\ell)}_{1,m}(\tau_d)$ for any $\tau_d\in\mathcal{T}_{\rm grid}$ is small with high probability conditioned on $\mathcal{E}_3$. Since $\mx{z}_j^i$ are zero mean and independent in \eqref{eq.I1rel}, $\mx{I}_{1,m}^{(\ell)}(\tau_d)$ is stated as a sum of independent zero mean vectors. Hence, the matrix Bernstein inequality can be then applied. First, we obtain $R$ parameter as follows:
\begin{align}\label{Rpar}
&R=\|\mx{z}_j^i\|_2\le \|\mx{Q}_j^i\|_{2\rightarrow 2}\le \|\mx{Q}\|_{2\rightarrow 2} \le {4\epsilon_2} 
\end{align}
where we used \eqref{eq.E3cond} and that $\mx{Q}_j^i$ is a sub-matrix of $\mx{Q}$. Conditioned on $\mathcal{E}_3$, we compute the variance parameter used in matrix Bernstein inequality:
\begin{align}\label{rl9}
&\sigma^2=|\sum_{i=1}^r\sum_{j=1}^{s_i}\mathds{E}{\mx{z}_j^i}^{\mathsf{H}}\mx{z}_j^i|=\nonumber\\
&|\mathds{E}\sum_{i=1}^r\sum_{j=1}^{s_i}\frac{1}{\|\mx{f}_i\|_2^2}\mx{f}_i^{\mathsf{H}}{sgn}^*(c_j^i){\mx{Q}_j^i}^{\mathsf{H}}\mx{Q}_j^i{sgn}(c_j^i)\mx{f}_i|=\nonumber\\
&\sum_{i=1}^r\sum_{j=1}^{s_i}{\rm trace}({\mx{Q}_j^i}^{\mathsf{H}}{\mx{Q}_j^i}\mathds{E}[\frac{\mx{f}_i\mx{f}_i^{\mathsf{H}}}{\|\mx{f}_i\|_2^2}])\le\sum_{i=1}^r\frac{1}{ k_i}\|\mx{Q}^i\|_F^2\le \frac{1}{k_-}\|\mx{Q}\|_F^2\nonumber\\
& \le \frac{1}{k_-}\min\{k_m, \sum_i s_i k_i\}\|\mx{Q}\|_{2\rightarrow 2}^2\le 16\epsilon_2^2 \frac{k_m}{k_-}\le  16\epsilon_2^2 
\end{align}
where we used that $\|\mathds{E}\mx{f}_i\mx{f}_i^{\mathsf{H}}\|_{2\rightarrow 2}\le \frac{1}{k_i}$.  As a result by \eqref{rl9},
\begin{align}\label{sigpar}
\sigma^2\le 16 \epsilon_2^2
\end{align}
Applying the matrix Bernstein inequality, we have
\begin{align}\label{rl4}
&\mathds{P}\{\sup_{\tau_d\in\mathcal{T}_{\rm grid}}\|\mx{I}_{1,m}^{(\ell)}(\tau_d)\|_2\ge \epsilon_4 \Big|\mathcal{E}_3\}\le\nonumber\\
&r|\mathcal{T}_{\rm grid}|
\mathds{P}\{\sum_{i=1}^r\sum_{j=1}^{s_i}\mx{z}_j^i\Big|\mathcal{E}_3\}\le |\mathcal{T}_{\rm grid}|(k_m +1)e^{-\frac{\epsilon_4^2}{\sigma^2+\frac{R \epsilon_4}{3}}}\le\nonumber\\
&\left\{\begin{array}{cc}
|\mathcal{T}_{\rm grid}|(k_m +1)e^{-\frac{3\epsilon_4^2}{8\sigma^2}}& \epsilon_4\le \frac{\sigma^2}{R}\\
|\mathcal{T}_{\rm grid}|( k_m +1)e^{-\frac{3\epsilon_4}{8R}}&\epsilon_4\ge \frac{\sigma^2}{R}.
\end{array}\right\}
\end{align}
Hence, by taking 
\begin{align}
&\epsilon_2^2= c_1 \tfrac{4^{\ell+1}}{M}\mu_+\sqrt{\sum_{i}s_ik_i}\sqrt{k_m}\log\Big(1+c_2\tfrac{4^{2\ell+2}\mu_{+}^2\sum_{i}s_ik_i}{\mu_{-}}\Big)\nonumber\\
&\log\Big(\tfrac{2\sum_is_ik_i+k_m}{\delta_2}\Big)
\end{align}
and using the relation \eqref{eq.prob.E3c} and \eqref{rl4}, 
\begin{align}
&\mathds{P}\{\sup_{\tau_d\in\mathcal{T}_{\rm grid}}\|\mx{I}_{1,m}^{(\ell)}(\tau_d)\|_2\ge \epsilon_4\}\le \nonumber\\
&\mathds{P}\{\sup_{\tau_d\in\mathcal{T}_{\rm grid}}\|\mx{I}_{1,m}^{(\ell)}(\tau_d)\|_2\ge \epsilon_4 \Big|\mathcal{E}_{1,\epsilon_1}\}+\mathds{P}\{\mathcal{E}_{1,\epsilon_1}^c\}\le\nonumber\\ &\mathds{P}\{\sup_{\tau_d\in\mathcal{T}_{\rm grid}}\|\mx{I}_{1,m}^{(\ell)}(\tau_d)\|_2\ge \epsilon_4 \Big|\mathcal{E}_3\}+\mathds{P}\{\mathcal{E}_{1,\epsilon_1}^c\}\le\nonumber\\
&\scalebox{.9}{$\left\{
\begin{array}{cc}
|\mathcal{T}_{\rm grid}|(k_m +1)e^{-\frac{3\epsilon_4^2}{8\sigma^2}}+|\mathcal{T}_{\rm grid}|4\delta_2+\mathds{P}\{\mathcal{E}_{1,\epsilon_1}^c\}& \epsilon_4\le \frac{\sigma^2}{R}\\
|\mathcal{T}_{\rm grid}|(k_m +1)e^{-\frac{3\epsilon_4}{8R}}+|\mathcal{T}_{\rm grid}|4\delta_2+\mathds{P}\{\mathcal{E}_{1,\epsilon_1}^c\}&\epsilon_4\ge \frac{\sigma^2}{R}
\end{array}
\right\}$}\label{rl6}
\end{align}
According to Lemma \ref{lem.V_concentrate}, when the number of samples satisfies
\begin{align}\label{rl5}
&M\ge c_1 \tfrac{4^{\ell+1}}{\epsilon_2^2}\mu_+\sqrt{\sum_{i}s_ik_i}\sqrt{k_m}\log\Big(1+c_2\tfrac{4^{2\ell+2}\mu_{+}^2\sum_{i}s_ik_i}{\mu_{-}}\Big)\nonumber\\
&\log\Big(\tfrac{4|\mathcal{T}_{\rm grid}|(2\sum_is_ik_i+k_m)}{\delta}\Big),
\end{align}
it holds that the second term of \eqref{rl6} is no more than $\delta$, i.e. $|\mathcal{T}_{\rm grid}|4\delta_2\le \delta$. For the first term of \eqref{rl6} to be less than $\delta$, by replacing \eqref{sigpar}, \eqref{Rpar} in \eqref{rl6}, we choose $\epsilon_2$ such that:
\begin{align}
\left\{\begin{array}{cc}
\frac{3\epsilon_4^2}{128\epsilon_2^2 \tfrac{k_m}{k_-}} =\log(\frac{|\mathcal{T}_{\rm grid}|(k_m +1)}{\delta})& \epsilon_4\le \frac{\sigma^2}{R}\\
\frac{3\epsilon_4 }{32\epsilon_2}=\log(\frac{|\mathcal{T}_{\rm grid}|(k_m +1)}{\delta})&\epsilon_4\ge \frac{\sigma^2}{R}.
\end{array}\right\}
\end{align}
Plugging the latter choice of $\epsilon_2$ into \eqref{rl5}, when $\epsilon_4< \frac{\sigma^2}{R}$, we require
\begin{align}\label{rl77}
&M\ge c_1 \tfrac{4^{\ell+1}}{\epsilon_2^2}\mu_+\sqrt{\sum_{i}s_ik_i}\sqrt{k_m}\nonumber\\
&\log\Big(1+c_2\tfrac{4^{2\ell+2}\mu_{+}^2\sum_{i}s_ik_i}{\mu_{-}}\Big)\log\Big(\tfrac{4|\mathcal{T}_{\rm grid}|(2\sum_is_ik_i+k_m)}{\delta}\Big) \nonumber\\
&\scalebox{.9}{$=\frac{c_1 4^{\ell+1} 128 \mu_{+}}{3\epsilon_4^2}\sqrt{\sum_i s_i k_i} \sqrt{k_m}\tfrac{k_m}{k_-} \log\Big(1+c_2\tfrac{4^{2\ell+2}\mu_{+}^2\sum_{i}s_ik_i}{\mu_{-}}\Big)$}
 \nonumber\\
 &\log\Big(\tfrac{4|\mathcal{T}_{\rm grid}|(2\sum_is_ik_i+k_m)}{\delta}\Big)\log(\frac{|\mathcal{T}_{\rm grid}|(k_m +1)}{\delta}).
\end{align}
 Also, when $\epsilon_4> \frac{\sigma^2}{R}$, we require
 \begin{align}\label{rl7}
 &M\ge \frac{c_1 4^{\ell+1} \mu_{+} 32^2}{9\epsilon_4^2} \sqrt{\sum_i s_ik_i} \sqrt{ k_m} \log\Big(1+c_2\tfrac{4^{2\ell+2}\mu_{+}^2\sum_{i}s_ik_i}{\mu_{-}}\Big)
 \nonumber\\
 &\log\Big(\tfrac{4|\mathcal{T}_{\rm grid}|(2\sum_is_ik_i+k_m)}{\delta}\Big)\log^2(\frac{|\mathcal{T}_{\rm grid}|(k_m +1)}{\delta}).
 \end{align}
 For the third term of \eqref{rl6} to be less that $\delta$, i.e. $\mathds{P}\{\mathcal{E}_{1,\epsilon_1}^c\}\le \delta$, according to Lemma \ref{lemma.concentrate1}, we need
 \begin{align}\label{rl8}
 M\ge \frac{c_3 \mu_{+} \sum_i s_i k_i}{\epsilon_1^2}\log\Big(1+\frac{c_4\mu_{+}^2\sum_i s_i k_i}{\mu_{-}}\Big)\log(\frac{4\sum_i s_i k_i}{\delta})
 \end{align}
 By setting an arbitrary value for $0<\epsilon_1\le\frac{1}{4}$, combining \eqref{rl8}, \eqref{rl77}, \eqref{rl7} and  absorbing all constants into certain $c_5, c_6>0$, the required number of samples reads as
 {\small{
 \begin{align}
 &M\ge \nonumber\\
 &c_5 \mu_{+} \max\Big\{{\sum_i s_i k_i}, \sqrt{\sum_i s_ik_i} \sqrt{\sum_i k_i}, \sqrt{\sum_i s_i k_i} \sqrt{ k_m}\tfrac{k_m}{k_-} \Big\}\nonumber\\
 &\log\Big(1+c_6\tfrac{\mu_{+}^2\sum_{i}s_ik_i}{\mu_{-}}\Big) \max\Big\{ \frac{1}{\epsilon_4^2} \log\Big(\tfrac{|\mathcal{T}_{\rm grid}|(\sum_is_ik_i)}{\delta}\Big)\times \nonumber\\
 &\log^2(\frac{|\mathcal{T}_{\rm grid}|(k_m)}{\delta}), \log(\frac{\sum_i s_i k_i}{\delta})\Big\}
 \end{align}}}
which could be further simplified to 
\begin{align}
&M\ge c_5 \mu_{+} (\sum_i s_i k_i )\tfrac{k_+}{k_-}\log\Big(1+c_6\tfrac{\mu_{+}^2\sum_{i}s_ik_i}{\mu_{-}}\Big) \nonumber\\&\max\Big\{ \frac{1}{\epsilon_4^2} \log\Big(\tfrac{|\mathcal{T}_{\rm grid}|(\sum_is_ik_i)}{\delta}\Big)\times 
\log^2(\frac{|\mathcal{T}_{\rm grid}|(k_+)}{\delta})\nonumber\\&, \log(\frac{\sum_i s_i k_i}{\delta})\Big\}.
\end{align}
By applying the union bound for $\ell=0,1,2,3$, we conclude the result.
\subsection{Proof of Lemma \ref{lem.I2concentrate}}\label{proof.lem.I2}
First, for any $\tau_d\in\mathcal{T}_{\rm grid}$, define
\begin{align}
&\mathds{E}{\mx{V}_{\ell}^m}^{\mathsf{H}}(\tau_d)(\mx{L}-\overline{\mx{L}})=:\widetilde{\mx{Q}}=\nonumber\\
&[\underbrace{\widetilde{\mx{Q}}^1_{1},...,\widetilde{\mx{Q}}_{s_1}^1}_{\widetilde{\mx{Q}}^1},\underbrace{\widetilde{\mx{Q}}_{1}^2,...,\widetilde{\mx{Q}}_{s_2}^2}_{\widetilde{\mx{Q}}^2},...,
\underbrace{\widetilde{\mx{Q}}_{1}^r,...,\widetilde{\mx{Q}}_{s_r}^r}_{\widetilde{\mx{Q}}^r}]
\end{align}
In the following lemma whose proof is provided in Appendix \ref{proof.lem.Qtildspec}, we find an upper-bound for the spectral norm of $\mx{Q}$, which is later required for our analysis.
\begin{lem}\label{lem.Qtildspectral}
Conditioned on the event $\mathcal{E}_{1,\epsilon}$ with $\epsilon_1\in (0,\frac{1}{4}]$, it holds that
\begin{align}\label{eq.Qtildspectralbound}
\|\mathds{E}[{\mx{V}_{\ell}^m}^{\mathsf{H}}(\tau)] (\mx{L}-\overline{\mx{L}})\|_F^2\le c ~\epsilon_1^2 k_m,
\end{align}
for some constant $c$.
\end{lem}

\begin{proof}\label{proof.lem.Qtildspec}
% \subsection{Proof of Lemma \ref{lem.Qtildspectral}}\label{proof.lem.Qtildspec}
First, according to \eqref{eq.EV}, we have
\begin{align}\label{rr1}
\|\mathds{E}\mx{V}_{\ell}^m(\tau)\|_F^2= \|\mx{v}_{\ell}^{m}(\tau)\otimes \mx{I}_{k_m}\|_F^2=k_m\|\mx{v}_{\ell}^{m}(\tau)\|_2^2
\end{align}
By exploiting the minimum separation condition between $\tau_{k}^i $s for each $i\in [r]$ and benefiting the result of \cite[Proof of Lemma IV.9]{tang2013compressed}, we may write
\begin{align}\label{eq.norm2v}
&\|\mx{v}^m_{\ell}(\tau)\|_2\le \|\mx{v}^m_{\ell}(\tau)\|_1=\nonumber\\
&\kappa^{\ell}\sum_{k=1}^{s_m}\Big(|K^{(\ell)}(\tau-\tau_k^m)|+\kappa 
|K^{(\ell+1)}(\tau-\tau_k^m)|\Big)\le \sqrt{c}
\end{align}
for some constant $c$.Then, by \eqref{rr1}, we have
\begin{align}\label{rr2}
\|\mathds{E}\mx{V}_{\ell}(\tau)\|_F^2\le c k_m
\end{align}
 When $\mathcal{E}_{1,\epsilon_1}$ holds with $0<\epsilon_1\le\frac{1}{4}$, leveraging Lemma \ref{lemma.concentrate1}, it follows that
\begin{align}
&\|\mathds{E}[{\mx{V}_{\ell}^m}^{\mathsf{H}}(\tau)] (\mx{L}-\overline{\mx{L}})\|_F^2\le \|\mathds{E}\mx{V}_{\ell}^m(\tau)\|_F^2\|\mx{L}-\overline{\mx{L}}\|_{2\rightarrow 2}^2\le\nonumber\\
&c \times k_m \|\mx{D}^{-1}-(\mathds{E}\mx{D})^{-1}\|_{2\rightarrow 2}^2\le  2c \times 1.568^2 \epsilon_1^2 k_m,
\end{align}
where in the first inequality, we used the fact that for any arbitrary matrices $\mx{A}$ and $\mx{B}$, $\|\mx{AB}\|_{F}\le \|\mx{A}\|_F\|\mx{B}\|_{2\rightarrow 2}$. The second inequality uses Lemma \ref{lemma.concentrate1} and that $\mx{L}-\overline{\mx{L}}$ is a sub-matrix of $\mx{D}^{-1}-(\mathds{E}\mx{D})^{-1}$ followed by a permutation which does not affect the spectral norm. By redefining $c$, we conclude the result.    
\end{proof}

To proceed, we write $\mx{I}_{2,m}^{\ell}(\tau_d)$ as follows:
\begin{align}\label{eq.I2rel}
&\mx{I}^{\ell}_{2,m}(\tau_d)=\mathds{E}{\mx{V}_{\ell}^m}^{\mathsf{H}}(\tau_d)(\mx{L}-\overline{\mx{L}})\widetilde{\mx{u}}=:\widetilde{\mx{Q}}\widetilde{\mx{u}}=\nonumber\\
&\sum_{i=1}^r\sum_{j=1}^{s_i}\widetilde{\mx{Q}}_j^i{\rm sgn}(c_j^i)\frac{\mx{f}_i}{\|\mx{f}_i\|_2}=:\sum_{i=1}^r\sum_{j=1}^{s_i}\widetilde{\mx{z}}_j^i \in \mathbb{C}^{k_m \times 1},
\end{align}
To apply matrix Bernstein inequality, we first obtain $R$ parameter as follows:
\begin{align}\label{eq:R_in_I2}
&R=\|\widetilde{\mx{z}}_j^i\|_2\le \|\widetilde{\mx{Q}}_j^i\|_2\le\|\widetilde{\mx{Q}}\|_{2\rightarrow 2}\le\nonumber\\
&\|\mx{L}-\overline{\mx{L}}\|_{2\rightarrow 2}\|\mathds{E}\mx{V}_{\ell}^m(\tau)\|_{2\rightarrow 2}\le  2\times 1.568^2\times \epsilon_1 \|\mx{v}^m_{\ell}(\tau)\|_2 ,
\end{align}
where in the last line, we used \eqref{eq.EV}, $\mx{L}-\overline{\mx{L}}$ is a sub-matrix of permuted  $\mx{D}^{-1}-(\mathds{E}\mx{D})^{-1}$ and Lemma \ref{lem.EDrel2}. By \eqref{eq.norm2v} and redefining constants, we find that
\begin{align}\label{Rterm1}
R\le c {\epsilon_1},
\end{align}
for some constant $c$.
For the variance term, we can write
\begin{align}\label{varterm1}
&\sigma^2=|\sum_{i=1}^r\sum_{j=1}^{s_i}
{\mathds{E}{\widetilde{\mx{z}}_j^i}}^{\mathsf{H}}
\widetilde{\mx{z}}_j^i|=\nonumber\\
&|\mathds{E}\sum_{i=1}^r\sum_{j=1}^{s_i}\frac{1}{\|\mx{f}_i\|_2^4}{\mx{f}_i}^{\mathsf{H}}{sgn}^*(c^i_j){{\widetilde{\mx{Q}}}_j^{i~\mathsf{H}}}{{\widetilde{\mx{Q}}}_j^i}{sgn}(c_j^i)\mx{f}_i|=\nonumber\\
&\sum_{i=1}^r\sum_{j=1}^{s_i}{\rm trace}({\widetilde{\mx{Q}}_j^{i~\mathsf{H}}}{\widetilde{\mx{Q}}_j^i}\mathds{E}[\frac{\mx{f}_i\mx{f}_i^{\mathsf{H}}}{\|\mx{f}_i\|_2^2}])\le \sum_{i=1}^r\frac{1}{k_i}\|\widetilde{\mx{Q}}^i\|_F^2\le \frac{1}{k_-}\|\widetilde{\mx{Q}}\|_F^2\nonumber\\
&\le \frac{k_m}{k_-}10 c_1 \epsilon_1^2
\end{align}
where the last line comes from the arguments in \eqref{eq:R_in_I2}. As a result, we have:
\begin{align}\label{varterm2}
\sigma^2\le 10~c_1 \epsilon_1^2 \frac{k_m}{k_-}  .  
\end{align}

Applying the matrix Bernstein inequality, we have
\begin{align}\label{rr3}
&\mathds{P}\{\sup_{\tau_d\in\mathcal{T}_{\rm grid}}\|\mx{I}_{2,m}^{(\ell)}(\tau_d)\|_2\ge \epsilon_5 \Big|\mathcal{E}_{1,\epsilon_1}\}\le\nonumber\\
& |\mathcal{T}_{\rm grid}|
\mathds{P}\{\sum_{i=1}^r\sum_{j=1}^{s_i}\widetilde{\mx{z}}_j^i\Big|\mathcal{E}_{1,\epsilon_1}\}\le|\mathcal{T}_{\rm grid}|( k_m +1)e^{-\frac{\epsilon_5^2}{\sigma^2+\frac{R \epsilon_5}{3}}}\le\nonumber\\
&\left\{\begin{array}{cc}
|\mathcal{T}_{\rm grid}|(k_m +1)e^{-\frac{3\epsilon_5^2}{8\sigma^2}}& \epsilon_5\le \frac{\sigma^2}{R}\\
|\mathcal{T}_{\rm grid}|(k_m +1)e^{-\frac{3\epsilon_5}{8R}}&\epsilon_5\ge \frac{\sigma^2}{R}.
\end{array}\right\}
\end{align}
Thus,
\begin{align}
&\mathds{P}\{\sup_{\tau_d\in\mathcal{T}_{\rm grid}}\|\mx{I}_{2,m}^{(\ell)}(\tau_d)\|_2\ge \epsilon_5\}\le \nonumber\\
&\mathds{P}\{\sup_{\tau_d\in\mathcal{T}_{\rm grid}}\|\mx{I}_{2,m}^{(\ell)}(\tau_d)\|_2\ge \epsilon_5 \Big|\mathcal{E}_{1,\epsilon_1}\}+\mathds{P}\{\mathcal{E}_{1,\epsilon_1}^c\}\le\nonumber\\ 
&\left\{
\begin{array}{cc}
|\mathcal{T}_{\rm grid}|( k_m +1)e^{-\frac{3\epsilon_5^2}{8\sigma^2}}+\mathds{P}\{\mathcal{E}_{1,\epsilon_1}^c\}& \epsilon_5\le \frac{\sigma^2}{R}\\
|\mathcal{T}_{\rm grid}|(k_m +1)e^{-\frac{3\epsilon_5}{8R}}+\mathds{P}\{\mathcal{E}_{1,\epsilon_1}^c\}&\epsilon_5\ge \frac{\sigma^2}{R}
\end{array}
\right\}\label{rr4}
\end{align}
For the first term of \eqref{rr4} to be less than $\delta$, by replacing \eqref{varterm2}, \eqref{Rterm1} in \eqref{rr4}, we choose $\epsilon_1$ as follows:
\begin{align}\label{epsilon_1choose}
\left\{\begin{array}{cc}
\frac{3\epsilon_5^2}{80 c_1 \epsilon_1^2\tfrac{k_m}{k_-}}=\log(\frac{|\mathcal{T}_{\rm grid}|(k_m +1)}{\delta})& \epsilon_5\le \frac{\sigma^2}{R}\\
\frac{3\epsilon_5}{c\epsilon_1}=\log(\frac{|\mathcal{T}_{\rm grid}|(k_m +1)}{\delta})&\epsilon_5\ge \frac{\sigma^2}{R}.
\end{array}\right\}.
\end{align}
To ensure that $\mathds{P}\{\mathcal{E}_{1,\epsilon_1}^c\}\le \delta$, according to Lemma \ref{lemma.concentrate1}, we require
\begin{align}
M\ge \frac{c_3 \mu_{+} \sum_i s_i k_i}{\epsilon_1^2}\log\Big(1+\frac{c_4\mu_{+}^2\sum_i s_i k_i}{\mu_{-}}\Big)\log(\frac{4\sum_i s_i k_i}{\delta})
\end{align}
which by replacing $\epsilon_1$ in \eqref{epsilon_1choose}, leads to
\begin{align}\label{m1rel}
&M\ge \frac{c_1 \mu_{+} }{\epsilon_5^2}(\sum_i s_ik_i) \tfrac{k_+}{k_-}\log\Big(1+\frac{c_2\mu_{+}^2\sum_i s_i k_i}{\mu_{-}}\Big)\nonumber\\&\log(\frac{\sum_i s_i k_i}{\delta})\log(\frac{|\mathcal{T}_{\rm grid}|(k_m +1)}{\delta})
\end{align}
for some redefined constants $c_1$ and $c_2$. 
Also, when $\epsilon_5> \frac{\sigma^2}{R}$, we require
\begin{align}\label{m2rel}
&M\ge \frac{c_3 \mu_{+} }{9\epsilon_5^2}\sum_i s_i k_i \log\Big(1+\frac{c_4\mu_{+}^2\sum_i s_i k_i}{\mu_{-}}\Big)\nonumber\\&\log(\frac{\sum_i s_i k_i}{\delta})\log^2(\frac{|\mathcal{T}_{\rm grid}|( k_m +1)}{\delta})
\end{align}
for redefined certain constants $c_3$ and $c_4$.
Combining \eqref{m1rel}, \eqref{m2rel} and using the union bound and redefining constants lead to the final result.

\subsection{Proof of Lemma \ref{lem.everywhere_off_suppoort}  }\label{proof.lem.everywhere_off_suppoort}
We first use MATLAB notation to denote the $p$-th column of $\mx{V}_{\ell}(\tau)$ by $\mx{V}_{\ell}(\tau)[:,p]$ and $p$-th entry of $\mx{q}^{\ell}(\tau)$ by $\mx{q}^{\ell}(\tau)[p]$. Provided that $\mathcal{E}_{1,\epsilon_1}$ holds with $0<\epsilon_1\le \frac{1}{4}$,
we have:
\begin{align}
\nonumber |\kappa^{\ell}\mx{q}^{\ell}(\tau)[p]|&=\big|\mx{V}_{\ell}(\tau)[:,p]^{\mathsf{H}}\mx{L}\widetilde{\mx{u}}\big|\\ 
& \le \|\mx{V}_{\ell}(\tau)[:,p]\|_2\|\mx{L}\|_{2\rightarrow 2}\|\widetilde{\mx{u}}\|_2. \label{q_p}
\end{align}
Moreover by \eqref{eq.V_l(tau)} and \eqref{V_l(tau)}, 
\begin{align}
&\|\mx{V}_{\ell}(\tau)[:,p]\|_2=\Big\|\frac{1}{M}\sum_{n=-2M}^{2M}g(n)(j2\pi n\tau)^{\ell}e^{-j2\pi n\kappa\tau}\mx{B}_n[:,p]\Big\|_2\nonumber \\ &\le
\frac{4M+1}{M} 4^{\ell} \|\mx{B}_n[:,p]\|_2\le \nonumber \\ 
& \frac{4M+1}{M} 4^{\ell}\sqrt{\sum_{i} \|\bs{\nu}_i\|_2^2\|\mx{b}_n^i\|_2^2\max_l |b^l(p)|^2}\nonumber\\
&\le
 \frac{4M+1}{M} 4^{\ell}  \sqrt{\sum_i 14 s_i \mu_{+} k_i \mu_{+}   } =c\mu_{+} \sqrt{\sum_{i}s_i k_i},
\end{align}
where in the last inequality, we used \eqref{eq.nu-i} and incoherence property \eqref{eq.incoherence}.
By plugging this into \eqref{q_p}, using Lemma \ref{lem.EDrel2} and \eqref{ui}, we have
\begin{align}
|\kappa^{\ell}\mx{q}^{\ell}(\tau)[p]|\le c \mu_{+} \sqrt{\sum_i s_i k_i} \sqrt{\sum_i s_i} 
\end{align}
By combining Bernstein's polynomial inequality \cite[Lemma IV.11]{tang2013compressed} and the latter relation, it then follows that, for any fixed $\tau_a, \tau_b \in [0,1]$
\begin{align}
&\Big|\kappa^{\ell}\mx{q}^{\ell}(\tau_a)[p]-\kappa^{\ell}\mx{q}^{\ell}(\tau_b)[p]\Big|\le |e^{j2\pi \tau_a}-e^{j2\pi \tau_b}|\nonumber\\
&\sup_{z=e^{j2\pi\tau}}\Big|\frac{\partial \kappa^{\ell}\mx{q}^{\ell}(z)[p]}{\partial z}\Big|\le 4\pi |\tau_a-\tau_b| 2M \sup_{\tau}|\kappa^{\ell} \mx{q}^{\ell}(\tau)[p]|\le\nonumber\\
& c M \mu_{+} \sqrt{\sum_{i}s_i k_i}\sqrt{\sum_i s_i}|\tau_a-\tau_b| ,
\end{align}
for a redefined constant $c>0$.
As a consequence
\begin{align}\label{rrr}
&\|\kappa^{\ell}\mx{q}_m^{\ell}(\tau_a)-\kappa^{\ell}\mx{q}_m^{\ell}(\tau_b)\|_2\le\nonumber\\
&c M \mu_{+} \sqrt{\sum_{i}s_i k_i}\sqrt{ k_m}\sqrt{\sum_i s_i}|\tau_a-\tau_b| \nonumber\\&\le c M^2 \frac{k_+}{k_-}|\tau_a-\tau_b|,
\end{align}
where the last line holds when $M\ge \sum_i s_i k_i \tfrac{k_+}{k_-}$. The reason for the latter relation is that
\begin{align}
    \frac{k_+}{k_-}\sqrt{\sum_i s_i k_i}\ge \frac{k_+}{\sqrt{k_-}}\sqrt{\sum_i s_i}\ge\frac{k_+}{\sqrt{k_+}}\sqrt{\sum_i s_i}\ge \sqrt{k_+}\sqrt{\sum_i s_i}.
\end{align}
We can select a grid size with length $|\mathcal{T}_{\rm grid}|\le \frac{3 c M^2}{\epsilon}$ such that $|\tau-\tau_d|\le \frac{\epsilon }{3 c M^2 }$ for any $\tau\in[0,1)$. With this selection, conditioned on the event $\mathcal{E}_{1,\epsilon_1}$ with $\epsilon_1\in (0,\frac{1}{4}]$ and for any $\tau\in [0,1)$,
\begin{align}
&\|\kappa^{\ell}\mx{q}_m^{\ell}(\tau)-\kappa^{\ell}\overline{\mx{q}}_m^{\ell}(\tau)\|_2\le \|\kappa^{\ell}\mx{q}_m^{\ell}(\tau)-\kappa^{\ell}\overline{\mx{q}}_m^{\ell}(\tau_d)\|_2+\nonumber\\&
\|\kappa^{\ell}\mx{q}_m^{\ell}(\tau_d)-\kappa^{\ell}\overline{\mx{q}}_m^{\ell}(\tau_d)\|_2+
\|\kappa^{\ell}\overline{\mx{q}}_m^{\ell}(\tau_d)-\kappa^{\ell}\overline{\mx{q}}_m^{\ell}(\tau)\|_2\nonumber\\
&\le c M^2 |\tau-\tau_d|+\frac{\epsilon}{3}+c M^2 |\tau-\tau_d|\le \epsilon,
\end{align}
where the last line is the result of Proposition \ref{prop.closeness on grid}, \eqref{rrr} and a redefinition of constant $c$.
With this grid size selection, a redefinition of $c_1, c_2>0$ and using Proposition \ref{prop.closeness on grid}, we get the sufficient bound
\begin{align}
&M\ge c_1 \mu_{+} (\sum_i s_i k_i )\tfrac{k_+}{k_-}\log\Big(1+c_2\tfrac{\mu_{+}^2\sum_{i}s_ik_i}{\mu_{-}}\Big) \nonumber\\&\max\Big\{ \frac{1}{\epsilon^2} \log\Big(\tfrac{M(\sum_is_ik_i)}{\epsilon\delta}\Big)\times 
\log^2(\frac{M(\sum_i k_i)}{\epsilon \delta})\nonumber\\&, \log(\frac{\sum_i s_i k_i}{\delta})\Big\}.
\end{align}
\subsection{Proof of Lemma \ref{lem.Tfar}}\label{proof.lem.Tfar}
Setting $\epsilon=10^{-5}$ in Lemma \ref{lem.everywhere_off_suppoort}, we find that
\begin{align}\label{re1}
&\max_{m}\|\mx{q}_m(\tau)\|_2\le  \|\mx{q}_m(\tau)-\overline{\mx{q}}_{m}(\tau)\|_2+\max_{m} \|\overline{\mx{q}}_m(\tau)\|_2\le\nonumber\\
& \|\mx{q}_m(\tau)-\overline{\mx{q}}_m(\tau)\|_2+\max_{m} \|\overline{\mx{q}}_m(\tau)\|_2\le\nonumber\\
&10^{-5}+\max_{m} \|\overline{\mx{q}}_m(\tau)\|_2,
\end{align}
where we used Lemma \ref{lem.everywhere_off_suppoort} in the last line and that $\mx{q}_m(\tau)-\overline{\mx{q}}_{m}(\tau)$ is a sub-matrix of $\mx{q}(\tau)-\overline{\mx{q}}(\tau)$. It remains then to estimate $\|\overline{\mx{q}}_m(\tau)\|_2$ which by \eqref{re} reads as
\begin{align}
&\|\overline{\mx{q}}_m(\tau)\|_2=\sup_{\mx{z}:\|\mx{z}\|_2=1}\langle \mx{z}, ({\mx{v}_{0}^m(\tau)}^{\mathsf{H}}\otimes \mx{I}_{k_m}) (\overline{\mx{L}}^{m,m}\otimes \mx{I}_{k_m})\mx{u}_m\rangle\nonumber\\&=\sup_{\mx{z}:\|\mx{z}\|_2=1} \sum_{j=1}^{s_m}({\mx{v}_{0}^m(\tau)}^{\mathsf{H}}\overline{\mx{L}}^{m,m})[j](\mx{z}^{\mathsf{H}}{\rm sgn}(c_j^m)\mx{f}_m)\le 0.99992\nonumber\\
& <1, \forall m=1,...,r
\end{align}  
where in the last line, we used $|\mx{z}^{\mathsf{H}}{\rm sgn}(c_j^m)\mx{f}_m|\le 1$ and \cite[Proofs of Lemmas 2.3 \& 2.4]{candes2014towards}. 

 \section{Proof of Lemma \ref{lem.Tnear}}\label{proof.lem.T_near}

First, due to the relation \eqref{eq.supp_cond_alt}, we have that
\begin{align}\label{eq:rel1}
    \frac{\partial \|\mx{q}_m(\tau)\|_2^2}{\partial \tau}\Big|_{\tau=\tau_j^m}=2{\rm Re}\langle \mx{q}_m^\prime(\tau_j^m),\mx{q}(\tau_j^m) \rangle=0 .
\end{align}
By writing the Taylor's expansion of $\|\mx{q}_m(\tau)\|_2^2$ around $\tau_j^m$, we may write:
\begin{align}
    &\|\mx{q}_m(\tau)\|_2^2=\|\mx{q}_m(\tau_j^m)\|_2^2+  \frac{\partial \|\mx{q}_m(\tau)\|_2^2}{\partial \tau}\Big|_{\tau=\tau_j^m}
    (\tau-\tau_j^m)+\nonumber\\
    &\frac{1}{2}  \frac{\partial \|\mx{q}_m(\tau)\|_2^2}{\partial \tau}\Big|_{\tau=\widetilde{\tau}}(\tau-\tau_j^m)^2
\end{align}
where $\widetilde{\tau}\in [\tau_j^m-\tau_{b,1}, \tau_j^m+\tau_{b,1}]$. Due to \eqref{eq:rel1}, the second term above is zero. A sufficient condition for $\|\mx{q}_{m}(\tau)\|_2<1, \tau\in\mathcal{T}^m_{\rm near}$ is to show that
\begin{align}\label{eq:rel4}
    \frac{1}{2}  \frac{\partial \|\mx{q}_m(\tau)\|_2^2}{\partial \tau}=\|\mx{q}_m^\prime(\tau)\|_2^2+{\rm Re}\langle\mx{q}{''}(\tau), \mx{q}_m(\tau) \rangle<0
\end{align}
for $\tau\in \mathcal{T}_{\rm near}^m$.
For the first term, it follows that
\begin{align}\label{eq:rel2}
 &\|\mx{q}_m{'}(\tau)\|_2^2=\|\mx{q}_m{'}(\tau)-\overline{\mx{q}}_m{'}(\tau)+\overline{\mx{q}}_m{'}(\tau)\|_2^2\le  \nonumber\\
 &
 \|\mx{q}{'}(\tau)-\overline{\mx{q}}{'}(\tau)\|_2^2+2\| \mx{q}{'}(\tau)-\overline{\mx{q}}{'}(\tau)\|_2 \|\overline{\mx{q}}_m{'}(\tau) \|_2 
 +\nonumber\\
 &\|\overline{\mx{q}}_m{'}(\tau)\|_2^2\le \frac{\epsilon^2}{\kappa^2}+ \frac{2\epsilon}{\kappa}1.6 M+2.56 M^2
\end{align}
where we used $\|\overline{\mx{q}}^{'}(\tau)\|_2\le 1.6 M$ \cite[Appendix I]{wakinsuperres}.
For the second term in \eqref{eq:rel2}, we have
\begin{align}\label{eq:rel3}
&\scalebox{.7}{$
    \langle\mx{q}_m{''}(\tau), \mx{q}_m(\tau) \rangle_{\mathbb{R}}=\langle\mx{q}_m{''}(\tau)-\overline{\mx{q}}_m{''}(\tau)+\overline{\mx{q}}_m{''}(\tau), \mx{q}_m(\tau)-\overline{\mx{q}}_m(\tau)+\overline{\mx{q}}_m(\tau) \rangle_{\mathbb{R}}$}\nonumber\\
    &\scalebox{.8}{$=\langle\mx{q}_m{''}(\tau)-\overline{\mx{q}}_m{''}(\tau), \mx{q}_m(\tau)-\overline{\mx{q}}_m(\tau) \rangle_{\mathbb{R}}
    +\langle \overline{\mx{q}}_m{''}(\tau), \overline{\mx{q}}_m(\tau)\rangle_{\mathbb{R}}+$}\nonumber\\
    &\langle\mx{q}_m{''}(\tau)-\overline{\mx{q}}_m{''}(\tau), \overline{\mx{q}}_m(\tau) \rangle_{\mathbb{R}}+\langle\overline{\mx{q}}_m{''}(\tau), \mx{q}_m(\tau)-\overline{\mx{q}}_m(\tau) \rangle_{\mathbb{R}}\nonumber\\
    &\le\frac{\epsilon^2}{\kappa^2}-2.9 M^2+\frac{1.04\epsilon}{\kappa^2}+21.15M^2{\epsilon}.
\end{align}
According to \eqref{eq:rel2} and \eqref{eq:rel3}, we may write \eqref{eq:rel4} as
\begin{align}
     &\frac{1}{2}  \frac{\partial \|\mx{q}_m(\tau)\|_2^2}{\partial \tau}\le \tfrac{2\epsilon^2}{\kappa^2}+\tfrac{1.04\epsilon }{\kappa^2}
     +\tfrac{3.2\epsilon M}{\kappa}-0.34M^2=\nonumber\\
     &\tfrac{2\epsilon^2+1.04\epsilon+0.087M^2\epsilon-0.34M^2}{\kappa^2}<0
\end{align}
 where in the last line we used the fact that $\epsilon<10^{-5}$ and $\kappa> \tfrac{0.27}{M}$ \cite[Appendix I]{wakinsuperres}.

% ===================
\section{Proof of Lemma \ref{lem.phi_bound}}\label{proof.lem.phi_bound}
% ===================
To find bounds for the norms of the matrices $\mathbf{\Phi}_i$, we first need to obtain an upper-bound on the convolution of two squared {\color{\change}Fej\'{e}r} kernels $K(t)$ in \eqref{eq.fejer_kernel}, $G(t):= K(t) * K(t)$ which is given by the following lemma:
  % ---------------
  \begin{lem}
 		\label{lem:UpKernel}
           Let $G^{(\ell)}(t)$ be the $\ell$-th derivative of the kernel $G(t)$ defined as
           \begin{align*}
               G(t) = \dfrac{1}{M}\sum_{n=-2M}^{2M} g^2(n)(1+|2\pi n\kappa |^2 ){\rm e}^{2{\rm j}\pi n t },
           \end{align*}
 		for $\ell \in \{0,1,2\}$. Then, $|G^{(\ell)}(t)|$  is bounded by $B_{\ell}(t)$, i.e., $|G^{(\ell)}(t)|\leq B_{\ell}(t)$ for $\frac{1}{4M}\leq t 	\leq{1}/{2}$, where 
 		\begin{align}
		\label{eq:UpKernel}
		\hspace{-5pt}B_{\ell}(t)= \begin{cases}B^{L}_{\ell}(t)=\frac{\pi^{\ell-4}H_{\ell}(t)}{6(M^2-1)M^2t^4},  &  t \in \Delta_{\rm near}  \\
		B^{R}_{\ell}(t) = \frac{\pi^{\ell}H^{\infty}_{\ell}}{96(M^2-1)M^{2-\ell}t^{4}},  & t \in \Delta_{\rm far} 
		\end{cases} 
		\end{align}
 		where $\Delta_{\rm near} := [\frac{1}{4M}, \frac{\sqrt{2}}{\pi}]$, $\Delta_{\rm far} := [\frac{\sqrt{2}}{\pi}, 0.5]$ and $H^{\infty}_{0} = 7$, $H^{\infty}_{1} = 9$, $H^{\infty}_{2} = 17$, and
           %------------------
           \begin{subequations}
 		\begin{align}
		H_{0}(t) & =  7a(t)^4, \\
		H_{1}(t) & =  24a(t)^4(b(t)+2M), \\
		H_{2}(t) & =  231a(t)^4(2M^2+Mb(t)+35b(t)^2).
		\end{align}               
           \end{subequations}
            %------------------
 		Here, $a(t) =  \frac{1}{(1-\pi^2t^2/6)}$ and $b(t) = \frac{a(t)}{t}$. Moreover, for all $\ell\in\{0,1,2\}$, $B^{L}_{\ell}(\Delta - t)+ B^{L}_{\ell}(\Delta + t)$ is increasing in terms of $t$ for $0 \leq  t\leq \Delta/2$, if $0 < t+\Delta \leq 0.5$.
 	\end{lem}
  \begin{proof}
      See Appendix~\ref{sec:proof_UpKernel}. 
  \end{proof}
  % ------------------
 Then, by using Lemma \ref{lem:UpKernel}, we can find an upper-bound on expressions of the form $\sum_{t_i \in \mathcal{T}^{m} \setminus \{\tau\}} |G^{(\ell)}(t-t_i)|$ for $\tau \in \mathcal{T}^{m}$.
 	\begin{lem}
 		\label{lem:UpDmatrix}
 		Suppose that $0 \in \mathcal{T}^{m}$. Then for all $t \in [0,\Delta/2]$, we have that
 		\begin{align}
 \nonumber		\sum_{t_i \in \mathcal{T}^{m} \setminus \{\tau\}} |G^{(\ell)}(t-t_i)| &\leq \mathfrak{G}_{\ell}(\Delta, t) \\
   &= \mathfrak{G}^{+}_{\ell}(\Delta, t) + \mathfrak{G}^{-}_{\ell}(\Delta, t), 
 		\end{align}
 		where
           \begin{subequations}
            \begin{align}
 			\mathfrak{G}^{+}_{\ell}(\Delta, t)& = \mathfrak{B}^{+}(\Delta, t)  + \sum_{j=2}^{32}B^{L}_{\ell}(j\Delta_{\min}-t) + C_{\ell},\\
 			\mathfrak{G}^{-}_{\ell}(\Delta, t)&= \mathfrak{B}^{-}(\Delta, t)+  \sum_{j=2}^{32}B^{L}_{\ell}(j\Delta_{\min}+t) + C_{\ell},
 	 	\end{align} 
           \end{subequations}
             where $ C_{\ell} =  \frac{\pi^{\ell}H^{\infty}_{\ell}}{96(M^2-1)M^{2-\ell}}\gamma$ with $\gamma =\sum_{j=33}^{\infty}\frac{1}{j^4}$, and
                \begin{align*}
                \mathfrak{B}^{+}(\Delta, t) & =  \max \big\{\max_{\Delta\leq t_{+}\leq 3\Delta_{\min}} \hspace{-10pt}|G^{(\ell)}(t-t_{+})|,B_{\ell}(3\Delta_{\min}-t)\big\}, \\
                \mathfrak{B}^{-}(\Delta, t) & = \max \big\{ \max_{\Delta\leq t_{-}\leq 3\Delta_{\min}}\hspace{-10pt} |G^{(\ell)}(t_{-})|,B_{\ell}(3\Delta_{\min})\big\},
             \end{align*}    
	  for $\ell =\{0,1,2,3\}$.	Moreover, $\mathfrak{G}_{\ell}(\Delta, t)$ is non-increasing with respect to $\Delta$ for all $t$, and $\mathfrak{G}_{\ell}(\Delta_{\min}, t)$ is non-decreasing with respect to $t$.
 	\end{lem}
         \begin{proof}
             See Appendix~\ref{sec:proof_UpDmatrix}. 
         \end{proof}
         Note that the non-increasing property of $\mathfrak{G}_{\ell}(\Delta, t)$ with respect to both $t$ and $\Delta$ allows us to set $\Delta = \Delta_{\rm min}$ when evaluating $\mathfrak{G}_{\ell}(\Delta_{\rm min}, 0)$ to obtain an upper-bound on the sum $\mathfrak{G}^{+}_{\ell}(\Delta, t)$. 
         
	To use Lemma \ref{lem:UpDmatrix} for obtaining the upper-bound on the spectral norm of matrix $\mathbf{\Phi}_i$, let us partition each matrix $\boldsymbol{\Phi}_{i}$ as follows. 

 $$\boldsymbol{\Phi}_{i} = \begin{bmatrix} \boldsymbol{\Phi}^{0}_{i} & \boldsymbol{\Phi}^{1}_{i}\\
	\boldsymbol{\Phi}^{1}_{i} & \boldsymbol{\Phi}^{2}_{i} \end{bmatrix},$$ 
 which leads to the following expressions for all $t_k,t_\ell\in \mathcal{T}^{m}$:
        \begin{subequations}
          \begin{align}
		[\boldsymbol{\Phi}^{0}_{i}]_{k,\ell} & = G(t_{k}-t_{\ell}),\\ 	[\boldsymbol{\Phi}^{1}_{i}]_{k,\ell} & = G^{(1)}(t_{k}-t_{\ell}), \\ 
   [\boldsymbol{\Phi}^{2}_{i}]_{k,\ell} &= G^{(2)}(t_{k}-t_{\ell}),
	\end{align}  
        \end{subequations}
	where $G(t): = \dfrac{1}{M}\sum_{n=-2M}^{2M} g^2(n)(1+|2\pi n\kappa |^2 ){\rm e}^{2{\rm j}\pi n t }$ is a deterministic kernel. 
    % ---------------------
    % ---------------------
     \begin{figure}[!t]
\centering
 \begin{tikzpicture} 
    \begin{axis}[
        width=0.5\textwidth,
        height=6cm,
        xmin=1.971, xmax=5.92,
         ymin=1e-5,% ymax=1,
        legend style={nodes={scale=0.75, transform shape}, at={(0.7,0.95)}}, 
        ticklabel style = {font=\footnotesize},
        ymajorgrids=true,
        xmajorgrids=true,
        grid style=dashed,
        grid=both,
        grid style={line width=.1pt, draw=gray!10},
        major grid style={line width=.2pt,draw=gray!30},
    ]
    \addplot[ smooth,
             thin,
        color=chestnut,
        line width=0.9pt,
        ]
    table[x=x,y=y1]
    {Kernel_upper.dat};
\addplot[ smooth,
             thin,
        color=airforceblue,
        line width=0.9pt,
        ]
    table[x=x,y=y2]
    {Kernel_upper.dat};
    \addplot[ 
        color=cadmiumorange,
        mark=star,
        dashed,
        line width=1pt,
        mark size=1pt,
        ]
    table[x=x,y=y3]
    {Kernel_upper.dat};
    \legend{$|G(t)|$, $|G^{(1)}(t)|/M$,$|G^{(2)}(t)|/M^2$};
    \end{axis}
\end{tikzpicture}
		\caption{Absolute value of the $G^{\ell}(t)$ in $t\in [\Delta_{\min},3\Delta_{\min}]$ for $\ell =\{0,1,2\}$.}
   \label{fig:Kernel}
\end{figure}
     % ---------------------
    % ----------------------
	Under the assumption that $\Delta > \frac{1}{M}$, from Lemma \ref{lem:UpDmatrix}, we can bound the sum of rows of each partition $\boldsymbol{\Phi}^{\ell}_{i}$ by the following inequalities. 
   %----------------
       \begin{subequations}
           \begin{align}
		\| \mathbf{I} - \boldsymbol{\Phi}_{i}^{0}\|_{\infty \rightarrow \infty} \leq \mathfrak{G}_{0}(\Delta_{\min},0)&\leq 7.77\times 10^{-2},\\
		\|\boldsymbol{\Phi}_{i}^{1}\|_{\infty \rightarrow \infty} \leq \mathfrak{G}_{1}(\Delta_{\min},0)&\leq \frac{1.95}{2}\times 10^{-1}M,\\
		\|\mathbf{I}-\boldsymbol{\Phi}^{2}_{i}\|_{\infty \rightarrow \infty} \leq \mathfrak{G}_{2}(\Delta_{\min},0)&\leq \frac{2.883}{4}\times 10^{-1}{M}^2,
	\end{align}
       \end{subequations}
       %----------------
	where  $\mathbf{I} - \boldsymbol{\Phi}_{i}$ is a symmetric matrix with zero diagonals. Consequently, invoking the Gershgorin circle theorem~\cite{salas1999gershgorin}, we can obtain an upper-bound on the spectral norm of $\mathbf{I} - \boldsymbol{\Phi}_{i}$  as 
	\begin{align}
           \nonumber
		& \|\mathbf{I} - \boldsymbol{\Phi}_{i}\|_{2 \rightarrow 2} \leq \|\mathbf{I} - \boldsymbol{\Phi}_{i}\|_{\infty \rightarrow \infty}\leq \max \Big\{	\| \mathbf{I} - \boldsymbol{\Phi}_{i}^{0}\|_{\infty \rightarrow \infty}\\ &  +	\kappa\|\boldsymbol{\Phi}_{i}^{1}\|_{\infty \rightarrow \infty} , \kappa\|\boldsymbol{\Phi}_{i}^{1}\|_{\infty \rightarrow \infty} + \|\kappa^2\mathbf{I}-\boldsymbol{\Phi}^{2}_{i}\|_{\infty \rightarrow \infty} \Big\},\nonumber \\
  &\quad \quad =  1.39\times 10^{-1}.
	\end{align}
	As a result, $\boldsymbol{\Phi}_{i}$ is invertible and 
	\begin{align}
		\| \boldsymbol{\Phi}_{i}	\|_{2 \rightarrow 2} &\leq 1 + \|\mathbf{I} - \boldsymbol{\Phi}_{i}\|_{2 \rightarrow 2}\leq  1.139,  \\
		\|\boldsymbol{\Phi}_{i}^{-1}	\|_{2 \rightarrow 2} &\leq \frac{1}{1 - \|\mathbf{I} - \boldsymbol{\Phi}_{i}\|_{2 \rightarrow 2}}\leq 1.161. 
		\end{align}
         This concludes the proof. 

\subsection{Proof of Lemma~\ref{lem:UpKernel}}\label{sec:proof_UpKernel}
       To obtain the upper-bound on the kernel $G(t)$, we expand $G(t)$ using trigonometric functions and simplify the terms in the summation as follows. 
       \begin{align}
        \nonumber
       G(t) &= \dfrac{1}{M}\sum_{n=-2M}^{2M} g^2(n)(1+|2\pi n\kappa |^2 ){\rm e}^{2{\rm j}\pi n t }, \\ 
         & =    \frac{R_1(t) + R_2(t) + R_3(t) + R_4(t)}{96(M^2-1)M^7\sin^9(\pi t)},
       \end{align}
       where 
       \begin{subequations}
          \label{eq:R1R4formula}
                \begin{align}
              R_1(t) &= b_1\sin(\pi t) + b_2\sin(3\pi t)+ b_3\sin(5\pi t), \\
              R_2(t) &= a_1\sin(\pi t(2M+1)) + a_2\sin(\pi t(2M-1)) + \nonumber  \\ & a_3\sin(\pi t(4M+1)) + a_4\sin(\pi t(4M-1)), \\
              R_3(t) & =a_5\sin(\pi t(2M+3)) + a_6\sin(\pi t(2M-3)) +  \nonumber \\ 
              &  a_7\sin(\pi t(4M+3)) + a_8\sin(\pi t(4M-3)),\\
              R_4(t) & = a_9\sin(\pi t(2M+5)) + a_{10}\sin(\pi t(2M-5)) + \nonumber  \\ & a_{11}\sin(\pi t(4M+5)) + a_{12}\sin(\pi t(4M-5)),
            \end{align}      
       \end{subequations}
    and the coefficients $b_1,b_2,b_3$  and $a_1, a_2 ,\ldots a_{12}$ are given in Table \ref{tab:coefficients}. 
%-----------------       
\begin{table}
    \centering
    \caption{Square Fejer kernel $G(t)$}
    \begin{tabular}{|c|c|}
   \toprule
Param. & value  \\
\midrule
$b_1$  & {\footnotesize $30M^5-105M^3-3165M $}  \\
\hline
$b_2$  & {\footnotesize $-15M^5+7.5M^3-2422.5M $}  \\
\hline
$b_3$  &  {\footnotesize $3M^5+16.5M^3-181.5M $  }\\
\hline
       $a_1$  & {\footnotesize $-20M^5-12M^4-5M^3-300M^2-1775M +4056 $  }\\
       \hline
       $a_2$ &  {\footnotesize$20M^5-12M^4+5M^3-300M^2+1775M +4056$ }\\
       \hline
       $a_3$ &  {\footnotesize $-39M^2-180M + 507$ } \\
       \hline
       $a_4$& {\footnotesize  $-39M^2+180M + 507$} \\ 
       \hline
       $a_5$& {\footnotesize$+10M^5 + 18M^4 -12.5M^3+255M^2-1347.5M +951$}\\
       \hline
       $a_6$ & {\footnotesize $-10M^5 + 18M^4 +12.5M^3+255M^2+1347.5M +951$}\\
       \hline
       $a_7$& {\footnotesize $ 34.125M^2 -135M + 118.875$ }\\
       \hline
       $a_8$  & {\footnotesize$34.125M^2 +135M + 118.875$}\\
       \hline
       $a_9$  & {\footnotesize$-2M^5-6M^4+8.5M^3+45M^2-96.5M+33$ }\\
       \hline
        $a_{10}$ & {\footnotesize$+2M^5-6M^4-8.5M^3+45M^2+96.5M+33$}\\
        \hline
        $a_{11}$ & {\footnotesize$4.875M^2-9M+4.125$}\\
        \hline
        $a_{12}$ & {\footnotesize$4.875M^2+9M+4.125$} \\
      \bottomrule
    \end{tabular}
    \label{tab:coefficients}
\end{table}
%-----------------
{\color{\change} By using the trigonometric identities for addition and subtraction formulas for terms involving parameter $M$ in their arguments in \eqref{eq:R1R4formula}, we obtain the following 
%-----------------
\begin{subequations}
\label{eq:Rtilde_equation}
 \begin{align}
    \Tilde{R}_{1}(t) & \triangleq{}  b_1\sin(\pi t) + b_2\sin(3\pi t) + b_3 \sin(5\pi t), \\
    %---------------
    \Tilde{R}_{2}(t) & \triangleq{} ((a_1-a_2)\sin(\pi t) + (a_5-a_6)\sin(3\pi t) + \nonumber\\
    &(a_9-a_{10}) \sin(5\pi t)) \cos{(2 M\pi t)}, \\ 
    %------------
    \Tilde{R}_{3}(t) & \triangleq{}  ((a_3-a_4)\sin(\pi t) + (a_7-a_8)\sin(3\pi t) + \nonumber\\
    &(a_{11}-a_{12}) \sin(5\pi t)) \cos{(4 M\pi t)} \\
    %------------
    \Tilde{R}_{4}(t) & \triangleq{} ((a_1+a_2)\cos(\pi t) + (a_5+a_6)\cos(3\pi t) + \nonumber\\
    &(a_9+a_{10}) \cos(5\pi t)) \sin{(2 M\pi t)}, \\ 
    %------------
    \Tilde{R}_{5}(t) & \triangleq{}  ((a_3+a_4)\cos(\pi t) + (a_7+a_8)\cos(3\pi t) +\nonumber\\
    &(a_{11}+a_{12}) \cos(5\pi t)) \sin{(4 M\pi t)}
 \end{align}
 \end{subequations}
%----------------- 
where we have  $\sum_{i=1}^4{R}_{i}(t) = \sum_{i=1}^5\tilde{R}_{i}(t).$ 
To obtain an upper-bound for the given terms in \eqref{eq:Rtilde_equation}, we take the absolute value of the expressions in \eqref{eq:Rtilde_equation}, i.e., 
%-----------------
\begin{subequations}
\label{eq:Rtilde_equation222}
 \begin{align}
    |\Tilde{R}_{1}(t)| & \leq | b_1\sin(\pi t) + b_2\sin(3\pi t) + b_3 \sin(5\pi t)|, \\
    %---------------
   | \Tilde{R}_{2}(t)| & \leq |((a_1-a_2)\sin(\pi t) + (a_5-a_6)\sin(3\pi t) + \nonumber\\
   &(a_9-a_{10}) \sin(5\pi t))| |\cos{(2 M\pi t)}|, \\ 
    %------------
    |\Tilde{R}_{3}(t)| & \leq |((a_3-a_4)\sin(\pi t) + (a_7-a_8)\sin(3\pi t) + \nonumber\\
    &(a_{11}-a_{12}) \sin(5\pi t)) ||\cos{(4 M\pi t)}| \\
    %------------
    |\Tilde{R}_{4}(t)| & \leq |((a_1+a_2)\cos(\pi t) + (a_5+a_6)\cos(3\pi t) + \nonumber\\
    &(a_9+a_{10}) \cos(5\pi t)) |\sin{(2 M\pi t)}|, \\ 
    %------------
    |\Tilde{R}_{5}(t) |& \leq  |((a_3+a_4)\cos(\pi t) + (a_7+a_8)\cos(3\pi t) + \nonumber\\
    &(a_{11}+a_{12}) \cos(5\pi t))|| \sin{(4 M\pi t)}|.
 \end{align}
 \end{subequations}
%----------------- 
Then, we take the envelope of each term by setting  $\sin{(2 M\pi t)}$, $\cos{(2 M\pi t)}$, $\sin{(4M\pi t)},$ and $\cos{(4 M\pi t)}$ to $1$ or $-1$ depending on the sign of their amplitude. In other words, we 
%-----------------
\begin{subequations}
\label{eq:Rtilde_equation_absolute}
 \begin{align}
 |\Tilde{R}_{1}(t)| & \leq \Tilde{R}_{1}^{e}(t) \triangleq{} b_1\sin(\pi t) + b_2\sin(3\pi t) + b_3 \sin(5\pi t), \\
    %---------------
    |\Tilde{R}_{2}(t)| & \leq \Tilde{R}_{2}^{e}(t) \triangleq{} (a_2-a_1)\sin(\pi t) + (a_6-a_5)\sin(3\pi t)\nonumber\\
    &+ (a_{10}-a_9) \sin(5\pi t), \\ 
    %------------
    |\Tilde{R}_{3}(t) |& \leq  \Tilde{R}_{3}^{e}(t) \triangleq{}(a_4-a_3)\sin(\pi t) + (a_8-a_7)\sin(3\pi t)\nonumber\\
    &+ (a_{12}-a_{11}) \sin(5\pi t), \\
    %------------
    |\Tilde{R}_{4}(t) |& \leq \Tilde{R}_{4}^{e}(t) \triangleq{} -(a_1+a_2)\cos(\pi t)  -(a_5+a_6)\cos(3\pi t)  \nonumber\\
    &-(a_9+a_{10}) \cos(5\pi t), \\ 
    %------------
    |\Tilde{R}_{5}(t)| & \leq \Tilde{R}_{5}^{e}(t) \triangleq{}  -(a_3+a_4)\cos(\pi t) - (a_7+a_8)\cos(3\pi t)\nonumber\\
    &- (a_{11}+a_{12}) \cos(5\pi t),
 \end{align}
 \end{subequations}
%---------------
where we have $\Tilde{R}_{i}(t)=|\Tilde{R}_{i}(t)|\geq 0$ for $t\in [0,0.5]$ and $i\in [5]$.  It is worth noting that for $M\geq 7$, the sign of all coefficients $a_i$ and $b_i$  in Table~\ref{tab:coefficients} remains constant, matching the sign of the high degree term in $M$ for each expression.   
Figure~\ref{fig:envelop} illustrates $\tilde{R}_{i}(t)$ along with the envelope   $\tilde{R}^{e}_{i}(t)$  representing the cumulative sum of all $\tilde{R}_{i}(t)$ terms. Therefore, by applying the triangle inequality to $\sum_{i=1}^4{R}_{i}(t)$, we obtain the following inequality. 
%---------------
\begin{align}
    |\sum_{i=1}^{5}\Tilde{R}_{i}(t)|\leq \sum_{i=1}^{5}|\Tilde{R}_{i}(t)| \leq \sum_{i=1}^{5}\Tilde{R}^{e}_{i}(t).
\end{align}
%---------------
Next, we rearrange the terms of $\sum_{i=1}^5\tilde{R}^{e}_{i}(t)$ to define new terms $\overline{R}_{i}$ such that we have $\sum_{i=1}^{5}\Tilde{R}^{e}_{i}(t) = \sum_{i=1}^{6}\overline{R}_{i}(t)$ in which  $\overline{R}_{i}$s are given by
%-----------------
\begin{subequations}
\label{eq:inqeqialityRtild}
\begin{align}
      \overline{R}_{1}(t) & \triangleq{}   (b_1 - a_1 + a_2 + a_3-a_4 )\sin(\pi t) = \nonumber\\
      &5M(5-19M^2+14M^4) \sin(\pi t), \\
    \overline{R}_{2}(t) & \triangleq{}  -(a_1+a_2 + a_3+a_4)\cos(\pi t) = \nonumber\\
    &(24M^4+678M^2-9126)\cos(\pi t), \\ 
    %------------
    \overline{R}_{3}(t) & \triangleq{} (b_2 - a_5+a_6 -a_7 +a_8 )\sin(3\pi t) =  \nonumber\\
    &(-35M^5+32.5M^3 +542.5M)\sin(3\pi t),  \\
    \overline{R}_{4}(t) & \triangleq{}  -(a_5+a_6 + a_7+a_8)\cos(3\pi t)= \nonumber\\
    &-(36 M^4  + 578.25 M^2+2139.75 )\cos(3\pi t),\\ 
    %------------
    \overline{R}_{5}(t) & \triangleq{} (b_3 - a_9+a_{10} - a_{11}+a_{12} )\sin(5\pi t) =\nonumber\\
    &(7 M^5 - 0.5 M^3 + 29.5 M)\sin(5\pi t), \\
    \overline{R}_{6}(t) & \triangleq{}  -(a_9+a_{10} + a_{11}+a_{12})\cos(5\pi t) = \nonumber\\
    &( 12 M^4 - 99.75 M^2-74.25)\cos(5\pi t). 
\end{align}
 \end{subequations}
%-----------------
We note that the individual terms in \eqref{eq:inqeqialityRtild} may be either positive or negative , but the total summation remains positive. Then, the right-hand side of the inequality in \eqref{eq:inqeqialityRtild} can be written as
%-----------------
\begin{subequations}
\label{eq:inqeqialityRtild2}
\begin{align}
      \overline{R}_{1}(t) & = 70M^5\Big(\frac{5-19M^2}{14M^4}+1\Big) \sin(\pi t) = \nonumber\\ &70M^5\sin(\pi t) 
      + \mathcal{O}(M^{-2}\sin(\pi t)) , \\
    \overline{R}_{2}(t) &  =24M^4\Big(\frac{678M^2-9126}{24M^4}+1\Big)\cos(\pi t) =\nonumber\\
    &24M^4\cos(\pi t) + \mathcal{O}(M^{-2}\cos(\pi t)), \\ 
    %------------
    \overline{R}_{3}(t) &  = -35M^5 \Big(\frac{32.5M^2 +542.5}{35M^4}+1\Big)\sin(3\pi t) =\nonumber\\
    &-35M^5\sin(3\pi t) + \mathcal{O}(M^{-2}\sin(3\pi t)),  \\
    \overline{R}_{4}(t) & = -36 M^4\Big(\frac{2139.75 + 578.25 M^2}{36M^4} + 1\Big)\cos(3\pi t) =\nonumber\\
    &-36 M^4\cos(3\pi t) +\mathcal{O}(M^{-2}\cos(3\pi t)) ,\\ 
    %------------
    \overline{R}_{5}(t) &  =7 M^5 \Big(\frac{- 0.5 M^2 + 29.5 }{7 M^4}+1 \Big)\sin(5\pi t) =\nonumber\\
    &7 M^5\sin(5\pi t)+\mathcal{O}(M^{-2}\sin(5\pi t)), \\
    \overline{R}_{6}(t) & = 12 M^4\Big( \frac{- 99.75 M^2-74.25}{12 M^4}+1\Big)\cos(5\pi t) =\nonumber\\
    &12 M^4\cos(5\pi t) + \mathcal{O}(M^{-2}\cos(5\pi t)). 
\end{align}
 \end{subequations}
%-----------------

%-----------------
\begin{figure*}[!t]
\centering
\subfigure[$M=20$]{
\begin{tikzpicture}
  \begin{scope}[spy using outlines={rectangle, magnification=10, 
        width=1.3cm, height=1.3cm, connect spies}]
        \begin{axis}[
            xlabel={$t$},
            ylabel ={Amplitude},
            width=0.45\textwidth,
            height=5cm,
            xmin=0.019, xmax=0.5,
            % ymin=1e-7, ymax=0.5e-3,
            legend style={nodes={scale=0.75, transform shape}, at={(0.3,0.95)}},
            ticklabel style={font=\footnotesize},
            ymajorgrids=true,
            xmajorgrids=true,
            grid style=dashed,
            grid=both,
            grid style={line width=.1pt, draw=gray!10},
            major grid style={line width=.2pt, draw=gray!30},
        ]
            \addplot[smooth, thin, color=chestnut, line width=0.5pt]
                table[x=t1, y=R1tild] {Envelop_approx1.dat};
            \addplot[smooth, thin, color=airforceblue, line width=0.1pt]
                table[x=t1, y=R1bar] {Envelop_approx1.dat};
            \legend{$\sum_{i=1}^5\overline{R}_{i}$, $\sum_{i=1}^5\hat{\overline{R}}_{i}$}
            \path (axis cs:0.15, 1e+4) coordinate (X); % Adjust as needed
        \end{axis}
              % Use the spy command to zoom in on the specified coordinate
        \spy [black] on (X) in node (zoom) [left] at ([xshift=1cm, yshift=1.1cm]X);

    \end{scope}
\end{tikzpicture}
}\subfigure[$M=72$]{
\begin{tikzpicture}
 \begin{scope}[spy using outlines={rectangle, magnification=10, 
        width=1.3cm, height=1.3cm, connect spies}]
        \begin{axis}[
            xlabel={$t$},
            ylabel ={Amplitude},
            width=0.45\textwidth,
            height=5cm,
            xmin=0.019, xmax=0.5,
            % ymin=1e-7, ymax=0.5e-3,
            legend style={nodes={scale=0.75, transform shape}, at={(0.3,0.95)}},
            ticklabel style={font=\footnotesize},
            ymajorgrids=true,
            xmajorgrids=true,
            grid style=dashed,
            grid=both,
            grid style={line width=.1pt, draw=gray!10},
            major grid style={line width=.2pt, draw=gray!30},
        ]
            \addplot[smooth, thin, color=chestnut, line width=0.5pt]
                table[x=t2, y=R2tild] {Envelop_approx2.dat};
            \addplot[smooth, thin, color=airforceblue, line width=0.1pt]
                table[x=t2, y=R2bar] {Envelop_approx2.dat};
            \legend{$\sum_{i=1}^5\overline{R}_{i}$, $\sum_{i=1}^5\hat{\overline{R}}_{i}$}
            \path (axis cs:0.1, 1e+4) coordinate (X); % Adjust as needed
        \end{axis}
               \spy [black] on (X) in node (zoom) [left] at ([xshift=1cm, yshift=1.1cm]X);
    \end{scope}
\end{tikzpicture}
}

\caption{ We observe that by increase of $M$ from $20$ to $72$ the approximation error reduces significantly. 
 }

\label{fig:approximte}
\end{figure*}
%-----------------

%-----------------
\begin{figure*}[!t]
\centering
\subfigure[]{
\begin{tikzpicture}
        \begin{axis}[
            xlabel={$t$},
            ylabel ={Amplitude},
            width=0.45\textwidth,
            height=5cm,
            xmin=0.019, xmax=0.5,
            % ymin=1e-7, ymax=0.5e-3,
            legend style={nodes={scale=0.75, transform shape}, at={(0.3,0.95)}},
            ticklabel style={font=\footnotesize},
            ymajorgrids=true,
            xmajorgrids=true,
            grid style=dashed,
            grid=both,
            grid style={line width=.1pt, draw=gray!10},
            major grid style={line width=.2pt, draw=gray!30},
        ]
            \addplot[smooth, thin, color=chestnut, line width=0.5pt]
                table[x=t, y=R2tild] {Envelop.dat};
            \addplot[smooth, thin, color=airforceblue, line width=0.5pt]
                table[x=t, y=R2bar] {Envelop.dat};
            \legend{$\Tilde{R}_{2}$, $\tilde{R}^{e}_{2}(t)$}
            
        \end{axis}
\end{tikzpicture}
}\subfigure[]{
\begin{tikzpicture}
        \begin{axis}[
            xlabel={$t$},
            ylabel ={Amplitude},
            width=0.45\textwidth,
            height=5cm,
            xmin=0.019, xmax=0.5,
            % ymin=1e-7, ymax=0.5e-3,
            legend style={nodes={scale=0.75, transform shape}, at={(0.3,0.95)}},
            ticklabel style={font=\footnotesize},
            ymajorgrids=true,
            xmajorgrids=true,
            grid style=dashed,
            grid=both,
            grid style={line width=.1pt, draw=gray!10},
            major grid style={line width=.2pt, draw=gray!30},
        ]
            \addplot[smooth, thin, color=chestnut, line width=0.5pt]
                table[x=t, y=R3tild] {Envelop.dat};
            \addplot[smooth, thin, color=airforceblue, line width=0.5pt]
                table[x=t, y=R3bar] {Envelop.dat};
            \legend{$\Tilde{R}_{3}$, $\tilde{R}^{e}_{3}(t)$}
        \end{axis}
\end{tikzpicture}
}

\subfigure[]{
\begin{tikzpicture}
        \begin{axis}[
            xlabel={$t$},
            ylabel ={Amplitude},
            width=0.45\textwidth,
            height=5cm,
            xmin=0.019, xmax=0.5,
            % ymin=1e-7, ymax=0.5e-3,
            legend style={nodes={scale=0.75, transform shape}, at={(0.3,0.95)}},
            ticklabel style={font=\footnotesize},
            ymajorgrids=true,
            xmajorgrids=true,
            grid style=dashed,
            grid=both,
            grid style={line width=.1pt, draw=gray!10},
            major grid style={line width=.2pt, draw=gray!30},
        ]
            \addplot[smooth, thin, color=chestnut, line width=0.5pt]
                table[x=t, y=R4tild] {Envelop.dat};
            \addplot[smooth, thin, color=airforceblue, line width=0.5pt]
                table[x=t, y=R4bar] {Envelop.dat};
            \legend{$\Tilde{R}_{4}$, $\tilde{R}^{e}_{4}(t)$}
        \end{axis}
\end{tikzpicture}
}\subfigure[]{\label{fig:envelop(d)}
\begin{tikzpicture}
        \begin{axis}[
            xlabel={$t$},
            ylabel ={Amplitude},
            width=0.45\textwidth,
            height=5cm,
            xmin=0.019, xmax=0.5,
            % ymin=1e-7, ymax=0.5e-3,
            legend style={nodes={scale=0.75, transform shape}, at={(0.3,0.95)}},
            ticklabel style={font=\footnotesize},
            ymajorgrids=true,
            xmajorgrids=true,
            grid style=dashed,
            grid=both,
            grid style={line width=.1pt, draw=gray!10},
            major grid style={line width=.2pt, draw=gray!30},
        ]
            \addplot[smooth, thin, color=chestnut, line width=0.5pt]
                table[x=t, y=R5tild] {Envelop.dat};
            \addplot[smooth, thin, color=airforceblue, line width=0.5pt]
                table[x=t, y=R5bar] {Envelop.dat};
            \legend{$\Tilde{R}_{5}$, $\tilde{R}^{e}_{5}(t)$}
        \end{axis}
\end{tikzpicture}
}

\caption{ We observe that terms involving $\sin(4M\pi t)$ and $\cos(4M\pi t)$ exhibit higher frequencies compared to terms such as $\sin(M\pi t)$, $\sin(\pi t)$, $\cos(M\pi t)$, and $\cos(\pi t)$ for $M=72$.
 }

\label{fig:envelop}
\end{figure*}
%-----------------
 Then, the sum of $\overline{R}_{i}(t)$ can be written as 
%-----------------
\begin{align}
 \nonumber
   & \sum_{i=1}^5\overline{R}_{i}(t)  =  70M^5\sin(\pi t) - 35M^5\sin(3\pi t) +7M^5\sin(5\pi t)\nonumber\\
    &+24M^4\cos(\pi t) - 36M^4\cos(3\pi t) +12M^4\cos(5\pi t), \\ &  + 
     \mathcal{O}(M^{-2}\max \{\sin (\pi t), \cos (\pi t)\}), \\ \label{eq:trigonometricsin}
     & = M^5 112\sin(\pi t)^5+\nonumber\\
     &192M^4 \big(\cos(\pi t)^5 +\cos(\pi t) -2\cos(\pi t)^3\big)  + \mathcal{O}(M^{-2}\pi t).
\end{align}
%-----------------
Here, the last equality comes from utilizing quintuple and triple angle trigonometric formulas for $\sin(5\pi t)$, $\cos(5\pi t)$, and $\sin(3\pi t)$, $\cos{(3 \pi t)}$. In Figure~\ref{fig:approximte}, we depict $\sum_{i=1}^6\overline{R}_{i}(t)$ and its approximation $\sum_{i=1}^6\hat{\overline{R}}_{i}(t)$ for $M=20$ and $M=72$. Then, by invoking \eqref{eq:trigonometricsin}, we obtain the following upper-bound
%-----------------
\begin{align}
    &|G(t)|  =  \frac{|\sum_{i=1}^5\Tilde{R}_{i}(t)|}{96(M^2-1)M^7\sin^9(\pi t)} \leq \nonumber\\
    &\frac{|\sum_{i=1}^6\hat{\overline{R}}_{i}(t)|}{96(M^2-1)M^7\sin^9(\pi t)} +  \underbrace{\mathcal{O}(M^{-11}(\pi t)^{-8} )}_{\mathcal{O}(M^{-3} )},
\end{align}
%-----------------
where the last inequality holds because each $\hat{\overline{R}}_{i}$ serves as an upper-bound on the magnitude of $\tilde{R}_{i}$ over time, and thus, the envelope sum provides a bound on the overall sum of the signals. Also,  the term $\mathcal{O}(M^{-11}(\pi t)^{-8})$ arises from the denominator’s dependence on $M^7(M^2 - 1)$ and $\sin^9(\pi t)$ (see Figure~\ref{fig:envelop(d)}). For small values of $t$, particularly when $t = 1/M$, the approximation error term attains its maximum due to the $\sin^9(\pi t)$ factor. Nevertheless, even in this case, the order remains at most $\mathcal{O}(M^{-3})$. For a sufficiently large $M$,   the kernel can be approximated as follows:  
%-----------------
 	\begin{align}\label{eq:2UpperSumR} 
 &\Big|\sum_{i=1}^6\overline{R}_{i}(t)\Big| \leq \Big|  M^5112\sin(\pi t)^5+\nonumber\\
 &192M^4 \big(\cos(\pi t)^5 +\cos(\pi t) -2\cos(\pi t)^3\big)    \Big|. 
 	\end{align}}
%-----------------
 	Finally, using the upper-bound in \eqref{eq:2UpperSumR} for $G(t)$ and with Taylor expansions around the origin for $\sin{(\pi t)}$ and $\cos{(\pi t)}$, we obtain a suitable approximation. To obtain a suitable approximation, we use Taylor expansions of the sine and cosine functions in two regions of $t$: near zero (denoted by $\Delta_{\rm near}:= \{t~|~t \leq \sqrt{2}/\pi\}$) and far from the origin (denoted by $\Delta_{\rm far}:= \{t |~\sqrt{2}/\pi < t \leq 0.5\}$).  For $t \in \Delta_{\rm near}$, we use the approximations $\sin(\pi t)\approx \pi t - \frac{\pi^3t^3}{6}$  and $\cos(\pi t) \approx 1$, which leads to the following upper-bound.
    \begin{align}
       |G(t)|\leq  B_{0}^{L}(t) := \frac{7}{6(M^2-1)M^2(\pi t - \pi^3 t^3/6)^4},
    \end{align}
    for $t \in \Delta_{\rm near}$.  Similarly,  the approximations $\sin(\pi t)\approx  \pi t $  and $\cos(\pi t) \approx 0$ are used for $ t\in \Delta_{\rm far}$ to obtain the following upper-bounds. 
     % ----------------
      \begin{align}
  |G(t)|\leq B_{0}^{R}(t) & := \frac{7}{96(M^2-1)M^2t^4},~t \in \Delta_{\rm far}.
 	\end{align} 
     % ---------------
     
     This concludes the proof for the upper-bound on the kernel $G(t)$.  For higher derivatives of the $G(t)$, we employed a similar strategy to find corresponding upper-bounds for $G^{(\ell)}(t)$ for  $\ell =\{1,2,3\}$; these proofs are omitted for brevity. 

     Note that the derivative of the term $b(t) := 1/(\pi t- \pi^3t^3/6)$ is negative for $t \in [0,\sqrt{2}/\pi]$. Furthermore, for $M\geq 71$, we can check that for all $\ell$,  $B_{\ell}^{R}(\sqrt{2}/\pi) < B_{\ell}^{L}(\sqrt{2}/\pi)$, which implies that the upper-bounds $B_{\ell}(t)$ are non-increasing with respect to $t$. Finally,  $B_{\ell}^{L}(\Delta-t') + B_{\ell}^{L}(\Delta+t')$ is increasing with respect to $t'$ because $b(t)$ and $b(t)^2$ are strictly convex for $t>0$, therefore, the derivative of $b(\Delta-t') + b(\Delta+ t')$ with respect to $t'$ is positive for $0\leq t'<\Delta /2$.  
%-----------------       
\begin{figure*}[!t]
\centering
\subfigure[$\Delta_{ \rm near}, M = 72$]{
\begin{tikzpicture}
    \begin{scope}[spy using outlines={rectangle, magnification=10, 
        width=1.8cm, height=1.8cm, connect spies}]
        \begin{axis}[
            xlabel={$t$},
            ylabel ={Absolute value},
            width=0.45\textwidth,
            height=5cm,
            xmin=0.019, xmax=0.35,
            ymin=1e-7, ymax=0.5e-3,
            legend style={nodes={scale=0.75, transform shape}, at={(0.9,0.95)}},
            ticklabel style={font=\footnotesize},
            ymajorgrids=true,
            xmajorgrids=true,
            grid style=dashed,
            grid=both,
            grid style={line width=.1pt, draw=gray!10},
            major grid style={line width=.2pt, draw=gray!30},
        ]
            \addplot[smooth, thin, color=chestnut, line width=0.5pt]
                table[x=t, y=UK0] {Kernel_Upper_arppox.dat};
            %\addplot[smooth, thin, color=airforceblue, line width=0.1pt]
             %   table[x=t, y=UKn0] {Kernel_Upper_arppox.dat};
            \addplot[smooth, thin, dashed, color=cssgreen, line width=1pt]
                table[x=t, y=UBLt] {Kernel_Upper_arppox.dat};
            \legend{$|G(t)|$, $B_0^{L}(t)$}
            
            % Define the coordinate to zoom in
            \path (axis cs:0.105, 8e-6) coordinate (X); % Adjust as needed
        \end{axis}

        % Use the spy command to zoom in on the specified coordinate
        \spy [black] on (X) in node (zoom) [left] at ([xshift=1.5cm, yshift=1.5cm]X);

    \end{scope}
\end{tikzpicture}
}
\subfigure[$\Delta_{ \rm far}, M =72$]{
\begin{tikzpicture}
    % \begin{scope}[spy using outlines={rectangle, magnification=20, 
    %     width=1.8cm, height=1.8cm, connect spies}]
        \begin{axis}[
            xlabel={$t$},
            ylabel ={Absolute value},
            width = 0.45\textwidth,
            height = 5cm,
            xmin=0.45, xmax=0.5,
            ymin=0, ymax=3.5e-7,
            restrict y to domain = 0:3.5e-7,
            legend style={nodes={scale=0.75, transform shape}, at={(0.9,0.95)}},
            ticklabel style={font=\footnotesize},
            ymajorgrids=true,
            xmajorgrids=true,
            grid style=dashed,
            grid=both,
            grid style={line width=.1pt, draw=gray!10},
            major grid style={line width=.2pt, draw=gray!30},
        ]
            \addplot[smooth, thin, color=chestnut, line width=0.5pt]
                table[x=t, y=UK0] {Kernel_Upper_arppox.dat};
            %\addplot[smooth, thin, color=airforceblue, line width=0.1pt]
             %   table[x=t, y=UKn0] {Kernel_Upper_arppox.dat};
            \addplot[smooth, thin, dashed, color=cssgreen, line width=1pt]
                table[x=t, y=UBRt] {Kernel_Upper_arppox.dat};
            \legend{$|G(t)|$, $B_0^{R}(t)$}
            % Define the coordinate to zoom in
            % \path (axis cs:0.035, 3e-4) coordinate (X); % Adjust as needed
        \end{axis}
        % % Use the spy command to zoom in on the specified coordinate
        % \spy [black] on (X) in node (zoom) [left] at ([xshift=3cm, yshift=-1cm]X);
    % \end{scope}
\end{tikzpicture}
}

\subfigure[$\Delta_{ \rm near}, M = 16$]{
\begin{tikzpicture}
    %\begin{scope}[spy using outlines={rectangle, magnification=10, 
     %   width=1.6cm, height=1.6cm, connect spies}]
        \begin{axis}[
            xlabel={$t$},
            ylabel ={Absolute value},
            width=0.45\textwidth,
            height=5cm,
            xmin=0.019, xmax=0.35,
            ymin=1e-7, ymax=3.5e-3,
            legend style={nodes={scale=0.75, transform shape}, at={(0.95,0.95)}},
            ticklabel style={font=\footnotesize},
            ymajorgrids=true,
            xmajorgrids=true,
            grid style=dashed,
            grid=both,
            grid style={line width=.1pt, draw=gray!10},
            major grid style={line width=.2pt, draw=gray!30},
        ]
            \addplot[smooth, thin, color=chestnut, line width=0.5pt]
                table[x=t, y=UK0] {Kernel_Upper_arppox2.dat};
            \addplot[smooth, thin, dashed, color=cssgreen, line width=1pt]
                table[x=t, y=UBLt] {Kernel_Upper_arppox2.dat};
            \legend{$|G(t)|$, $B_0^{L}(t)$}
            
            % Define the coordinate to zoom in
            %\path (axis cs:0.127, 3.5e-4) coordinate (X); % Adjust as needed
        \end{axis}

        % Use the spy command to zoom in on the specified coordinate
        %\spy %[black] on (X) in node (zoom) [left] at ([xshift=3.5cm, yshift=0.75cm]X);

    %\end{scope}
\end{tikzpicture}
}\subfigure[$\Delta_{ \rm far}, M = 16$]{
\begin{tikzpicture}
    % \begin{scope}[spy using outlines={rectangle, magnification=20, 
    %     width=1.8cm, height=1.8cm, connect spies}]
        \begin{axis}[
            xlabel={$t$},
            ylabel ={Absolute value},
            width = 0.45\textwidth,
            height = 5cm,
            xmin=0.45, xmax=0.5,
            ymin=0, ymax=1e-4,
            restrict y to domain = 0:1e-4,
            legend style={nodes={scale=0.75, transform shape}, at={(0.9,0.95)}},
            ticklabel style={font=\footnotesize},
            ymajorgrids=true,
            xmajorgrids=true,
            grid style=dashed,
            grid=both,
            grid style={line width=.1pt, draw=gray!10},
            major grid style={line width=.2pt, draw=gray!30},
        ]
            \addplot[smooth, thin, color=chestnut, line width=0.5pt]
                table[x=t, y=UK0] {Kernel_Upper_arppox2.dat};
            \addplot[smooth, thin, dashed, color=cssgreen, line width=1pt]
                table[x=t, y=UBRt] {Kernel_Upper_arppox2.dat};
            \legend{$|G(t)|$, $B_0^{R}(t)$}
            % Define the coordinate to zoom in
            % \path (axis cs:0.035, 3e-4) coordinate (X); % Adjust as needed
        \end{axis}
        % % Use the spy command to zoom in on the specified coordinate
        % \spy [black] on (X) in node (zoom) [left] at ([xshift=3cm, yshift=-1cm]X);
    % \end{scope}
\end{tikzpicture}
}

\caption{Comparison between the squared Fejér kernel  $G(t)$. The dashed line represents the bound  $B_0(t)$. The magnified region, zoomed in $20$ times, highlights the detailed behavior of the functions.  }

\label{fig:Kernel_approx}
\end{figure*}
%-----------------
\subsection{Proof of Lemma~\ref{lem:UpDmatrix}}\label{sec:proof_UpDmatrix}

 	Let $t_{+}\leq 2\Delta_{\min}$ be the first positive element in $\mathcal{T}^{m}$ closest to $0$. Then, for the sum of the kernel over all positive $t_i \in \mathcal{T}^{m}$, we have 
 	\begin{align}
 	\nonumber
 		 \sum_{t_i \in \mathcal{T}^{m} :0<t_i\leq \tfrac{1}{2} } &|G^{(\ell)}(t-t_i)|  =  |G^{(\ell)}(t-t_{+})| + \\
   & \sum_{t_i \in \mathcal{T}^{m} \backslash \{t_{+}\} :0<t_i\leq \tfrac{1}{2} } |G^{(\ell)}(t-t_i)|.  \label{eq:FirstUpper}
 	 \end{align}
          The assumptions $\Delta_{\min} = \frac{1}{M}$ and $M\geq 72$ imply that $32\Delta_{\min}<\frac{\sqrt{2}}{\pi}$. Then,  we use the upper-bounds in Lemma \ref{lem:UpKernel} to obtain an upper-bound on the second term on the right-hand side as follows. 
 	\begin{align*}
 		 \sum_{j=2}^{32}B_{\ell}(j\Delta_{\min}-t) + \frac{\pi^{\ell}H^{\infty}_{\ell}}{96(M^2-1)M^{2-\ell}}\sum_{j=32}^{\infty}\frac{1}{(j\Delta_{\min}-t)^{4}}.
 	\end{align*}
	 The second term can be bounded as 
 	\begin{align*}
 	\sum_{j=33}^{\infty}\frac{1}{(j\Delta_{\min}-t)^{4}} \leq \sum_{j=33}^{\infty}\frac{1}{(j\Delta_{\min})^{4}}  = \frac{1}{\Delta_{\min}^4}\sum_{j=33}^{\infty}\frac{1}{j^4}, 
 	\end{align*}
    where the last summation is lower than $8.98\times10^{-5}$. Additionally, for the first term in \eqref{eq:FirstUpper}, we have the following upper-bound 
 	\begin{align*}
 	% \label{eq:K+Upper}
 		|G^{(\ell)}(t-t_{+})|  \leq \begin{cases} 
   \underset{\Delta\leq t_{+}\leq 3\Delta_{\min}}{\max}|G^{(\ell)}(t-t_{+})|,    t_{+} \leq  3\Delta_{\min},  \\
 		B_{\ell}(3\Delta_{\min}-t),~~t_{+} > 3\Delta_{\min}.
 		\end{cases}
 	\end{align*}
 	Hence, by considering $C_{\ell} = \frac{\pi^{\ell}H^{\infty}_{\ell}}{96(M^2-1)M^{2-\ell}} \gamma$ the kernel $G^{\ell}(t)$ is bounded by $\mathfrak{G}^{+}_{\ell}(\Delta, t)  +  C_{\ell}$. With similar arguments, we can see that the sum over negative $t_i \in \mathcal{T}^{m}$ would be bounded by $\mathfrak{G}^{-}_{\ell}(\Delta, t)  +  C_{\ell}$.
    To  show that  $\mathfrak{G}^{+}_{\ell}(\Delta, t)$ is  non-increasing with respect to $\Delta$, we need to show $\mathfrak{B}^{+}(\Delta, t)$ is non-increasing in terms of $\Delta$. In this regard, we follow the same arguments as stated in the proof of \cite[Lemma 2.7]{candes2014towards} by rewriting the $\mathfrak{B}^{+}(\Delta, t)$ as follows.     
    \begin{align*}
        \max \big\{\max_{\Delta_{\rm min}-t \leq \mu \leq 3\Delta_{\min}-t} \hspace{-10pt}|G^{(\ell)}(\mu)|,B_{\ell}(3\Delta_{\min}-t)\big\}
    \end{align*}
    Therefore, invoking Lemma~\ref{lem:UpKernel}, we have that
       \begin{align*}
    B_{\ell}(3\Delta_{\rm min} - t') \geq \begin{cases}
        B_{\ell}(3\Delta_{\rm min}-t), \\
        |G(\mu)|, \quad \mu \geq 3 \Delta_{\rm min} -t'. 
    \end{cases}
    \end{align*}
    Consequently, this yields the following inequality for  $t' > t$. 
           \begin{align*}
    \max_{\Delta_{\rm min}-t' \leq \mu \leq 3 \Delta_{\rm min} - t'} \Big | G^{\ell}(\mu) \Big| \geq    \max_{\Delta_{\rm min}-t \leq \mu \leq 3 \Delta_{\rm min} - t'} \Big | G^{\ell}(\mu) \Big|.
    \end{align*}
     Note that based on Lemma \ref{lem:UpKernel}, for $j\leq 32$ with $32\Delta_{\min}\leq \sqrt{2}/\pi$,  we know that $B_{\ell}(j\Delta -t) + B_{\ell}(j\Delta +t)$ is increasing in terms of $t$. Therefore, $\mathfrak{G}^{+}_{\ell}(\Delta, t)$ is generally non-increasing with respect to $\Delta$, allowing us to set $\Delta = \Delta_{\min}$. This completes the proof.

 \ifCLASSOPTIONcaptionsoff
  \newpage
\fi

\bibliographystyle{ieeetr}
\bibliography{Ref}
\begin{IEEEbiographynophoto}{Sajad~Daei}%
(Member, IEEE) received the B.Sc.\ degree in Electronic Engineering from
Guilan University, Iran, in 2011, the M.Sc.\ degree in Telecommunications Engineering
from Sharif University of Technology, Tehran, in 2013, and the Ph.D.\ degree in
Telecommunications Engineering from Iran University of Science and Technology,
Tehran, in 2020.  
From 2020 to 2021 he was with the Electronics Research Institute of Sharif University of Technology, Tehran, and
subsequently a Post-Doctoral Researcher at EURECOM, Sophia Antipolis, working
on semantic and goal-oriented signal processing.  
He is currently a Post-Doctoral Researcher with KTH Royal Institute of
Technology, Stockholm, focusing on integrated sensing and communications.  
Dr.~Daei received the 2020 IEEE Iran Section Outstanding Ph.D.\ Thesis Award.
His interests include inverse problems, compressed sensing, blind
deconvolution, super-resolution, optimization, massive MIMO and millimeter wave communications, random access, and
goal-oriented inverse problems.
\end{IEEEbiographynophoto}

\begin{IEEEbiographynophoto}{Saeed~Razavikia}
(Member, IEEE) is currently
pursuing the Ph.D. degree with the School of Electrical Engineering and Computer Science, KTH
Royal Institute of Technology, funded by the Wallenberg AI, Autonomous Systems and Software
Program (WASP). His current research interests
include optimization, the theory of machine learning, and statistical signal processing. He received
several awards and scholarships, such as the Scholarship from the Hans Werthén Foundation, the Royal
Academy of Engineering Sciences, and the IEEE
Sweden VT-COM-IT Joint Chapter Best Student Journal Paper Award in 2024
\end{IEEEbiographynophoto}

\begin{IEEEbiographynophoto}{ Mikael~Skoglund}
    (S'93-M'97-SM'04-F'19) received the Ph.D. degree in
1997 from Chalmers University of Technology, Sweden.  In 1997, he
joined the Royal Institute of Technology (KTH), Stockholm, Sweden,
where he was appointed to the Chair in Communication Theory in 2003.
At KTH he heads the Division of Information Science and Engineering.
Dr. Skoglund has worked on problems in source-channel coding, coding
and transmission for wireless communications, Shannon theory,
information-theoretic security, information theory for statistics and
learning, information and control, and signal processing. He has
authored and co-authored more than 200 journal and 450 conference
papers.
Dr. Skoglund is a Fellow of the IEEE. During 2003--08 he was an
associate editor for the IEEE Transactions on Communications.  In the
interval 2008--12 he was on the editorial board for the IEEE
Transactions on Information Theory and starting in the Fall of 2021 he
he joined it once again. He has served on numerous technical program
committees for IEEE sponsored conferences, he was general co-chair for
IEEE ITW 2019 and TPC co-chair for IEEE ISIT 2022. He is an elected
member of the IEEE Information Theory Society Board of Governors.
\end{IEEEbiographynophoto}

\begin{IEEEbiographynophoto}{ Gabor Fodor}
(IEEE Fellow ‘25) received the Ph.D. degree in electrical engineering from the Budapest University of Technology and Economics in 1998 and the D.Sc. degree from the Hungarian Academy of Sciences (Doctor of MTA) in 2019. He is currently a Master Researcher at Ericsson Research and a Docent and an Adjunct Professor at the KTH Royal Institute of Technology, Stockholm, Sweden. He has authored or coauthored more than 200 refereed journal articles and conference papers and seven book chapters and holds more than 250 European and U.S. granted patents. He was a co-recipient of the IEEE Communications Society Stephen O. Rice Prize in 2018 and the Best Student Conference Paper Award from the IEEE Sweden VT/COM/IT Chapter in 2018. From 2017 to 2020, he was also a member of the Board of the IEEE Sweden joint Communications, Information Theory and Vehicle Technology Chapter. He served as Editor for IEEE TRANSACTIONS ON WIRELESS COMMUNICATIONS between 2017 and 2022. He is currently serving as an Associate Editor-in-Chief for IEEE COMMUNICATIONS MAGAZINE and Editor for IEEE WIRELESS COMMUNICATIONS. Dr. Fodor is a Fellow of the IEEE and the Asia-Pacific Artificial Intelligence Association (AAIA). 
\end{IEEEbiographynophoto}

\begin{IEEEbiographynophoto}{Carlo Fischione} (IEEE Fellow) received
the Laurea degree (summa cum Laude) in electronic engineering and the Ph.D. degree in electrical
and information engineering from the University of
L’Aquila, Italy, in April 2001 and May 2005, respectively. He is currently a Full Professor of electrical
engineering and computer science with the Division
of Network and Systems Engineering (NSE), KTH
Royal Institute of Technology, Stockholm, Sweden.
He is also the Director of the KTH-Ericsson Data
Science Micro Degree Program directed to Ericsson
globally, the Director of undergraduate education at NSE, the Chair of the
IEEE Machine Learning for Communications Emerging Technologies Initiative, and the Founding General Chair of the IEEE International Conference
on Machine Learning for Communications and Networking—IEEE ICMLCN
2024. He is a Distinguished Lecturer of the IEEE Communication Society.
He received the Starting Grant of the Swedish Research Council in 2008.
He has held research positions at Massachusetts Institute of Technology,
Cambridge, MA, USA, in 2015, as a Visiting Professor; Harvard University,
Cambridge, in 2015, as an Associate Professor; and the University of
California at Berkeley, CA, USA, from 2004 to 2005, as a Visiting Scholar
and a Research Associate from 2007 to 2008. He is also an Honorary
Professor with the Department of Mathematics, Information Engineering, and
Computer Science, University of L’Aquila, Italy. He is also the Co-Founder
and the Scientific Advisor of ELK.Audio. His research interests include
applied optimization, wireless Internet of Things, and machine learning.
He is an Ordinary Member of Italian Academy of History Deputazione
Abruzzese di Storia Patria (DASP). He received several awards, such as
the “IEEE Communication Society S. O. Rice” Award for the Best IEEE
TRANSACTIONS ON COMMUNICATIONS Paper of 2018 and the Best Paper
Award of IEEE TRANSACTIONS ON INDUSTRIAL INFORMATICS in 2007.
He is an Editor of IEEE TRANSACTIONS ON COMMUNICATIONS (Machine
Learning for Communications area) and IEEE TRANSACTIONS ON MACHINE
LEARNING IN COMMUNICATIONS AND NETWORKING. He has served as an
Associated Editor for Automatica (IFAC) from 2014 to 2019.
\end{IEEEbiographynophoto}
\end{document}